\documentclass[a4paper,fleqn,usenatbib]{mnras}

\usepackage{newtxtext,newtxmath}

\usepackage[T1]{fontenc}
\usepackage{ae,aecompl}

\usepackage{graphicx}	
\usepackage{amsmath}	
\usepackage{amssymb}	
\usepackage{color}
\usepackage{comment}
\usepackage{threeparttable}
\usepackage{subfig}

\title[Hydrodynamic convection in accretion discs]{Hydrodynamic convection in accretion discs}

\author[L.E. Held et al.]{
Loren E. Held,$^{1}$\thanks{E-mail: leh50@cam.ac.uk (LEH)}
Henrik N. Latter$^{1}$
\\
$^{1}$Department of Applied Mathematics and Theoretical Physics, University of Cambridge, Centre for Mathematical Sciences,\\
 Wilberforce Road, Cambridge CB3 0WA, UK\\
}

\date{Accepted XXX. Received YYY; in original form ZZZ}

\pubyear{2018}

\begin{document}
\label{firstpage}
\pagerange{\pageref{firstpage}--\pageref{lastpage}}
\maketitle

\begin{abstract}
The prevalence and consequences of convection perpendicular to the
plane of accretion discs have been discussed for several
decades.
Recent simulations combining
convection and the magnetorotational instability have given fresh
impetus to the
debate, as the interplay of the two processes can enhance angular
momentum transport, at least in the optically thick outburst
stage of dwarf novae. In this paper we seek to isolate and understand the most
generic features of disc convection, and so undertake its study in purely hydrodynamical models. First, we
investigate the linear phase of the instability, obtaining estimates
of the growth rates both semi-analytically, using one-dimensional
spectral computations, as well as analytically, using WKBJ methods.
Next we perform three-dimensional, vertically stratified,
shearing box simulations with the conservative, finite-volume code
\textsc{PLUTO}, both with and without explicit diffusion
coefficients. We find that hydrodynamic convection can, in general,
drive outward angular momentum transport, a result that we confirm
with \textsc{ATHENA}, an alternative
finite-volume code. Moreover, we establish that the sign of the
angular momentum flux is sensitive to the diffusivity of the numerical
scheme. Finally, in sustained convection, whereby the system is continuously
forced
to an unstable state, we observe the formation of
various coherent structures,  
including large-scale and oscillatory convective cells, zonal
flows, and small vortices.\end{abstract}




\section{Introduction}
\label{Introduction}

Thermal convection -- the bulk motion of fluid due to an entropy
gradient -- transports heat, mass, and angular
momentum. There is compelling analytical, numerical, experimental, and
observational evidence for convection both in the inner regions of
stars and in the outer core and mantle of the Earth
\citep{spruit1990solar, aurnou2015rotating}, but its application to
astrophysical discs is less clear. In particular, convection
perpendicular to the plane of a disc bears marked
differences to convection in most stars and planets. For example, the
gravitational acceleration reverses sign at the disc midplane, and the
gas supports a strong background shear flow. A third difference
concerns the heat source necessary to drive the unstable entropy
gradient. In stars this heat is provided by nuclear fusion, 
while in the Earth's outer core convection is
driven by both thermal and compositional gradients. In discs
heating at the mid-plane is normally assumed to be
supplied by accretion (and the associated viscous dissipation of
heat). Turbulence arising from the magnetorotational 
 instability 
and dissipation of large-scale density waves are two
mechanisms that might convert orbital energy into the required heat. 
Note importantly that the fluid motions associated with
these heating processes may in fact impede the onset of convection, and at the
very least interact with it in non-trivial ways.

It has long been speculated that convection perpendicular to the plane
of the disc influences the optically thick, high-accreting outburst
phase of dwarf novae (for a succinct review, see
\cite{2011cannizzo}). More recently, simulations of magnetorotational
turbulence in stratified discs have shown that the MRI is capable of
generating a convectively unstable entropy gradient. Moreover,
three-dimensional, shearing box simulations with
zero-net vertical magnetic flux reveal an interplay between convection
and the MRI that might enhance angular momentum transport, typically
quantified by the dimensionless parameter $\alpha$
\citep{bodo2013fully, hirose2014convection}. A similar interplay
between magnetorotational instability and convection might operate in
the ionized inner regions of  protoplanetary discs, or during FU Orionis outbursts \citep{bell427jun, hirose2015magnetic}.

In addition to dwarf novae and the inner regions of protoplanetary
discs, hydrodynamic convection might drive activity in the weakly
ionized `dead zones' of protoplanetary discs where the
magnetorotational instability is unlikely to operate
\citep{armitage2011dynamics}. Although irradiation by the central star
usually results in an isothermal vertical profile in protoplanetary
discs \citep{chiang1997spectral, d1998accretion}, heating by spiral
density waves driven by a planet might nevertheless generate regions 
exhibiting a convectively unstable background equilibrium
\citep{boley2006hydraulic, lyra2016shocks}. So too might the heating
from the dissipation of the strong zonal magnetic fields associated with
the Hall effect (Lesur, Kunz and Fromang 2014).

As mentioned, accretion may instigate convection. But
convection might drive accretion itself. Motivated by
considerations of the primordial solar nebula, \cite{lin1980structure}
constructed a simple hydrodynamic model of a cooling young protoplanetary disk
contracting towards the
mid-plane. They showed that vertical convection
arises if the opacity increases sufficiently fast with temperature, with the
heating source initially provided
by the gravitational contraction. But if convection is modeled
via a mixing length theory, with an effective viscosity, the process
might \emph{self-sustain}:
convective eddies could extract energy from the background orbital shear, and viscous
dissipation of that energy might replace the initial
 gravitational contraction as a source of heat for maintaining convection.

In order to go beyond mixing length theory, \cite{ruden1988axisymmetric} analysed  linear axisymmetric modes in a thin, polytropic disc in the shearing box approximation and estimated that mixing of gas within convective eddies could result in values of $\alpha \sim 10^{-3}-10^{-2}$. They cautioned, however, that axisymmetric convective cells could not by themselves exchange angular momentum: such an exchange would have to be facilitated either by non-axisymmetric modes, or by viscous dissipation of axisymmetric modes.

Following these early investigations, a debate ensued that centered
not so much on the size of $\alpha$ but rather on its
sign. Dissipation of \textit{non-linear axisymmetric} convective cells
was investigated by \cite{kley1993angular} who performed
quasi-global simulations (spanning about
100 stellar radii) of axisymmetric viscous discs with
radiative heating and cooling. They measured an inward flux of angular
momentum, but warned that this might be due to the imposed axial
symmetry and to their relatively high viscosity. Convective shearing
waves were first investigated by \cite{ryu1992convective}, who 
concluded that linear \emph{non-axisymmetric} perturbations would result in a net \textit{inward} angular momentum flux at sufficiently large time. 
These results were questioned however by \cite{lin1993nonaxisymmetric}, who examined analytically and 
numerically a set of \textit{localised} linear non-axisymmetric disturbances in
global geometry. They demonstrated that these modes could transport angular momentum outwards in some cases, and opined 
that the inward transport of angular momentum reported by \cite{ryu1992convective} was an artifact of the shearing box approximation.

Interest in hydro convection waned after local non-linear 3D
compressible simulations showed that it resulted in \textit{inward}
rather than outward angular momentum transport. Using the
finite-difference code \textsc{ZEUS} and rigid, isothermal vertical
boundaries, Stone and Balbus (1996, heareafter SB96) initialized inviscid, fully compressible, and vertically stratified shearing box simulations with a convectively unstable vertical temperature profile, and measured a time-averaged value of $\alpha \sim -4.2\times10^{-5}$. Inward angular momentum transport was also reported by \cite{cabot1996numerical} who ran simulations similar to those of SB96 but included full radiative transfer and a relatively high explicit viscosity. Analytical arguments for the inward transport of angular momentum by convection were presented by SB96 which, crucially, assumed axisymmetry, especially in the pressure field. However, in a rarely cited paper \cite{klahr1999azimuthal} presented fully compressible, three-dimensional, global simulations including explicit viscosity and radiative transfer of hydrodynamic convection in discs that gave some indication that non-linear convection in discs actually assumes non-axisymmetric patterns. 

Some fifteen years later, the claims of SB96 and of \cite{cabot1996numerical} were called into question: 
fully local shearing box simulations of Boussinesq hydrodynamic convection in discs using the spectral code \textsc{SNOOPY} and employing explicit diffusion coefficients  indicated that the sign of angular momentum transport due to vertical convection can be reversed provided the Rayleigh number (the ratio of buoyancy to kinematic viscosity and thermal diffusivity) is sufficiently large. It would appear then
that vertical convection can possibly drive outward angular momentum transport after all \citep{lesur2010angular}. 

Our aim in this paper is to explore the competing results and claims
concerning hydrodynamic convection in accretion discs, in particular
to determine the sign and magnitude of angular momentum transport. We
also isolate and characterize other generic features of convection
that should be shared by multiple disc classes (dwarf novae,
protoplanetary disks, etc) and by different driving mechanisms (MRI
turbulence, spiral shock heating, etc). We do so both analytically and
through numerical simulations, working in the fully compressible, vertically stratified shearing box approximation. We restrict ourselves to an idealised hydrodynamic set-up, omitting magnetic fields and complicated opacity transitions, and in so doing ensure our results are as general as possible.

The structure of the paper is as follows: first, in  Section
\ref{methods} we introduce the basic equations and provide a brief
overview of the numerical code, the numerical parameters and set-up,
and our main diagnostics. In Section \ref {eigensolver} we investigate
the linear behavior of the convective instability, employing both
analytical WKBJ and semi-analytical spectral methods to calculate the
growth rates and the eigenfunctions. In Section \ref{unforcedsims} we
explore the non-linear regime through \textit{unforced} simulations,
in which the convection is not sustained and is permitted to decay
after non-linear saturation. Because this unforced convection is a
transient phenomenon which might depend on the initial conditions, in
Section \ref{forcedsims} we explore the non-linear regime through
simulations of \textit{forced} convection, in which the internal
energy is relaxed to its initial, convectively unstable, state to
mimic, possibly, the action of background MRI turbulent dissipation
and optically thick radiative cooling. Finally, Section
\ref{conclusions} summarizes and discusses the results.

\section{Methods}
\label{methods}

\subsection{Governing equations}
\label{governingequations}
We work in the shearing box approximation
\citep{goldreichlyndenbell1965, hawley1995local, latter2017local},
which treats
a local region of a disc as a Cartesian box located at some fiducial
radius $r = r_0$ and orbiting with the angular frequency of the disc
at that radius $\Omega_0 \equiv \Omega(r_0)$. A point in the box has
Cartesian coordinates $(x, y, z)$ which are related to the cylindrical
coordinates $(r, \phi, z)$ through $x = r - r_0$, $y = r_0(\phi -
\phi_0 - \Omega_0 t)$ and $z  = z$. The equations of gas dynamics are now
\begin{equation}
\partial_t \rho + \nabla \cdot (\rho \mathbf{u}) = 0,
\label{SB1}
\end{equation}
\begin{equation}
\partial_t \mathbf{u} + \mathbf{u}\cdot\nabla \mathbf{u} = -\frac{1}{\rho} \nabla P - 2\Omega_0 \mathbf{e}_z \times \mathbf{u} + \mathbf{g}_{\text{eff}} + \frac{1}{\rho}\nabla \cdot \mathbf{T},
\label{SB2}
\end{equation}
\begin{equation}
\partial_t (\rho e) +  \mathbf{u}\cdot \nabla (\rho e) = -\gamma \rho e \nabla\cdot\mathbf{u} + \mathbf{T}:\nabla \mathbf{u} +\xi \nabla^2 T + \Lambda_\text{relax} ,
\label{SB3}
\end{equation}
corresponding to conservation of mass, momentum and thermal energy,
respectively. These equations are closed with a perfect gas equation
of state $P = (\mathcal{R}/\mu)\rho T$, or, equivalently, $P = e
(\gamma - 1) \rho$ in terms of the specific internal energy $e$. Here
$P$ is the thermal pressure of the fluid, $T$ is the temperature,
$\rho$ is mass density,  $\mathcal{R}$ is the gas constant, and $\mu$
is the mean molecular weight. The adiabatic index (ratio of specific
heats) is denoted by $\gamma$ and is typically taken to be $\gamma =
5/3$ in all our simulations.\footnote{Technically $\gamma = 5/3$ is
  valid for \textit{ionized} hydrogen. Since all our simulations are
  hydrodynamic it might be more appropriate to use $\gamma = 7/5$,
  which is valid for molecular hydrogen, but the differences are small
  and do not affect our results.} The effective gravitational potential is embodied
in the tidal acceleration
$\mathbf{g}_{\text{eff}} = 2q\Omega_0^2x\hat{\mathbf{x}} -
\Omega_0^2z\hat{\mathbf{z}}$, where $q$ is the shear rate. For Keplerian
disks $q=3/2$ a value we adopt throughout the paper.

Other terms in the equations include the viscous stress tensor
$\mathbf{T} \equiv 2\rho \nu \mathbf{S}$. Here $\nu$ is the kinematic
viscosity, and $\mathbf{S} \equiv (1/2)[\nabla \mathbf{u} + (\nabla
\mathbf{u})^\text{T}] - (1/3)(\nabla\cdot\mathbf{u})\mathbf{I}$ is the
traceless shear tensor. The thermal conductivity is $\xi$, but
in our simulations we specify thermal diffusivity $\chi$ rather than
$\xi$. The former is related to the latter via $\chi = \xi / (c_p \rho)$ where $c_p$ is the specific heat capacity at constant pressure.

In some of our simulations we include heating and cooling through a \textit{thermal relaxation} term $\Lambda_\text{relax} \equiv (\rho e -\rho_0 e_0)/\tau_\text{relax}$. This relaxes the thermal energy in each cell back to its initial state $e_0(z) \equiv e(z,\,t=0)$ on a time-scale given by $\tau_\text{relax}$. We refer to simulations run with $\Lambda_\text{relax} \neq 0$ as simulations of \textit{forced} compressible convection. Conversely, simulations with $\Lambda_\text{relax} = 0$ are referred to as \textit{unforced}. The thermal relaxation time-scale $\tau_\text{relax}$ is taken to be the inverse of the linear growth rate of the convective instability.

\subsection{Important parameters and instability criteria}
\label{diffusioncoefficientsdefinitions}
The onset of thermal convection is controlled by the local \textit{buoyancy frequency}, defined as 
\begin{equation}
N^2_B = z \left[\frac{1}{\gamma} \frac{\partial \ln{P}}{\partial z} - \frac{\partial \ln{\rho}}{\partial z}\right]\Omega_0^2.
\end{equation}
But convection is opposed by the effects of viscosity and thermal
diffusivity, with the ratio of destabilising and stabilising processes quantified by the \textit{Rayleigh number}
\begin{equation}
\text{Ra} \equiv \frac{N_{B}^2 H^4}{\nu \chi},
\end{equation}
where $H$ is the disk's scaleheight (formally defined later).
On the other hand, the ratio of buoyancy to shear is expressed through the \textit{Richardson number}
\begin{equation}
\text{Ri} \equiv \frac{N_{B}^2}{q^2\Omega_0^2}.
\end{equation}
 But for convection perpendicular to the plane of an accretion
 disc, the buoyancy force is perpendicular to the direction of the
 background shear. Thus the Richardson number in our set-up is more an
 expression for the scaled intensity of $N^2_B$.
Finally, the \textit{Prandtl number} is defined as the ratio of kinematic viscosity to thermal diffusivity
\begin{equation}
\text{Pr} \equiv \frac{\nu}{\chi}.
\end{equation}

 When the explicit diffusion coefficients are neglected ($\nu, \chi =
 0$), convective instability occurs when the buoyancy frequency is
 negative (i.e. in regions where $N_B^2 < 0$). When explicit viscosity and thermal
 diffusivity are included, however, convective instability requires
 both that $N_B^2 < 0$, \textit{and} that the Rayleigh number exceed
 some critical value $\text{Ra}_c$. Though in real astrophysical disks
 the viscosity is negligible, the inclusion of
 magnetic fields complicates the stability criterion, 
 as does the presence of pre-existing turbulence (as might
 be supplied by the MRI) which itself may diffuse momentum and
 heat. In the latter case it may be possible to define a turbulent
 `Rayleigh number',  which must be sufficiently large so that convection
resists the disordered background flow. In any case, 
the sign of $N_B^2$ is certainly insufficient to assign convection
to MRI-turbulent flows, as is often done in recent work. 
 
\subsection{Numerical set-up}
\label{numericalsetup}

\subsubsection{Codes}
\label{PLUTO}
Most of our simulations are run with the conservative, finite-volume
code \textsc{PLUTO} \citep{mignone2007pluto}. Unless stated otherwise, we employ the
Roe Riemann-solver, 3rd order in space WENO interpolation, and the 3rd
order in space Runge-Kutta algorithm. Other
configurations are explored in Section 4. 
To allow for longer time-steps,
we employ the \textsc{FARGO} scheme
\citep{mignone2012conservative}.
When explicit viscosity $\nu$ and thermal diffusivity $\chi$ are
included, we further reduce the 
computational time by using the Super-Time-Stepping (STS) scheme \citep{alexiades1996super}.
Note that due to the code's
conservative form kinetic energy is not lost to the grid but converted
directly into thermal energy.
Ghost zones are used to implement the boundary conditions. 

We use the built-in shearing box module in \textsc{PLUTO}
\citep{mignone2012conservative}. Rather than solving Equations \eqref{SB1}-\eqref{SB3} (primitive
form), \textsc{PLUTO} solves the governing equations in \textit{conservative form}, evolving the total energy equation rather than the thermal energy equation, where the total energy density of the fluid is given by $E = (1/2)\rho u^2 + \rho e$ (kinetic + internal). In \textsc{PLUTO} the thermal relaxation term $\Lambda_\text{relax} = (\rho e - \rho_0 e_0)/\tau_\text{relax}$ is not implemented directly in the total energy equation. Instead, it is included on the right-hand-side of the thermal energy equation, which is then integrated (in time) analytically.

Finally, we have verified the results of our fiducial simulation (presented in Section \ref{fiducialsimulations}) with the conservative finite-volume code \textsc{ATHENA} \citep{stone2008athena}.

\begin{figure*}
\centering
\captionsetup[subfigure]{labelformat=empty}
\subfloat[]{\includegraphics[scale=0.12]{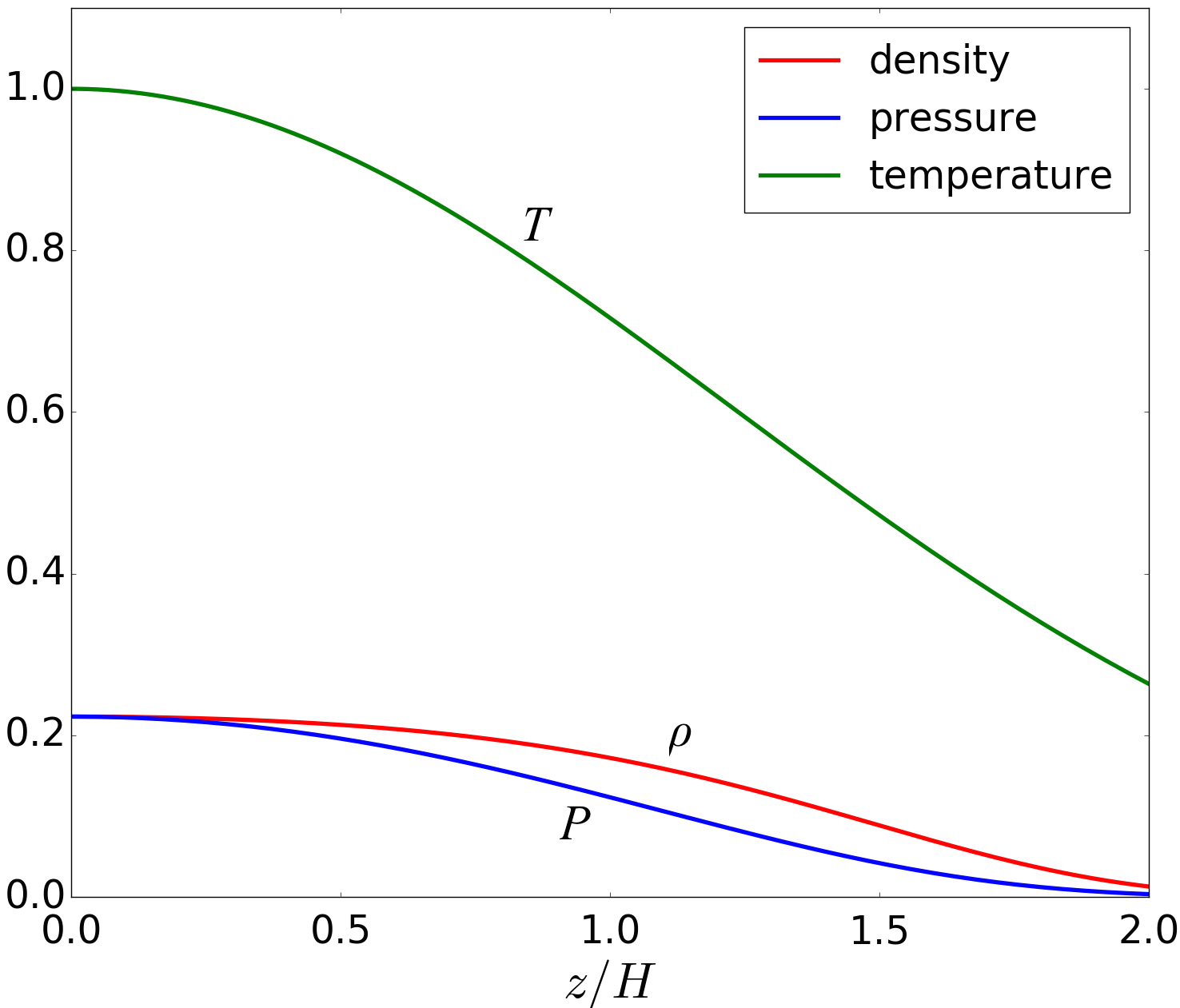}}\hspace{1.0em}
\subfloat[]{\includegraphics[scale=0.12]{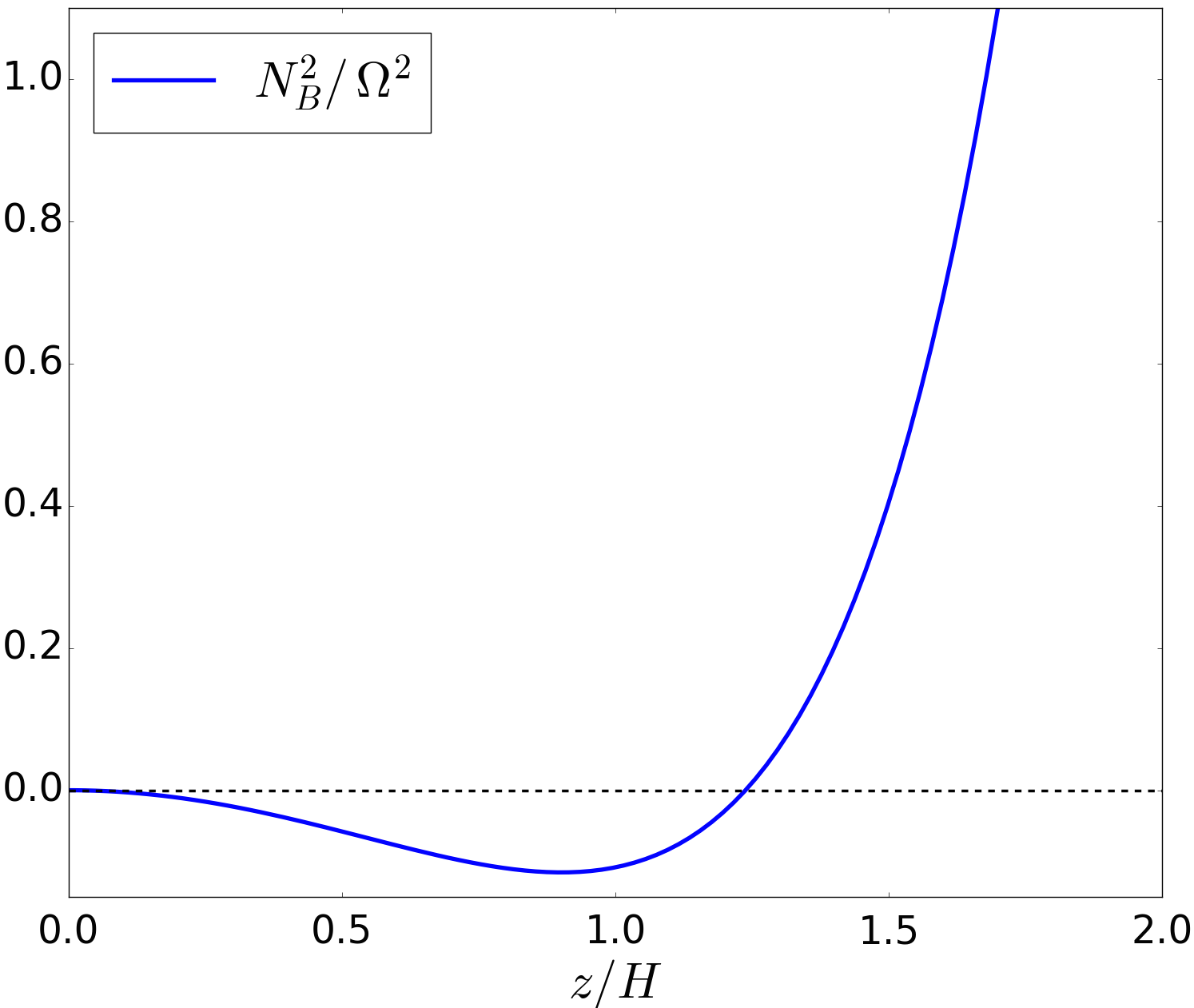}}
\caption{Vertical disc structure for Gaussian temperature profile in a disc of height $2H$ about the mid plane. \textit{Left}: vertical profiles for density $\rho$, pressure $P$, and temperature $T$. \textit{Right}: vertical profile of the buoyancy frequency $N_B^2/\,\Omega^2$. For clarity, only half of the vertical domain is shown. The profile parameters are $\{T_0 = 1.0, \rho_0 = 1.0, \beta = 3.0\}$ and the adiabatic index is $\gamma = 5/3$.}
\label{Gaussianprofileverticalstructure1}
\end{figure*}

\begin{figure*}
	\centering
   \subfloat[]{\includegraphics[scale=0.12]{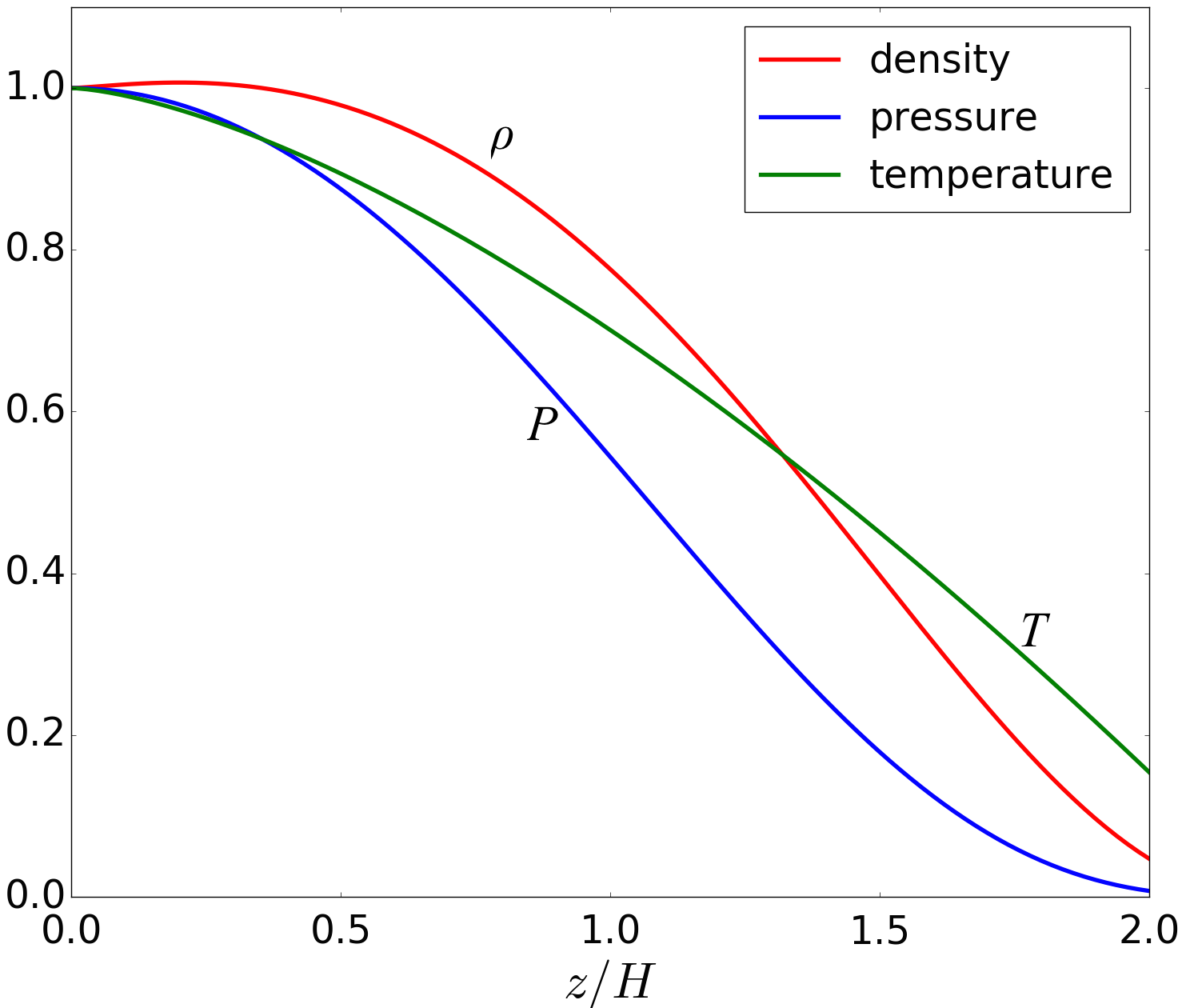}}\hspace{1.0em}
\subfloat[]{\includegraphics[scale=0.12]{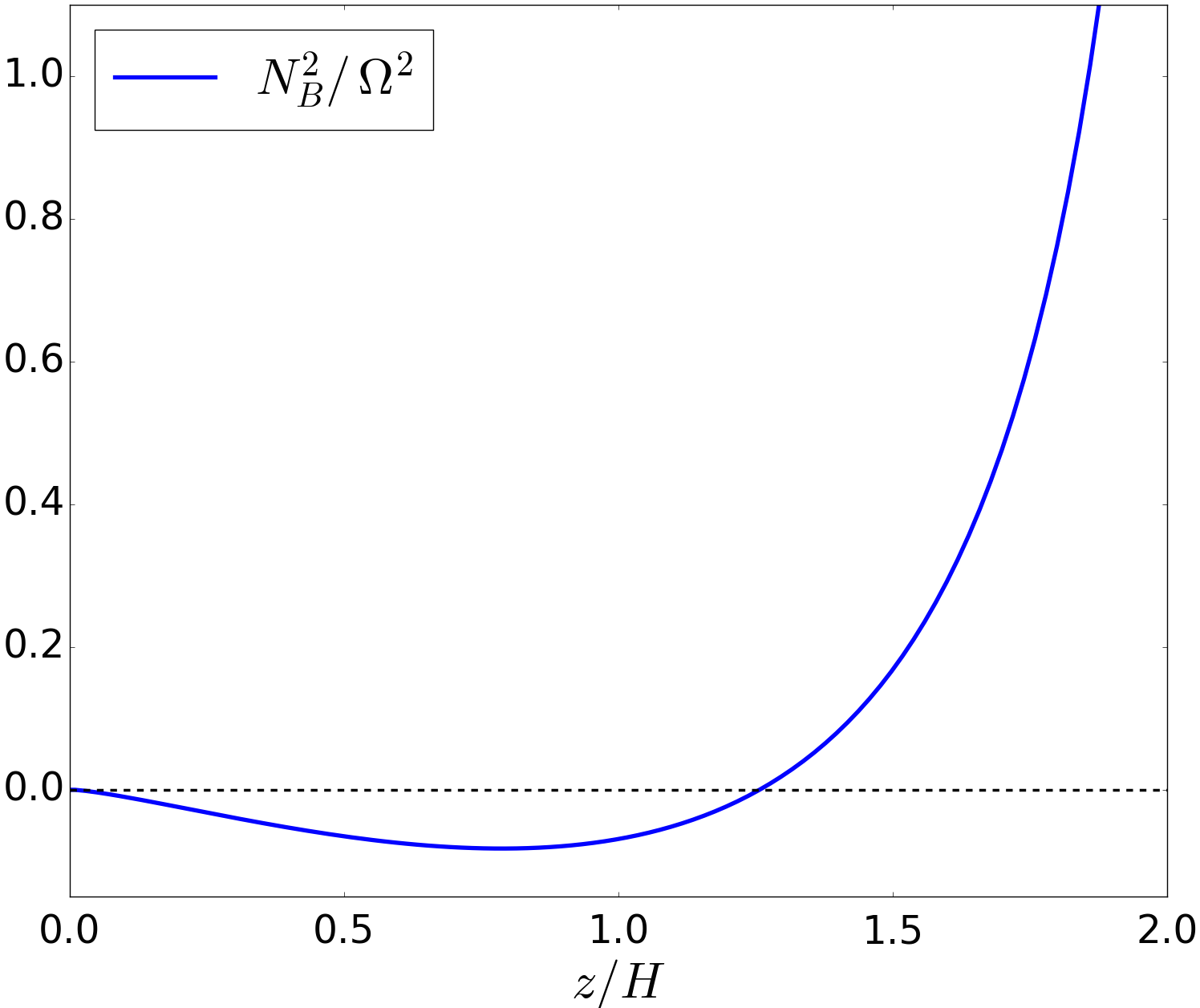}}
    \caption{Vertical disc structure of Stone and Balbus (1996) in a disc of height $2H$ about the mid-plane. \textit{Left}: vertical profiles for density $\rho$, pressure $P$, and temperature $T$. \textit{Right}: vertical profile of the buoyancy frequency $N_B^2/\,\Omega^2$. For clarity, only half of the vertical domain is shown. The profile parameters are $\{T_0 = 1.0, \rho_0 = 1.0, g = 5.0\}$ and the adiabatic index is $\gamma = 5/3$.}
    \label{SB96verticalprofile}
\end{figure*}

\subsubsection{Initial conditions and units}
\label{initial conditions}
We use two different convectively unstable profiles to initialize our
simulations. Nearly all employ an equilibrium exhibiting a Gaussian
temperature profile 
\begin{equation}\label{Tgauss}
 T = T_0 \text{exp}\left[-\frac{z^2}{\beta H^2}\right],
\end{equation}
where $T_0$ is midplane temperature and $\beta$ is a dimensionless
tuning parameter. See Figure
\ref{Gaussianprofileverticalstructure1} for all associated thermal
profiles. For comparison we also
use the profile introduced by SB96
in which the temperature follows a power law
\begin{equation}\label{Tpower}
 T = T_0 - A z^p,
\end{equation}
where $A$ and $p$ are parameters, with $p=3/2$ usually (see Figure
\ref{SB96verticalprofile}).
Further details of both profiles are
given in Appendix \ref{verticaldiskprofiles}.
 The
equilibria are convectively unstable within a confined region (of size
$L_c$) about the
disc mid-plane, and convectively stable outside this region. 
We have checked that
these vertical profiles
are indeed good equilibrium solutions in our numerical set-up by running a series of 
 simulations without perturbations; these show that after 55 orbits
 the initial profiles are unchanged with velocity fluctuation amplitudes
typically less than $10^{-3}$ at the vertical boundaries and
less than $10^{-7}$ at the midplane.

The background velocity is given by $\mathbf{u} = -(3/2) \Omega_0 x \,
\mathbf{e}_y$. At initialization we usually perturb all the
velocity components with random noise exhibiting a flat power
spectrum. The perturbations $\delta \mathbf{u}$ have maximum
relative amplitude of about $5\times10^{-5}$ and can be either positive or
negative. In order to investigate specifically the nature of linear
axisymmetric convective modes, we initialize several \textsc{PLUTO}
simulations with linear axisymmetric modes calculated semi-analytically
rather than with random noise (see Section
\ref{spectralmethodscomparison}).

Time units are selected so that $\Omega_0 = 1$. From now the subscript
on the angular frequency is dropped if it appears. The length unit
is chosen so that the initial midplane sound speed $c_s = 1$, which in turn defines a constant reference scaleheight $H\equiv c_s / \Omega_0=1$.
Note, however, that the sound speed is generally a function of both space and time.

\subsubsection{Box size and resolution}
\label{boxsizeandresolution}
We measure box size in units of scaleheight $H$, defined above. We
employ resolutions of $64^{3}, 128^{3}$, $256^{3}$  and $512^3$ in
boxes of size $4H\times4H\times4H$, which correspond to resolutions
of 16, 32, 64 and 128 grid cells per scaleheight $H$ in all
directions. 
However, in simulations in which we force convection we employ a `wide-box' ($6H\times6H\times4H$) with a resolution of $256\times256\times256$ which corresponds to about $43$ grid cells per $H$ in the horizontal directions, and 64 grid cells per $H$ in the vertical direction.

\subsubsection{Boundary conditions}
\label{boundaryconditions}
We use shear-periodic boundary conditions in the $x$-direction (see
\cite{hawley1995local}) and periodic boundary conditions in the
$y$-direction. In the vertical direction, we keep the ghost zones
associated with the thermal variables in
isothermal hydrostatic equilibrium (the temperature of the ghost zones
being kept equal to the temperature of the vertical boundaries at
initialization) in the manner described in
\cite{zingale2002mapping}. For the velocity components we use mostly
standard \textit{outflow} boundary conditions in the vertical
direction, whereby the vertical gradients of all velocity components
are zero, and variables in the ghost zones are set equal to those in
the active cells bordering the ghost zones.  
In a handful of our simulations we also use free-slip and periodic 
boundary conditions to test the robustness of our results.

\subsection{Diagnostics}
\label{diagnostics}
\subsubsection{Averaged quantities}

We follow the time-evolution of various volume-averaged quantities (e.g. kinetic energy density, thermal energy density, Reynolds stress, mass density, thermal pressure, and temperature). For a quantity $X$ the volume-average of that quantity is denoted $\langle X \rangle$ and is defined as 
\begin{equation}
\langle X \rangle(t) \equiv \frac{1}{V} \int_V X(x, y, z, t) dV
\end{equation}
where $V$ is the volume of the box.

If we are interested only in the vertical structure of a quantity $X$ then we average over the $x$- and $y$-directions, only. The horizontal average of that quantity is denoted $\langle X \rangle_{xy}$ and is defined as
\begin{equation}
\langle X \rangle_{xy}(z, t) \equiv \frac{1}{A} \int_A X(x, y, z, t) dA,
\end{equation}
where $A$ is the horizontal area of the box. Horizontal averages over different coordinate directions (e.g. over the $y$- and $z$-directions) are defined in a similar manner.

We are also interested in averaging certain quantities (e.g. the Reynolds stress) over time. The temporal average of a quantity $X$ is denoted $\langle{X}\rangle_t$ and is defined as
\begin{equation}
\langle X \rangle_t (x, y, z) \equiv \frac{1}{\Delta t} \int_{t_i}^{t_f} X(x, y, z, t) dt,
\end{equation}
where we integrate from some initial time $t_i$ to some final time $t_f$ and $\Delta t \equiv t_f - t_i$.

\subsubsection{Reynolds stress, alpha, and energy densities}
\label{reynoldsstressandalpha}

In accretion discs, the radial transport of angular momentum is
related to the $xy$-component of the Reynolds stress, defined as
\begin{equation}
R_{xy} \equiv \rho u_x \delta u_y
\end{equation}
where $\delta u_y \equiv u_y + q\Omega x$ is the fluctuating part of the
y-component of the total velocity $u_y$.
 $R_{xy} > 0$ corresponds to positive (i.e. radially outward) angular momentum transport, while $R_{xy} < 0$ corresponds to negative (i.e. radially inward) angular momentum transport. 

The Reynolds stress is related to the classic dimensionless
parameter $\alpha$, but the reader should note that various definitions of $\alpha$ exist in the literature and care must be taken when comparing measurements of $\alpha$ quoted in different papers. We define $\alpha$ as
\begin{equation}
\alpha \equiv \frac{\Omega_0 \langle R_{xy} \rangle}{q\Omega \langle P \rangle}.
\label{alphadefinition}
\end{equation}

The kinetic energy density of a fluid element is defined as
\begin{equation}
E_{\text{kin}} \equiv \frac{1}{2} \rho u^2,
\end{equation}
where $u$ is the magnitude of the \textit{total} velocity of a fluid
element. Often we will plot the vertical kinetic energy density, in
which case $u$ in the above is replaced by $u_z$.

The total energy (kinetic plus thermal) is \textit{not} conserved in the shearing box even when there are no (explicit or numerical) dissipative effects: work done by the effective gravitational potential energy on the fluid and flux of angular momentum through the $x$-boundaries due to the Reynolds stress act as sources for the total energy. The upper and lower vertical boundaries also let energy (and other quantities) leave
the box.

\subsubsection{Vertical profiles of horizontally averaged quantities}

Finally, we track the vertical profiles of horizontally averaged pressure, density,
temperature, and Reynolds stress. Of special interest is the buoyancy
frequency, which is calculated from the pressure and density data by finite differencing the formula
\begin{equation}
\frac{N_B^2}{\Omega^2} = z\left [\frac{1}{\gamma}\frac{d\ln{\langle P \rangle_{xy}}}{dz} - \frac{d\ln{\langle \rho \rangle_{xy}}}{dz}\right].
\end{equation}
 We also calculate the (horizontally averaged) vertical profiles of mass and heat flux. We  define the mass flux as
\begin{equation}
F_{\text{mass}} = \langle \rho  u_z \rangle_{xy},
\end{equation}
and the heat flux as
\begin{equation}
F_{\text{heat}} = \langle \rho u_z  T \rangle_{xy}.
\end{equation}

\section{Linear theory}
\label{eigensolver}

In this section we investigate the linear phase of the axisymmetric
convective instability both semi-analytically by solving a 1D boundary
value / eigenvalue problem using spectral methods, and also
analytically using WKBJ methods.

Our aim is to determine the eigenvalues (growth rates) and eigenfunctions (density, velocity and pressure perturbations) of the axisymmetric convective instability in the shearing box as a function of radial wavenumber $k_x$. The convectively unstable background vertical profile used in all calculations is shown in Figure \ref{Gaussianprofileverticalstructure1}. The profile is discussed in detail in Section \ref{gaussiantempprofile}. All calculations were carried out with profile parameters $T_0 = 1.0, \rho_0 = 1.0, \beta = 3.0$ and an adiabatic index of $\gamma = 5 / 3$. For our chosen background equilibrium, this corresponds to a  Richardson number of $\text{Ri} \sim  0.05$.

We  proceed as follows: first, we produce linearized equations for the 
perturbed variables; second, these
 are Fourier analyzed so that
$X'(x,z,t) = e^{\sigma t} e^{\text{i}k_x x}X'(z)$ for each of the perturbed
fluid variables $\rho', \mathbf{u}'$ and $p'$.
Here $\sigma$ can be complex (if $\sigma$ is real and positive it is referred to as a \textit{growth rate}). The linearized equations are given by
\begin{equation}
\sigma \rho' = -i k_x \rho_0 u'_x - \partial_z(\rho_0 u'_z),
\label{linearizedgasdynamcis9}
\end{equation}
\begin{equation}
\sigma u_x' = 2\Omega u'_y - \frac{ik_x}{\rho_0} P',
\label{linearizedgasdynamcis10}
\end{equation}
\begin{equation}
\sigma u'_y = (q-2)\Omega u'_x,
\label{linearizedgasdynamcis11}
\end{equation}
\begin{equation}
\sigma u_z' = -\frac{1}{\rho_0}\partial_z P' + \frac{\rho'}{\rho_0^2}\partial_z P_0,
\label{linearizedgasdynamcis12}
\end{equation}
\begin{align}
\begin{split}
\sigma P' &=  - u'_z\partial_z P_0 - \gamma P_0(i k_x u'_x + \partial_z u'_z).
\label{linearizedgasdynamcis13}
\end{split}
\end{align}
Background fluid
quantities are denoted $X_0 = X_0(z)$. 
For equilibrium $\rho_0$ and $P_0$ see Appendix \ref{verticaldiskprofiles}.

\subsection{Spectral methods}
\label{spectralmethodsconvectiveinstability}

We discretize the fluid variables on a vertical grid containing
$N_z$ points by expanding each eigenfunction as a linear superposition
of Whittaker cardinal (sinc) functions. These are approprate as all
perturbations should decay to zero far from the convectively unstable
region. This choice also saves us from explicitly imposing boundary
conditions. The discretised equations are next
gathered up into a single algebraic eigenvalue problem
\citep{boyd2001chebyshev}. 
Finally, we solve this matrix equation numerically using the
QR algorithm to obtain
 the growth rates $\sigma$ and eigenfunctions for a given radial
 wavenumber $k_x$. The code which computes the eigensolution 
we often refer to as our `spectral eigensolver'.

\subsection{WKBJ approach}

Alternatively, Equations \eqref{linearizedgasdynamcis9}-\eqref{linearizedgasdynamcis13} can be combined into a second-order ODE of the form
\begin{equation}
\frac{d^2 U}{d z^2} + k_z^2(z) U = 0,
\label{WKB1_main}
\end{equation}
where
\begin{equation}
U =\left(\frac{\gamma P_0}{\sigma^2+\kappa^2+k_x^2c_0^2}\right)^{1/2}u_z',
\end{equation}
in which $c_0$ is the equilibrium sound speed,
and the `vertical wavenumber' of the perturbations is given by
\begin{equation}
k_z^2(z) \approx -k_x^2 \frac{N_{B}^2(z) + \sigma^2}{\kappa^2 + \sigma^2},
\label{WKB2_main}
\end{equation}
with  $\kappa$ the epicyclic frequency \citep{ruden1988axisymmetric}.

We obtain approximate solutions to Equation \eqref{WKB1_main} analytically
using WKBJ methods in the limit $k_x$ large.  
Equation \eqref{WKB1_main} has two turning points at
$k_z^2(z) = 0$, which occur when
$-N_{B}^2(z) = \sigma^2$. Physically, one turning point occurs at the
disc mid-plane ($z = 0$), while the other turning point occurs at the
boundary of the convectively unstable region $z = L_c$. When
$-N_{B}^2(z) > \sigma^2$ then $k_z^2(z) > 0$ so the solutions of
Equation \eqref{WKB1_main} are spatially oscillatory, otherwise they are
evanescent. The unstable modes, in which we are interested,
  are confined and oscillatory within the convectively
 unstable region, and spatially decay exponentially outside.

The standard WKBJ solution is given by
\begin{equation*}
U(z) \sim c_1 \exp{\left(+\text{i}\int^z k_z(z') dz'\right)} + c_2 \exp{\left(-\text{i}\int^z k_z(z') dz'\right)},
\end{equation*}
where $c_1$ and $c_2$ are constants. By matching the interior (oscillatory) solution onto the exterior (exponentially decaying) solution,
we obtain the eigenvalue equation (dispersion relation):
\begin{equation}
 \int_{z_0}^{z_1} k_z(z) dz = \left(n + \frac{1}{2}\right)\pi \,\,\,\,\,\,\,\, n = 0, 1, 2, 3, \cdots,
\label{WKB4a}
\end{equation} 
where lower and upper bounds of integration $z_0$ and $z_1$,
respectively, are related by the implicit equation $N_{B}^2(z_0) =
N_{B}^2(z_1) = -\sigma^2$, and $n$ is the mode number.
Substituting
for $k^2_z(z)$ using Equation \eqref{WKB2_main}, we obtain an implicit
equation relating the radial wavenumber $k_x$ and the growth rate
$\sigma$ which we solve numerically via a root-finding algorithm.

\subsection{Eigenvalues}

\begin{figure*}
	\includegraphics[scale = 0.3]{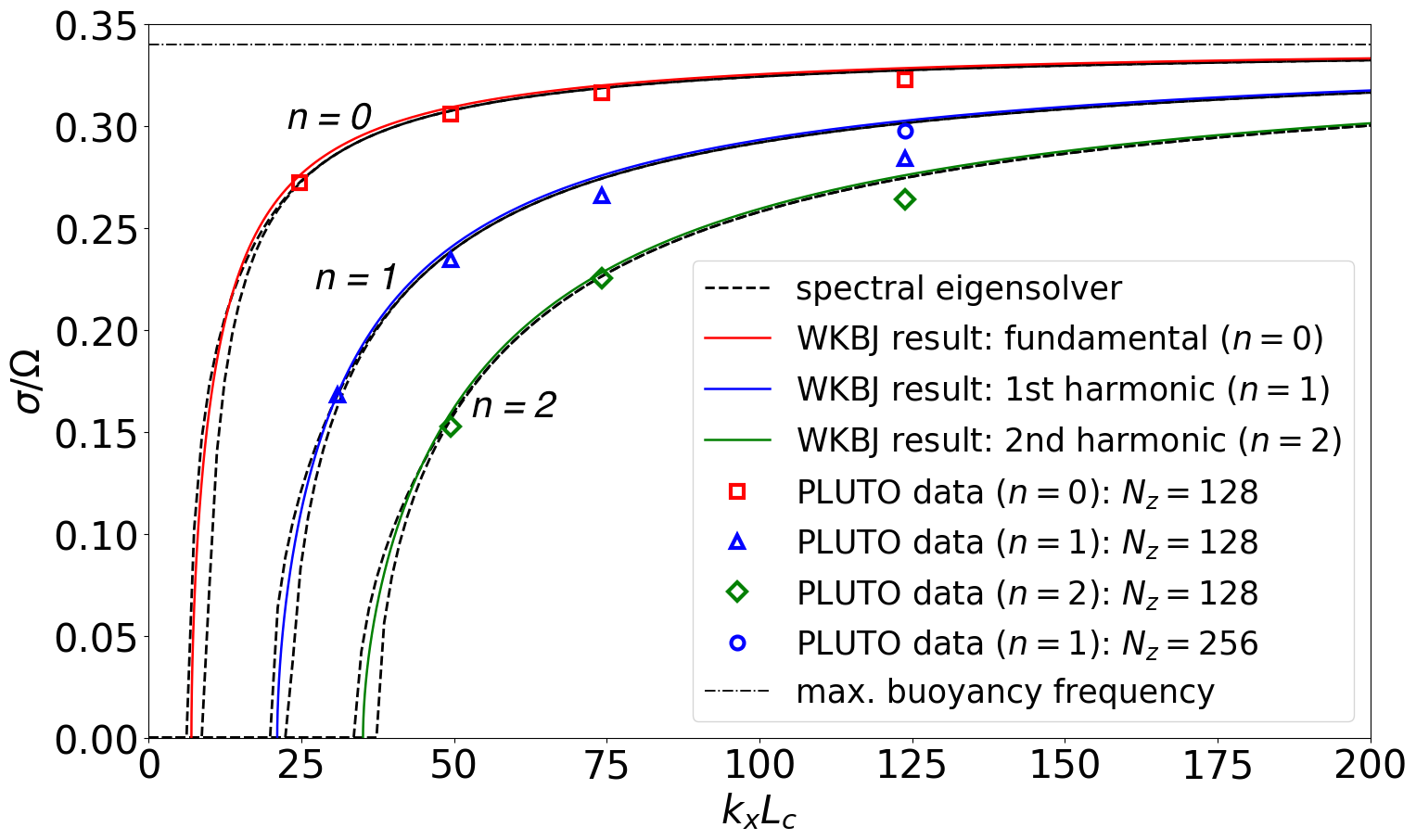}
    \caption{Growth rates as a function of radial wavenumber $k_x$ (scaled by the width of the convectively unstable region $L_c$). The solid lines were calculated analytically using WKBJ methods, and the dashed lines were calculated semi-analytically using pseudo-spectral methods. The squares, triangles, and diamonds correspond to measurements taken from $\textsc{PLUTO}$ simulations run at a vertical resolution of $N_z = 128$. The blue circle was taken from a \textsc{PLUTO} simulation run at $N_z = 256$.}
    \label{WKBNumericalComparison}
\end{figure*}

For a given radial wavenumber $k_x$, the eigenvalues calculated with
the spectral eigensolver come in pairs that lie close together in the
complex plane. These
correspond physically to the \textit{even} and \textit{odd} modes of
the convective instability. As described in Section
\ref{physicalinterpretationofeigenfunctions},
even modes are symmetric about the mid-plane, whereas the odd modes
are antisymmetric about the mid-plane. In
Fig.~\ref{WKBNumericalComparison}
we plot as black dashed curves the first few eigenvalue branches as functions of $k_x$, these
corresponding to vertical quantum numbers $n=0,\,1,$ and $2$. 
Note that in the figure $k_x$ has been normalized by the width of 
the convectively unstable region $L_c$ (given by Equation 
\eqref{LcGaussianTemp} in Appendix \ref{gaussiantempprofile}).

As the radial wavenumber $k_x$ is increased, the difference between
the growth rates of the even and odd modes vanishes, and the growth
rates tend asymptotically to the maximum absolute value of the
buoyancy frequency (indicated by a horizontal dot-dashed line). In
other words, the maximum growth rate is limited by the depth of the
 buoyancy-frequency profile. The fastest
growing modes are at arbitrarily large $k_x$ (small radial wavenumber
$\lambda_x$), and thus manifest as thin elongated structures. Thicker
structures are not favoured  
as they comprise greater radial displacements that are resisted by the
radial angular momentum gradient. 

Superimposed on the spectrally computed values are the WKBJ growth
rates (solid lines). WKBJ methods cannot distinguish between even or
odd modes, nonetheless we find very close agreement 
between the WKBJ and semi-analytical results for all radial
wavenumbers, even at low $k_x$ where the WKBJ approximation is,
strictly speaking, invalid. As the radial wavenumber increases, the
semi-analytical and WKBJ results converge. Note that associated with
each mode branch is a minimum radial wavenumber
 $k_{x, \text{min}}$ below which both the numerical and WKBJ growth
 rates are zero. 
This feature is just visible in the bottom left-hand side of Figure \ref{WKBNumericalComparison}.

We must stress that, because the fastest growing modes
are on the shortest possible scales, inviscid simulations of convection 
are problematic, at least in the early (linear) phase of the
evolution. It is here that the simulated behaviour may depend on the
varying ways that different numerical schemes deal with grid
diffusion, something we explore in Section 4.3. Only with resolved physical
viscosity can this problem be overcome. However,
in the fully
developed nonlinear phase of convection the short scale 
linear axisymmetric modes may be subordinate and this is less of an issue.

\begin{figure}
\centering
\includegraphics[scale=0.09]{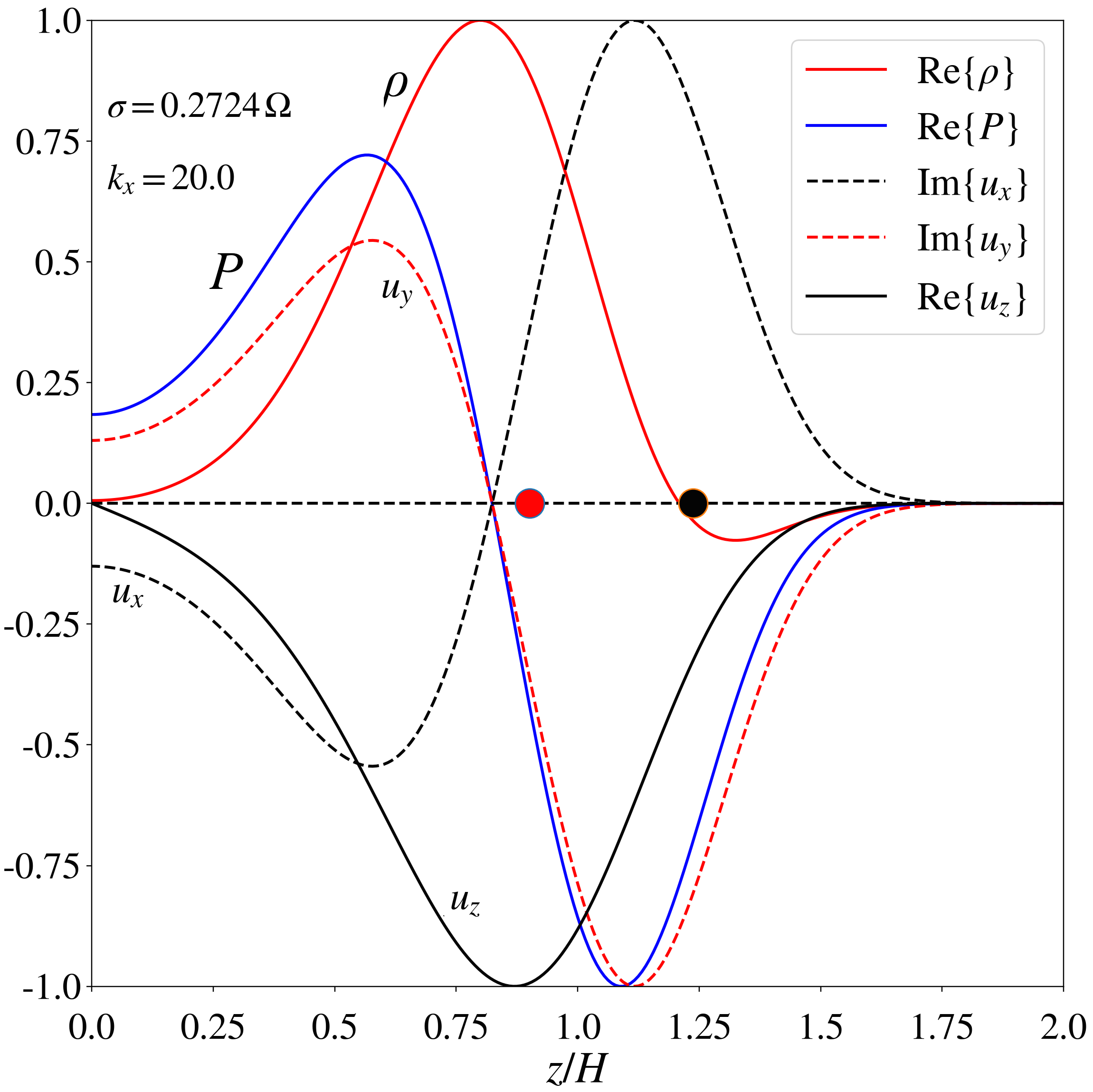}
\caption{Vertical profiles of the perturbations Re$\{\rho'(z)\}$,
  Re$\{u_z'(z)\}$, Re$\{P'(z)\}$, and of Im$\{u_x'(z)\}$ and
  Im$\{u_y'(z)\}$ in the upper half plane $z\in [0,2]H$, for the $n =
  0$ odd mode at $k_x = 20.0$. The imaginary and real parts not shown
  are effectively zero. 
 The black dot marks the upper extent of the convectively unstable region $L_c$, while the red dot marks the most convectively unstable points (see Equation \eqref{LcGaussianTemp}). The eigenfunctions have been rescaled so that the maximum of each eigenfunction is unity.}
    \label{Eigenfunctions1D}
\end{figure}

\subsection{Vertical structure of the eigenfunctions}
\label{physicalinterpretationofeigenfunctions}

\begin{figure*}
\captionsetup[subfigure]{labelformat=empty}
\centering
\subfloat[]{\includegraphics[scale=0.26]{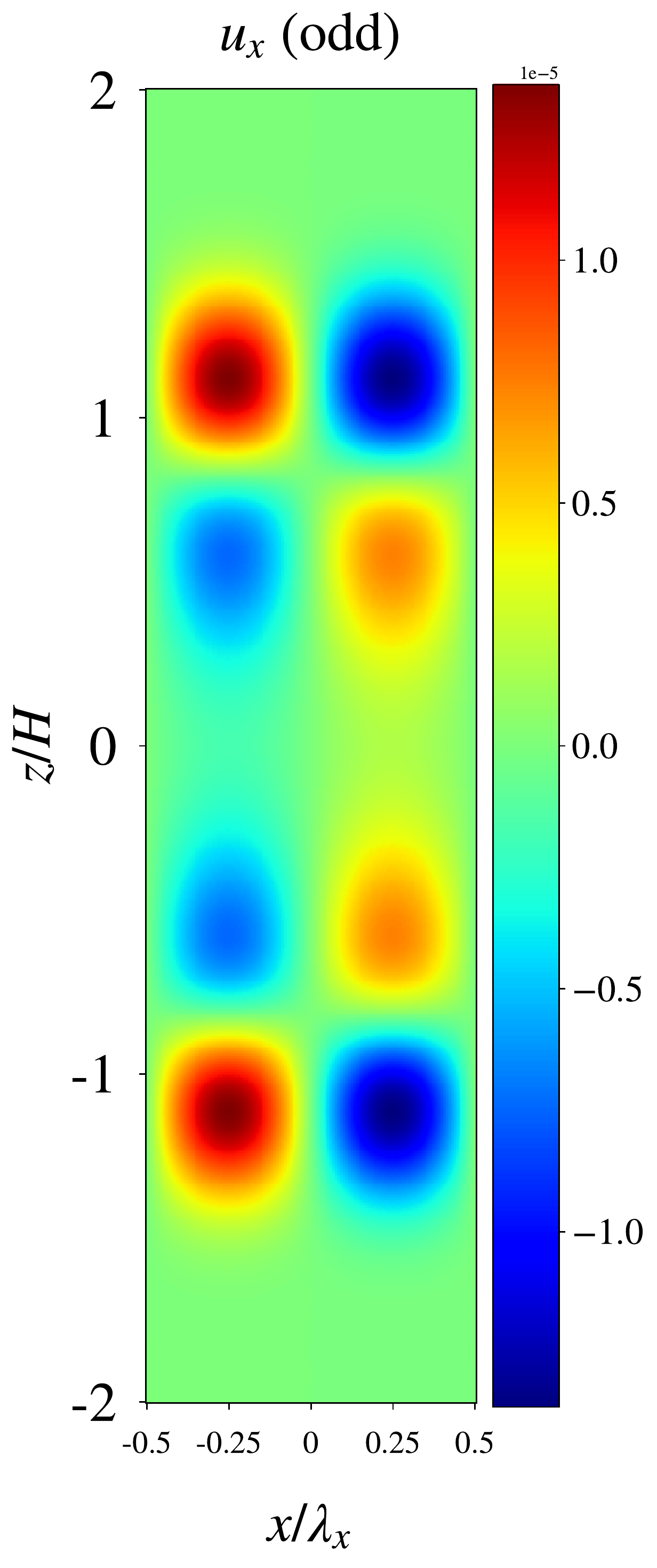}}\hspace{0.2em}
\subfloat[]{\includegraphics[scale=0.26]{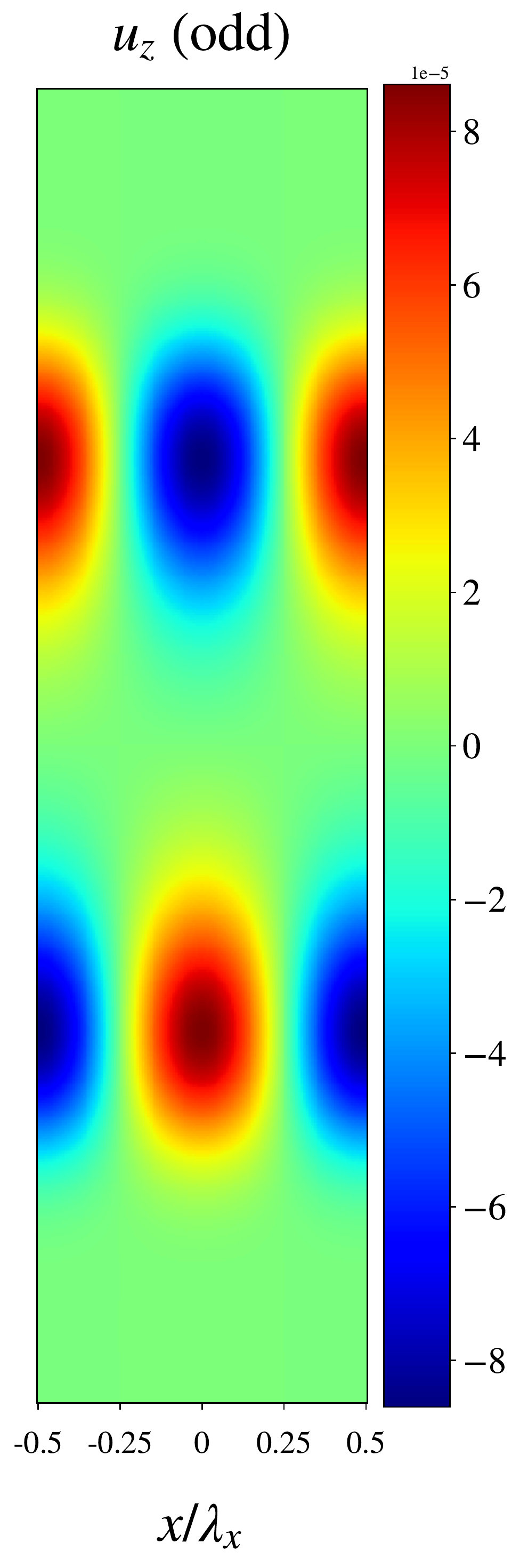}}\hspace{0.2em}
\subfloat[]{\includegraphics[scale=0.26]{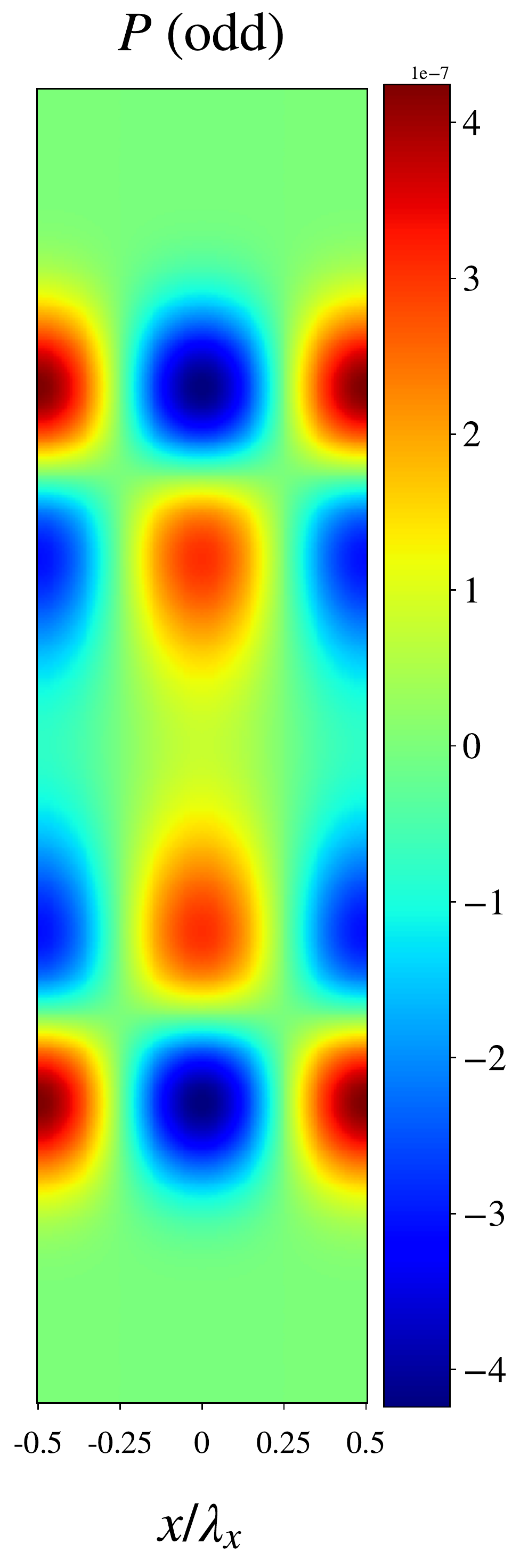}}\hspace{0.2em}
\subfloat[]{\includegraphics[scale=0.26]{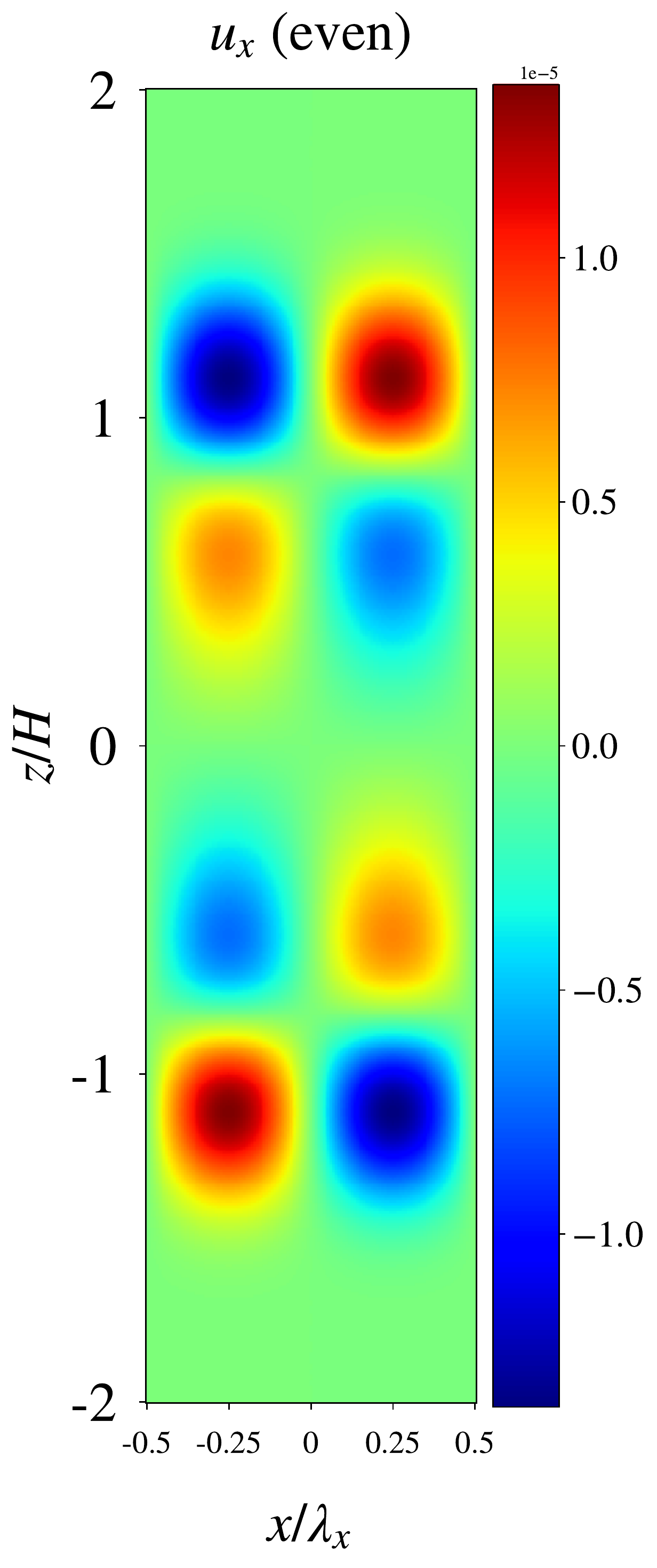}}\hspace{0.2em}
\subfloat[]{\includegraphics[scale=0.26]{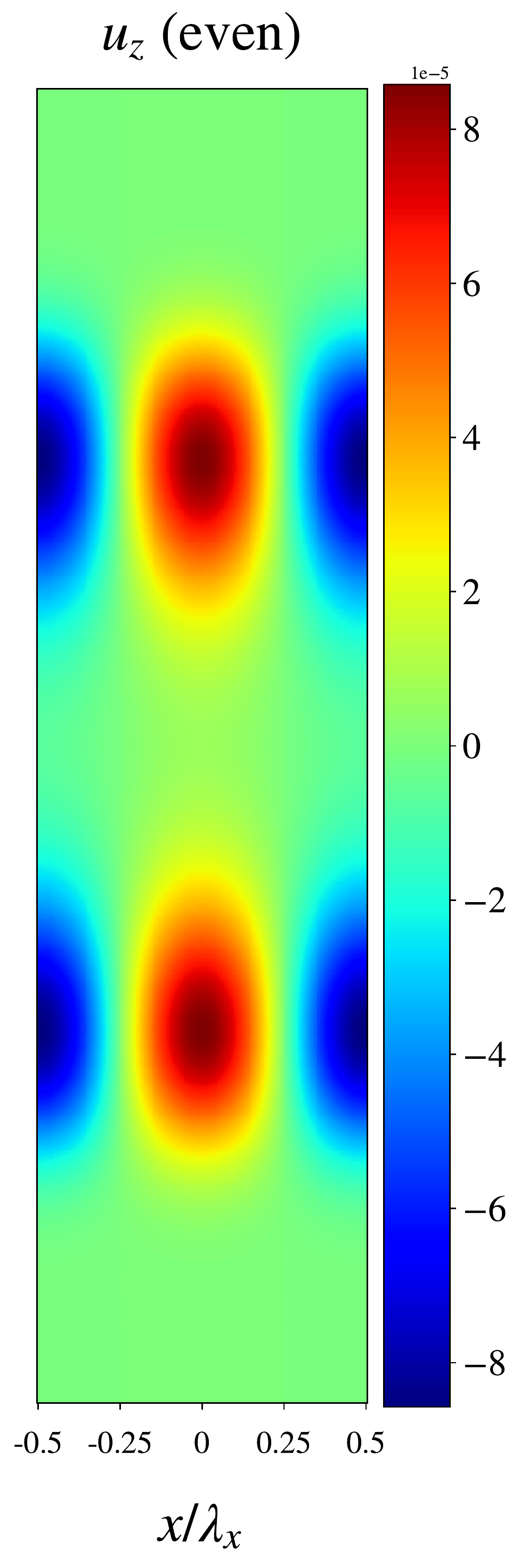}}\hspace{0.2em}
\subfloat[]{\includegraphics[scale=0.26]{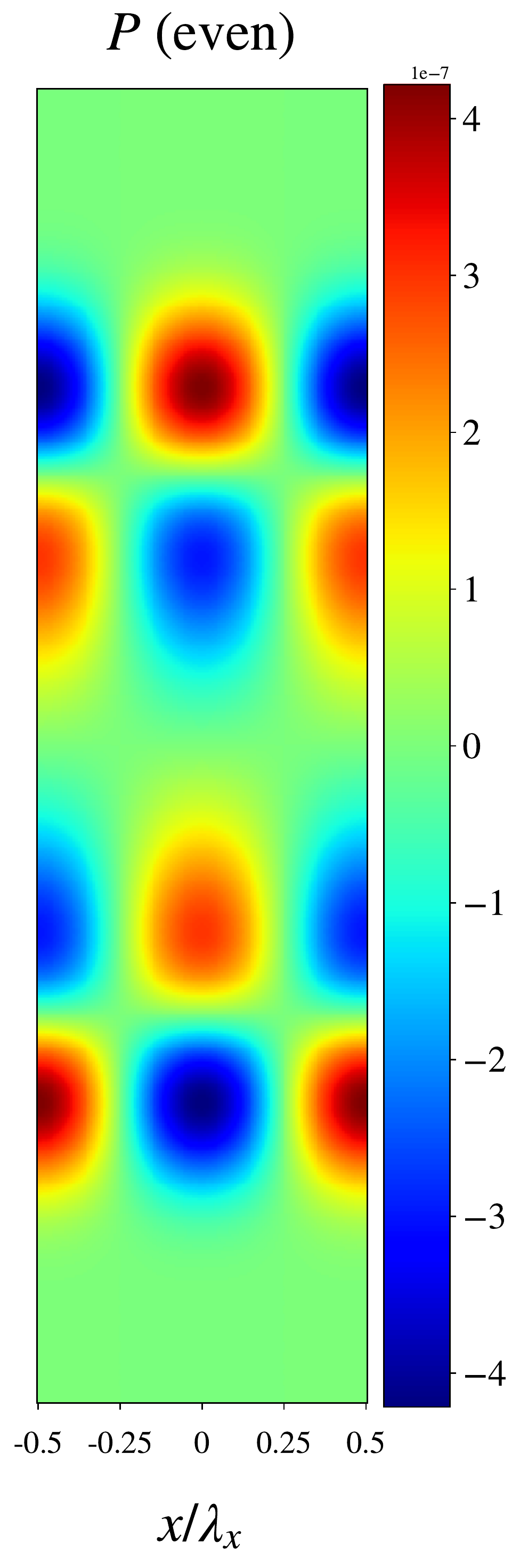}}
\caption{Two-dimensional profiles in the $xz$-plane of the real parts
  of the selected components of odd and even eigenfunctions for $n = 0$ and $k_x = 20.0$.}
    \label{Eigenfunctions2D}
\end{figure*}

The reader is reminded that here we have an unstably stratified fluid
in a medium in which the gravitational field smoothly changes
sign. This contrasts to the more commonly studied situation of
convection at the surface of the Earth (or within the solar
interior). It hence is of interest to study the structure of the
convective cells in some detail.  We  consider first
the fundamental ($n=0$) odd mode with radial wavenumber $k_x =
20$. 

In Figure \ref{Eigenfunctions1D} we plot the vertical profiles of the
nonzero components of the eigenfunction.
 To provide greater visual clarity, we have rescaled the perturbations
 so that the maximum amplitude of each perturbation is unity: thus we
 can more easily observe the vertical structure 
of the perturbations, but not
deduce their relative amplitudes. In Figure
\ref{Eigenfunctions2D} we plot their real parts in the $xz$-plane with
the correct amplitudes. We find that $|u_z'| > |u_y'| \sim
|u_x'| \gg |\rho'|  \gg |P'|$. 
Thus vertical velocity perturbations are greatest in magnitude, while
pressure perturbations are smaller than vertical velocity
perturbations by two-orders of magnitude, indicating that the
convective cells are roughly 
 in pressure balance with the surroundings, i.e that they are behaving adiabatically.

In Figure \ref{Eigenfunctions1D}, $u_z'$ (solid black curve) is
antisymmetric about the mid-plane, whereas the remaining
perturbations are symmetric about the mid-plane. Figs
\ref{Eigenfunctions1D} and \ref{Eigenfunctions2D} clearly show 
 that both odd and even modes consist of a chain of convective cells above and
 below the midplane and localized near
the most convectively unstable point (denoted by the large red
dot in the former Figure). The peak in the amplitude of $u_z'$ that occurs near the
this point tells us that the fluid elements reach
their maximum acceleration where the buoyancy force is
greatest. As the fluid perturbations behave adiabatically, they
adjust their pressure to maintain a balance with the background
pressure: thus where cool elements begin to rise (higher
background pressure), the pressure perturbations (solid blue line)
increase (i.e. $P' > 0$), and where hot elements begin to sink (lower
background pressure), the pressure perturbations decrease (i.e. $P'
<0$). The vertical velocity perturbation (black solid line) is out of
phase with the radial and azimuthal velocity perturbations (dashed
lines), since vertical  
motion is converted into radial motion where
 the convective cells turn over (picture fluid motion at the top of a fountain).

The even modes possess a nonzero, but relatively small, $u_z'$ at the
midplane,
and thus weakly couple the two sides of the disk. Such modes may
permit some exchange of
mass, momentum and thermal energy across the mid-plane when reaching
large amplitudes.

 As the radial wavenumber is increased from $k_x = 20.0$ to $k_x =
 200.0$ (very small radial wavelengths),
 the perturbations become increasingly localized about the most
 convectively unstable point(s). The localisation in $z$ however
 is not as narrow as the radial wavelength. For larger $k_x$, in fact,
each pair of even and odd modes change their character and become
 entirely localised to either the upper or lower disk. Activity in the two
 halves of the disk are hence completely decoupled for such
 small-scale (but fast growing) disturbances.

\subsection{Comparison of theory with simulations}
\label{spectralmethodscomparison}

In this section we compare the growth rates previously calculated with
those measured from \textsc{PLUTO} simulations initialized with the
exact linear modes taken from the spectral eigensolver.

\begin{table}
	\centering
	\caption{Comparison of eigensolver growth rates with growth rates measured in \textsc{PLUTO} simulations. The simulations were initialized with the eigenfunctions corresponding to each mode $n$ and radial wavenumber $k_x$. All simulations except the one marked with a footnote were run at a  resolution of $64\times10\times128$. The standard deviation on the growth rates $\sigma_{\text{PLUTO}}$ measured in the simulations is typically in the range $\pm0.0002$ to $\pm0.0004$.}
	\label{EigensolverWKBPLUTOgrowthrates}
	\begin{threeparttable}
	\begin{tabular}{lccccr} 
		\hline
		Mode & Parity & $k_x$ &  $\sigma_{\text{EIG}}$  & $\sigma_{\textsc{PLUTO}}$ & $\%$ error\\
		\hline
		$n = 0$ & even & $20.0$ & 0.2727& 0.2719 & 0.3 \\	
		 & odd & $20.0$  & 0.2724 & 0.2715 &  0.3 \\
		& even & $40.0$ & 0.3076 & 0.3059 & 0.6 \\	
		 & even & $60.0$ & 0.3186 & 0.3160& 0.8 \\	
		  & even & $100.0$ & 0.3273 &  0.3223&  1.6 \\	
		\hline
		$n = 1$ & even & $25.0$ & 0.1694& 0.1678 & 0.9  \\	
		 & odd &  $25.0$ & 0.1643& 0.1624  &0.9 \\
		 & even & $40.0$ & 0.2392& 0.2343 &  1.1\\	
		& even & $60.0$ & 0.2747& 0.2655 &  3.3\\	
		 & even & $100.0$ & 0.3016& 0.2840&  5.8\\
		 &  &  & & 0.2974 & 1.4$^*$\\	
		 \hline
		$n = 2$ & even & $40.0$ & 0.1593 & 0.1525 & 4.2  \\	
		 & odd &  $40.0$ & 0.1580&  0.1517 & 3.9 \\
		& even & $60.0$ &0.2272 &  0.2251& 0.9  \\	
		 & even & $100.0$ & 0.2749& 0.2637& 4 \\	
		 \hline
	\end{tabular}
	\begin{tablenotes}
	\item[*]  Simulation run with a resolution of $N_z = 256$ grid points in the vertical direction.
	\end{tablenotes}
	\end{threeparttable}
\end{table}

Each \textsc{PLUTO} simulation is run in a box of size $\lambda_x \times
0.625H\times4H$. We maintain a fixed number of $32$ grid cells per
scaleheight in the $z$-direction, $16$ grid cells per scaleheight in
the $y$-direction and $64$ grid cells per \textit{radial wavelength}
in the $x$-direction. Thus the radial wavelength of the modes is
always well resolved in our simulations. 
No explicit diffusion coefficients (i.e. viscosity or thermal
diffusivity) are used.

The results are plotted in Figure \ref{WKBNumericalComparison} as
squares (and one circle). Numerical values appear in Table
\ref{EigensolverWKBPLUTOgrowthrates}. We find excellent agreement
between simulation and theory: the relative error is $<1\%$ for the
fundamental mode in the interval $k_x \in [20.0, 60.0]$ and for the
first harmonic at $k_x = 25.0$. As the radial wavenumber is increased
the simulation growth rates begin to deviate from the theoretical
curves, although the percentage error remains less than $6\%$ even at
large wavenumbers. This behavior is expected due to the effects of
numerical diffusion which acts to \textit{decrease} the
growth rates as the radial wavelength approaches the size of the
vertical grid. Although the radial wavenumber $\lambda_x$ is always well
resolved in the simulations, 
the vertical wavelength becomes increasingly less well resolved. 

The numerical diffusion could be reduced by increasing the vertical
resolution of the simulations. To check this we have rerun the
\textsc{PLUTO} simulation initialized with eigenfunctions
corresponding to the first harmonic $(n=1)$ even mode with $k_x = 100$
at \textit{twice} the vertical resolution, i.e. $N_z = 256$ instead of
$N_z = 128$. At a resolution of $N_z = 128$ we measured a growth rate
of $\sigma_{\text{PLUTO}} = 0.2840\,\Omega$, corresponding to a
percentage error of $5.8\%$ with the growth rate calculated
semi-analytically ($\sigma_{\text{EIG}} = 0.3016\,\Omega)$. At a
vertical resolution of $N_z = 256$, however, we measured a growth rate
of $\sigma_{\text{PLUTO}} = 0.2974\,\Omega$, corresponding to a
percentage error of just $1.4\%$.
 Thus, the growth rates measured in the simulations  appear to converge to those calculated analytically and semi-analytically as the resolution of the simulations is increased.

Finally, in all simulations 
of the axisymmetric modes we observed \textit{inward} angular momentum
transport.
This is in agreement with the analytical argument presented in
\citep{stone1996angular}
regarding axisymmetric flow.

\section{Simulations of Unforced Compressible Convection}
\label{unforcedsims}

In this section we describe fully compressible, three-dimensional simulations of hydrodynamic convection in the shearing box carried out with \textsc{PLUTO}. These simulations are `unforced' in that they are initialized with a convectively unstable profile, but this unstable profile is not maintained by a separate process (e.g. turbulent heating via the magnetorotational instability). Thus convection rearranges the profile, shifting it to a marginally
stable state, and thus ultimately hydrodynamical activity dies out. 
While the unforced convection studied in this section is essentially a transient phenomenon that depends on initial conditions, 
and is sensitive to the numerical scheme (as we discuss in Sections
\ref{athenaplutocomparison} and
\ref{AMtransportnumericaldiffusivity}), it serves as a good starting
point for investigating the problem and illuminates a number of
interesting features.

The key result of this section is that, even without the inclusion of
explicit viscosity, we observe that hydrodynamic convection in the
shearing box generally produces a \textit{positive Reynolds stress},
and thus can drive outward transport of angular momentum. This is in
direct contradiction to simulations carried out in \textsc{ZEUS} by
SB96. In Section \ref{fiducialsimulations} we
describe our fiducial simulations, and in Section
\ref{athenaplutocomparison} we compare \textsc{ATHENA} and
\textsc{PLUTO} simulations. In Section
\ref{AMtransportnumericaldiffusivity} we describe the sensitivity of
the sign of angular momentum transport to the numerical
scheme. Finally, in Section \ref{simswithexplicitdiffusioncoeffs} we
investigate the effects of including explicit diffusion coefficients
in our simulations, thus connecting to the Boussinesq simulations of \cite{lesur2010angular}.

\begin{figure*}
	\centering
	\includegraphics[scale=0.35]{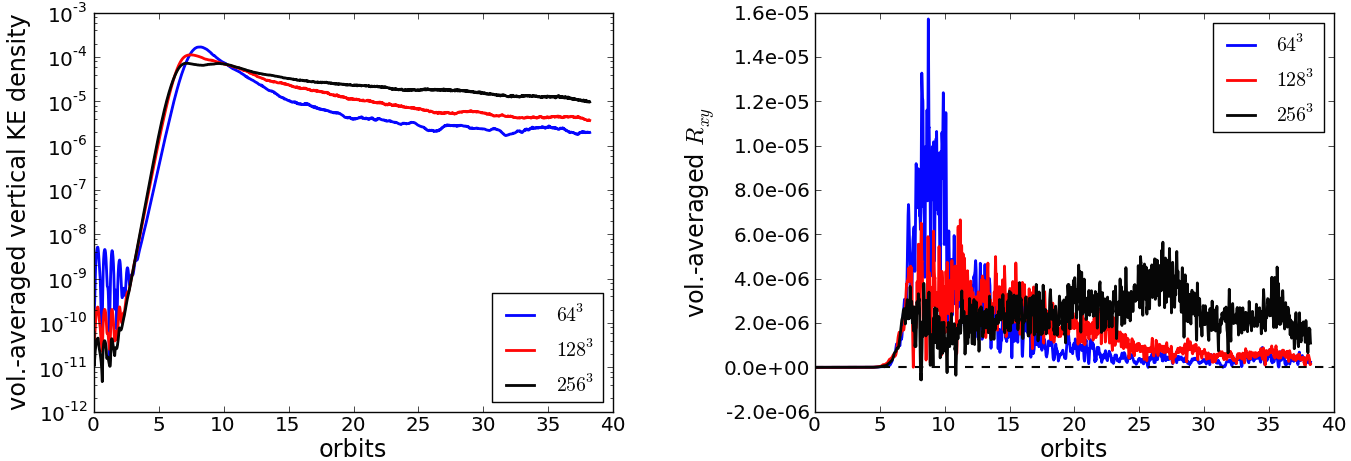}
    \caption{Left: semi-log plot of the time-evolution of the volume-averaged vertical kinetic energy density in \textsc{PLUTO} simulations at resolutions of $64^{3}$ (blue line), $128^{3}$ (red line) and $256^{3}$ (black line). Right: time-evolution of volume-averaged $xy$-component of the Reynolds stress tensor.}
    \label{fiducialsims01}
\end{figure*}

\subsection{Fiducial simulations}
\label{fiducialsimulations}

\subsubsection{Set-up and initialization of fiducial simulations}
\label{setupandinitializationoffiducialsimulations}

The simulations described in this section were run at resolutions of
$64^{3}$, $128^{3}$ and $256^{3}$ in boxes of size
$4H\times4H\times4H$, where $H$ is the
scaleheight.\footnote{A simulation run at a resolution
    $512^3$ is included in the list of fiducial simulations in Table
    \ref{fiducialKERxy} (see Appendix \ref{tables}) but is not
    described in this section for brevity.}  All the simulations
were initialized with the convectively unstable vertical profiles for
density and pressure described in Appendix \ref{gaussiantempprofile}
with profile parameters $T_0 = 1.0,\, \rho_0 = 1.0,\, \beta = 3.0$ and
an adiabatic index of $\gamma = 5/3$ (see Figure
\ref{Gaussianprofileverticalstructure1}). The Stone and Balbus
vertical profile SB96 was also trialled yielding very similar results, but
we have
omitted most of these in the interests of space.
 Vertical outflow conditions
were employed for our fiducial simulations
but periodic and free-slip boundary conditions in the vertical
direction produced the same behaviour.
Random perturbations to the velocity components were seeded
at initialization,  with a maximum amplitude $|\delta \mathbf{u}| \sim 10^{-5}$.
Finally, we adopt a very small but finite thermal diffusivity of $\chi = 2\times10^{-6}$ to facilitate conduction of thermal energy through the vertical boundaries and to aid code stability.  No explicit non-adiabatic heating, cooling, or thermal relaxation are included in the simulations described in this section: therefore convection gradual dies away after non-linear saturation.

\subsubsection{Time-evolution of averaged quantities}

In the left panel of Figure
\ref{fiducialsims01} we track the time-evolution of the
volume-averaged kinetic energy density associated with the vertical
velocity for simulations run at
resolutions of $64^{3}, 128^{3}$ and $256^{3}$,
respectively. 

Initially all simulations exhibit small-amplitude
oscillations due to internal gravity
waves excited in the convectively stable region at initialization. 
After some three orbits,
the linear phase of the convective instability begins in earnest, 
characterised by exponential growth of the perturbation amplitudes. 
During this phase, internal energy is converted into the
kinetic energy of the rising and sinking fluid motion that
comprises the convective cells. As the resolution is increased,
shorter scale modes are permitted to grow. Because they are the most
vigorous the growth rate (proportional to the slope of the
kinetic energy density) is slightly larger at better resolution.

\begin{figure*}
\captionsetup[subfigure]{labelformat=empty}
\centering
\subfloat[]{\includegraphics[scale=0.17]{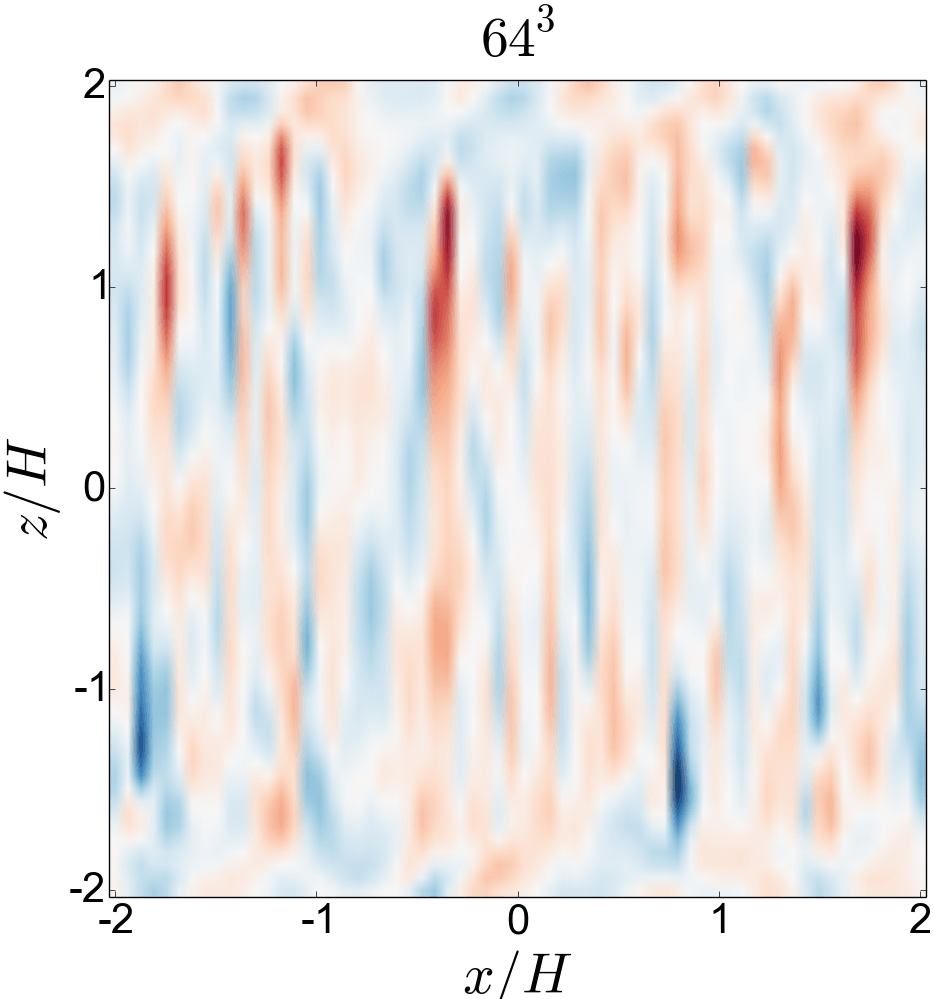}}\hspace{0.2em}
\subfloat[]{\includegraphics[scale=0.17]{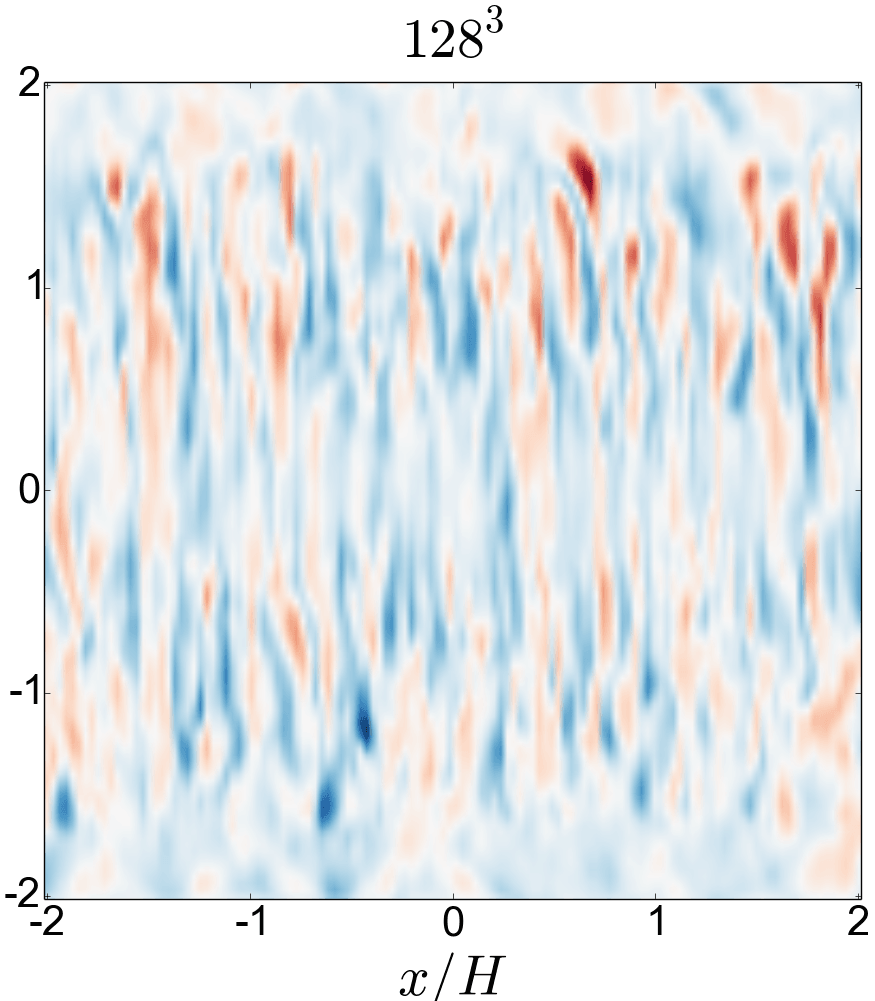}}\hspace{0.2em}
\subfloat[]{\includegraphics[scale=0.17]{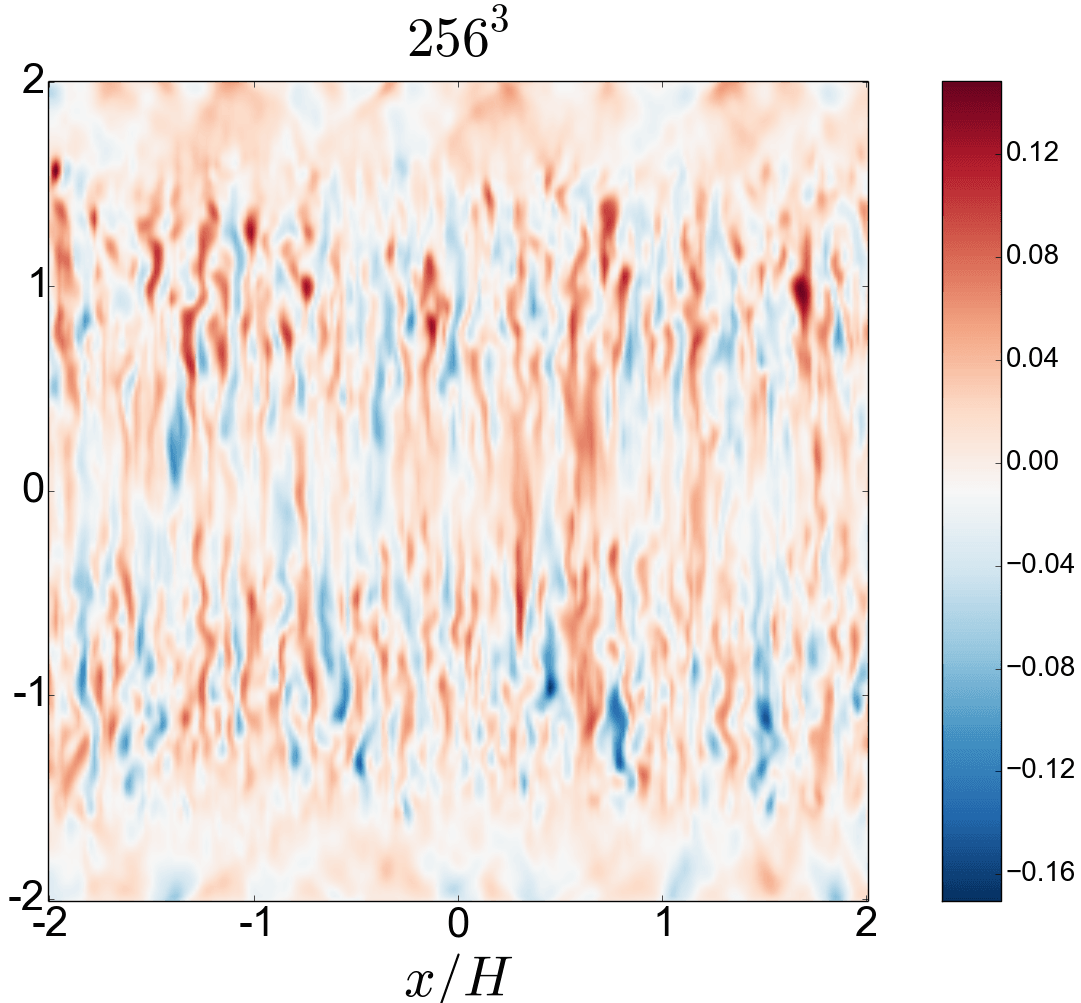}}\hspace{0.2em}
\caption{From left to right: snapshots of $u_z$ in the $xz$-plane taken at non-linear saturation (just after the linear phase) at resolutions of $64^{3}$, $128^{3}$ and $256^{3}$, respectively. Blue denotes $u_z < 0$ (sinking fluid) and red denotes $u_z > 0$ (rising fluid).}
\label{fiducialsims02}
\end{figure*}

The linear phase ends some 10 orbits into the simulation, and the flow
becomes especially disordered. The peak in the vertical
kinetic energy density occurs at this point, but this peak decreases 
 with resolution, something we discuss below. After non-linear
 saturation,
 the kinetic energy decreases gradually: the convective cells
 redistribute thermal energy and mass, thus shifting the thermal
 profile from a convectively unstable to a marginal state, and
 ultimately the convective motions die out. After about 38 orbits, the
 kinetic energy density has dropped to about one hundredth of its
 value at non-linear saturation in the lower resolution runs. 
The level of numerical diffusivity has an appreciable 
effect in damping activity after non-linear saturation, with the decrease in vertical kinetic energy successively smaller over the same period of time in the $64^{3}$, $128^{3}$  and $256^{3}$ simulations, respectively.

The behaviour of the kinetic energy aligns relatively well with our
expectations. The Reynolds stress, on the other hand, is more interesting. The evolution 
of the $xy$-component of the Reynolds stress is plotted
in the right panel of Fig.~\ref{fiducialsims01}. The stress is small,
but perhaps its most striking
feature is its positivity over the entire duration of all three
simulations. 

Perhaps most surprising is the positivity of the stress during
the \emph{linear} phase of the instability. We might expect that the
axisymmetric modes (which send angular momentum inwards) dominate this period of the evolution, as
non-axisymmetric disturbances only have a finite window of growth
(about an orbit) before they are sheared out and dissipated by
the grid. But the
positivity of the stress suggests that instead it is the shearing waves
that are the dominant players in the linear phase. Visual inspection of the 
velocity fields confirms that
the flow is significantly non-axisymmetric, and we find
several examples of strong shearing waves `shearing through'
$k_x=0$ during this phase. It would appear these waves transport angular
momentum primarily outward as they evolve from leading to
trailing, behaviour that is in fact consistent with \cite{ryu1992convective},
who  find that inward transport only occurs at sufficiently long times
when the waves are strongly trailing and hence effectively
axisymmetric. (Even then the stress is extremely oscillatory.)

Why do the shearing waves dominate over the axisymmetric modes? 
One hypothesis is that
3D white noise seeds fast growing shearing waves
that can outcompete the axisymmetric convective instability over their
permitted window of growth. If this
were to be true then the simulated nonlinear regime is achieved
before the dominant shearing waves become strongly trailing and begin
to send angular momentum inward. Given that the typical timescale for shearing
out is roughly a few orbits at best, this is marginal but not impossible.
An alternative, more plausible, and more troubling hypothesis is that our
inviscid
numerical code misrepresents trailing shearing waves: more
specifically, the code is artifically reseeding
fresh leading waves from strongly trailing waves (`aliasing'; Geoffroy Lesur, private
communication). As a consequence,
shearing waves 
outcompete the axisymmetric modes because they can shear through several
times. If true, this is certainly concerning. But we hasten to add that
this phenomena should only be problematic in the low amplitude
linear phase; once the perturbations achieve large amplitudes the
aliasing will be subsumed under \emph{physical} mode-mode interactions.

The linear phase ends in a spike in the stress.  Time-averages of
the volume-averaged value of $R_{xy}$ from orbit 5 to orbit 15, a
period spanning non-linear saturation, show that the Reynolds stress
decreases as the resolution increases (see Table \ref{fiducialKERxy}
in Appendix \ref{tables}). 
The time- and volume-averaged
value of $R_{xy}$  over a period spanning non-linear saturation
(orbits 5 to 15) is $+4.8\times10^{-6}$, $+2.7\times10^{-6}$ and
$+1.7\times10^{-6}$ in the  $64^{3}$, $128^{3}$  and $256^{3}$
simulations, respectively. In terms of the turbulent alpha-viscosity
this corresponds 
(from Equation \eqref{alphadefinition}) to $\alpha \sim
+2.3\times10^{-5}$, $+1.5\times10^{-5}$ and $9.6\times10^{-6}$,
respectively. The dependence on the numerical viscosity can be
explained by appealing to secondary shear instabilities (see next subsection).

The volume-averaged density remains roughly constant during the linear
phase of the instability, followed by a drop after non-linear
saturation as mass is lost through the lower and upper boundaries.
Although the convectively unstable region at initialization
 is confined to the
region $|z| < L_c \sim1.11H$, and is therefore well within the box, convective overshoot deposits mass (and
heat) outside the convectively unstable region. The overall decrease
in density over the duration of the simulation is small, but appears
to increase with greater resolution. The overall percentage change in
the density is $-1.4\%$, $-3.4\%$ and $-8.8\%$ in the $64^{3}$,
$128^{3}$ and $256^{3}$ simulations, respectively.

\subsubsection{Structure of the flow}

The development of convective instability and the associated
convective cells are best observed through snapshots of the vertical
component of the velocity in the $xz$-plane, shown at resolutions of
$64^{3}, 128^{3}$ and $256^{3}$ just after non-linear saturation in
Figure \ref{fiducialsims02}. A full set of convective cells is clearly
visible at all resolutions. They are thin,
filamentary structures several grid cells wide comprising
alternating negative and positive velocities (updrafts and
downdrafts).

The maximum vertical Mach number of the flow around non-linear
saturation in the $256^3$ run is about $M_z \sim 0.15$, with the
largest vertical Mach numbers generally being measured near the
vertical boundaries (where the temperature -- and therefore the sound
speed -- is lowest).

The higher resolution simulations indicate
that the plumes develop a wavy or buckling structure as they rise or
sink, indicating the onset of a secondary shear instability. It is 
likely that the  buckling of the convective plumes by these `parasitic
modes' limits the amplitudes of the linear modes, and ultimately leads
to their breakdown. 
At lower resolution numerical diffusion smooths out the
secondary shear modes, and so the convective plumes reach large
amplitudes before breaking down (blue curve in right panel of
Fig.~\ref{fiducialsims01}). 
At high resolution the shear modes
are not so impeded and make short work of these structures (black
curve in the same figure).

\subsubsection{Vertical heat and mass flux}

\begin{figure}
\centering
\subfloat[]{\includegraphics[scale=0.11]{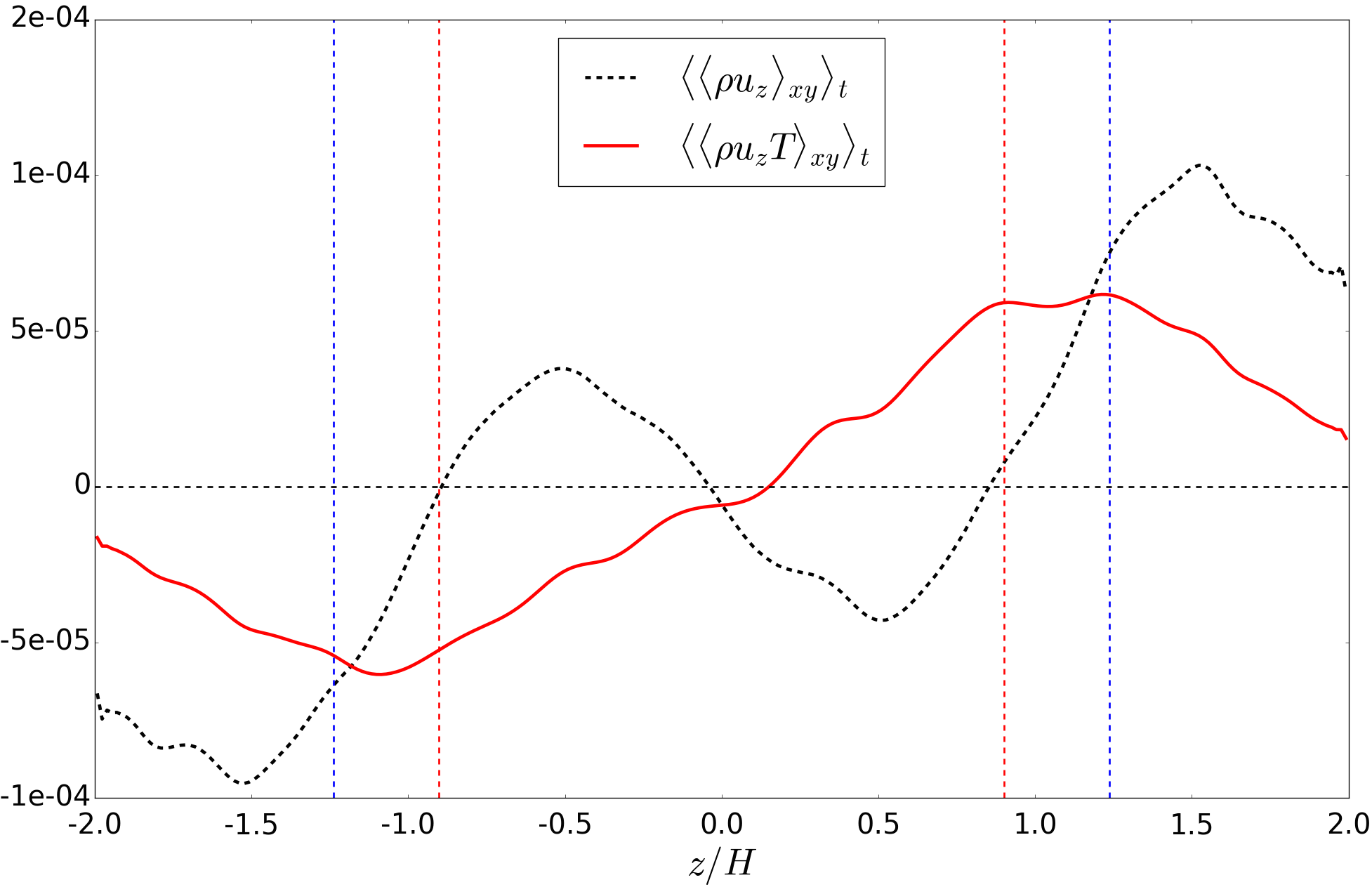}}
\qquad
\subfloat[]{\includegraphics[scale=0.11]{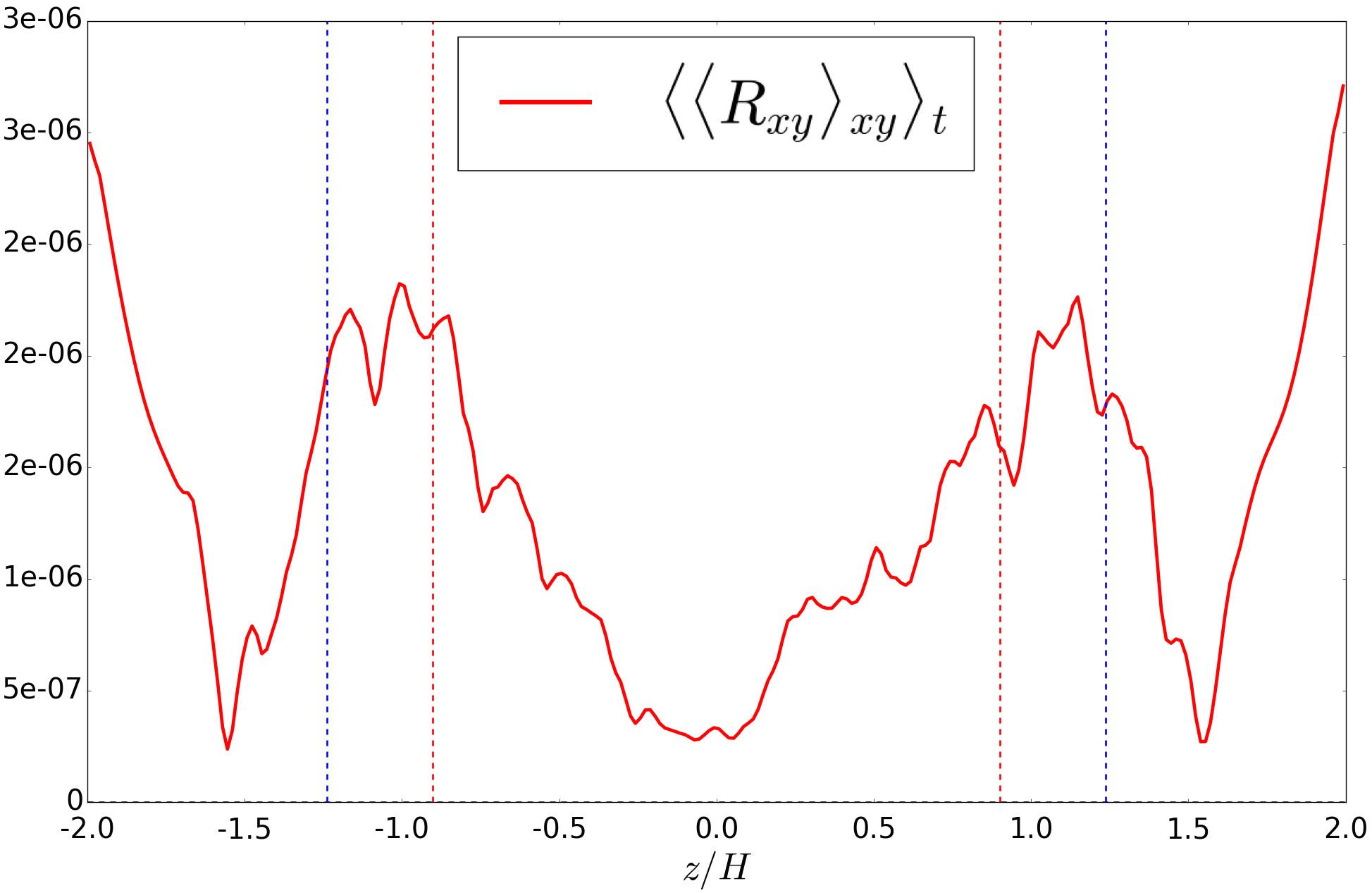}}
   \caption{(a) Vertical profiles of the vertical heat (solid red line) and mass flux (dashed black line).   The outer vertical dashed (blue) lines mark the boundaries of the convectively unstable region at initialization, while the inner vertical dashed (red) lines mark the most convectively unstable points at initialization. (b) Vertical profile of the $xy$-component of the Reynolds stress tensor $R_{xy}$. In all cases, the results have been time-averaged between orbits 5 and 10, spanning non-linear saturation.}
    \label{fiducialsim08}
\end{figure}

 Figure \ref{fiducialsim08} (a) shows the vertical profiles of
 horizontally-averaged heat and mass fluxes  taken from snapshots just
 after non-linear saturation from the simulation with resolution
 $256^3$. For clarity, we have time-averaged the horizontally-averaged
 vertical mass and heat flux profiles between orbits 5 and 
10, a period spanning non-linear saturation (see Figure \ref{fiducialsims01}). 

Negative (positive) values for the fluxes for $z < 0$ and positive
(negative) values for the fluxes for $z > 0$ correspond to transport
of heat and mass away from (towards) the mid-plane.  Overall, the heat
flux is away from the mid-plane, peaking in the vicinity of the most
convectively unstable points at initialization
(indicated by the vertical dashed red lines in Figure
\ref{fiducialsim08}). In addition, there is some mass flux towards the
mid-plane within $-H < z < H$.
Thus,
convection is transporting mass and heat such as to erase the
convectively unstable stratification, as expected.
Note that the positive peaks in the
heat flux near the most convectively unstable points are similar to
those observed by  \cite{hirose2014convection} in the thermally
dominated 
region of the disc (i.e. $P_\text{thermal} > P_\text{magnetic}$; c.f. the middle-panel of Figure 5 in their paper).

Beyond $|z| > H$,
the mass flux is directed away from the mid-plane, peaking
just beyond the point which marks the boundary of the convectively
unstable region. This outward mass flux might be due to convective
overshoot, although we expect this effect to vanish if averaged over a
suitable time interval. Alternatively, it is possible  
that heat transported towards the corona by the convective cells
causes 
matter in the corona to heat up and become buoyant, or else that this heat generates a weak thermal wind.

Finally, in Figure \ref{fiducialsim08} (b) we show the vertical
profile of the horizontally-averaged $R_{xy}$ taken from the $256^3$
simulation
and time-averaged between orbits 5 and
10. The Reynolds stress
is clearly positive over all of the vertical domain, peaking
just beyond the most convectively unstable points.\footnote{Note
  however that at any given instant in time, we generally do observe
  regions where the Reynolds stress is negative.} Thus the bulk of
outward angular momentum transport occurs where the convective cells
begin to turn-over, resulting in radial mixing of the
gas. Note, however, that we also observe rather large positive
stresses near the vertical boundaries, with the stress at  
the vertical boundaries about 1.5 times the peak stress in the remainder of the domain.

\subsection{Simulation of compressible hydro convection in \textsc{ATHENA}.}
\label{athenaplutocomparison}

In the previous section we found that hydrodynamic convection in a
vertically stratified shearing box in \textsc{PLUTO} (without explicit
viscosity) can drive outward angular momentum transport. Here we
verify this result using the finite-volume code \textsc{ATHENA}
\citep{stone2008athena, stone2010implementation}. To facilitate as
close a comparison as possible between the two codes and also to the
\textsc{ZEUS} runs of SB96, we initialize both
codes with Stone and Balbus's convectively unstable profile (see
Appendix \ref{SB96profile}). Explicit diffusion coefficients were
omitted and vertical periodic boundary conditions implemented. Both
simulations were run in a shearing box of size $4H\times4H\times4H$ and
at a resolution of $64\times64\times64$. \textsc{PLUTO} and
\textsc{ATHENA} offer somewhat different suites of numerical schemes:
here we settle on a combination that is slightly more
diffusive than that employed in our fiducial simulations because this
combination allows for as close a match as possible. Specifically, 
the numerical scheme employed in \textsc{ATHENA} is
(second-order) piecewise linear interpolation on primitive variables,
the HLLC Riemann solver and MUSCL-Hancock integration. In
\textsc{PLUTO} we use (second-order) piecewise linear interpolation on
primitive variables together with a Van Leer limiter function, the
HLLC Riemann solver and MUSCL-Hancock integration. The angular
frequency, and the  sound speed at initialization, 
in both simulations were set to $\Omega = 10^{-3}$ and $c_s = 10^{-3}$, respectively.

\begin{figure}
\centering
\includegraphics[scale=0.115]{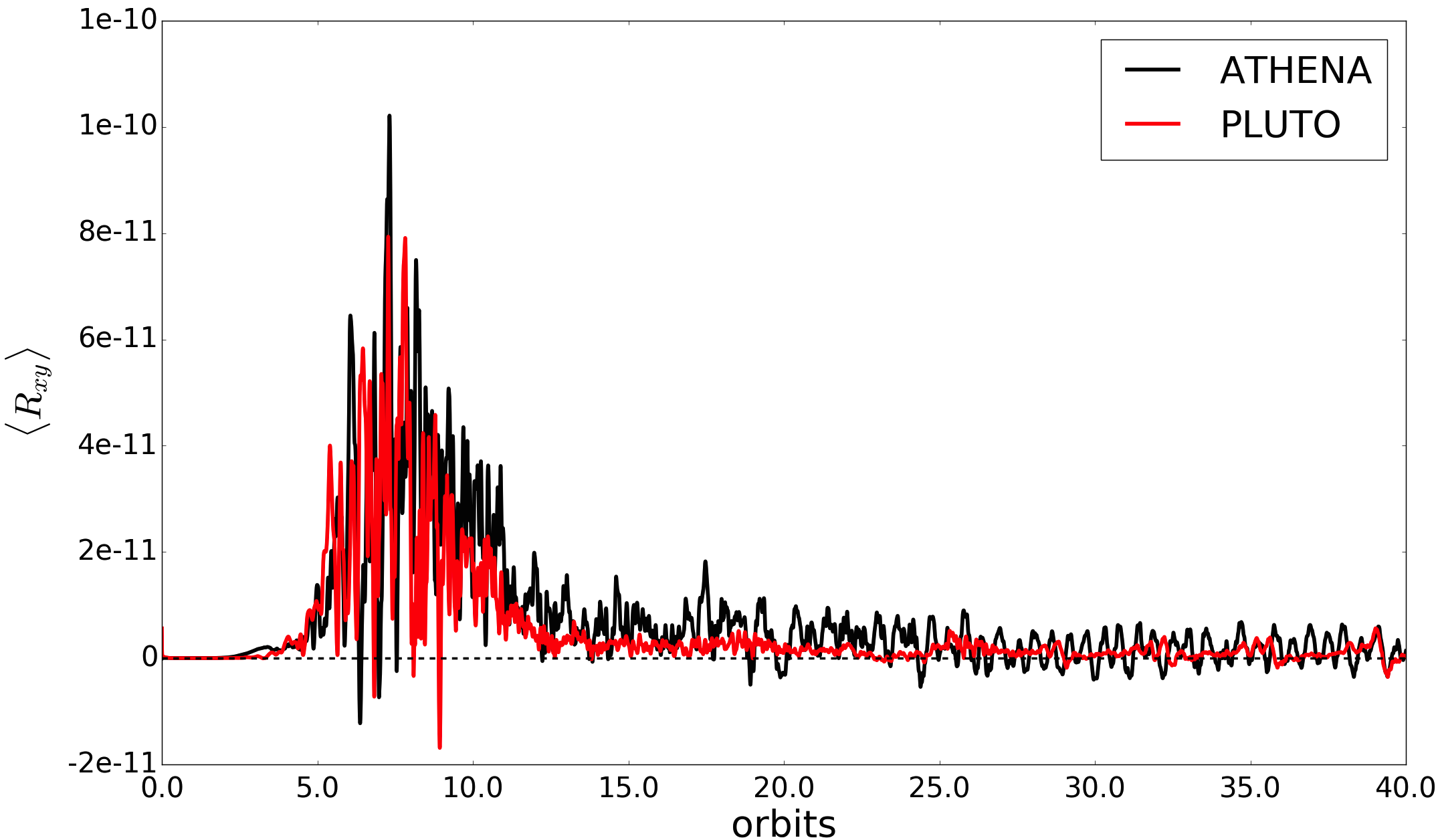}
\caption{Comparison of time-evolution of volume-averaged Reynolds stress taken from a simulation run in \textsc{ATHENA} (black) with the same quantity taken from a simulation run in \textsc{PLUTO} (red).}
\label{athenaplutocomparisonfigure}
\end{figure}

\begin{figure*}
	\centering
	\includegraphics[scale=0.35]{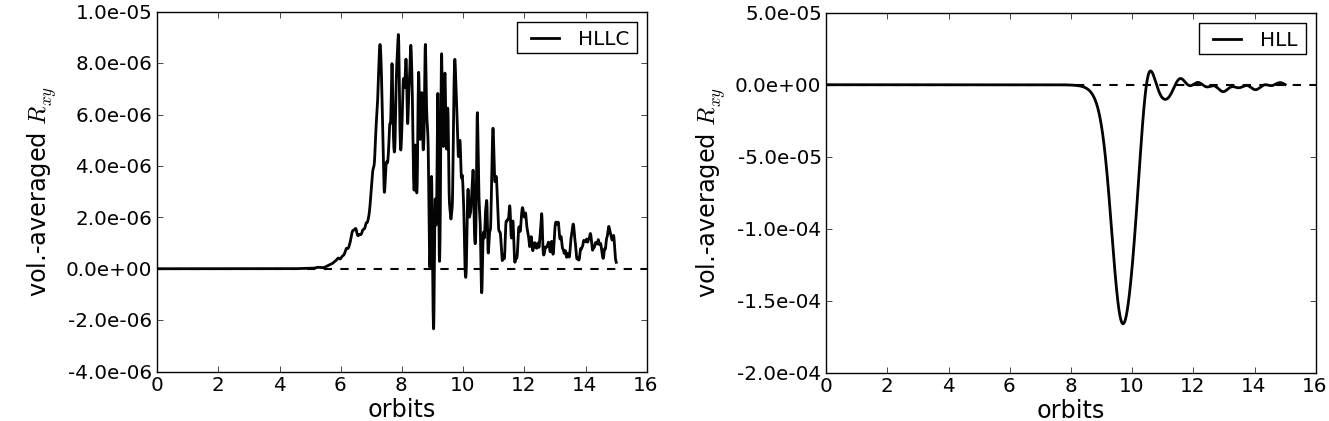}
    \caption{Left: time-evolution of volume-averaged Reynolds stress tensor taken from a \textsc{PLUTO} simulation run with the less diffusive HLLC solver. Right: the same, but taken from a simulation run with the more diffusive HLL solver. The change in the sign of the stress tensor provides compelling evidence that whether convection can transport angular momentum inwards or outwards depends on the diffusivity of the underlying numerical scheme.}
    \label{HLLHLLCfigure}
\end{figure*}

We find that angular momentum transport is directed \textit{outwards} in both codes, demonstrating that the outward transport of angular momentum by hydrodynamic convection in the non-linear phase is robust to a change of code. Figure \ref{athenaplutocomparisonfigure} compares the time-evolution of the volume-averaged Reynolds stress taken from the \textsc{ATHENA} simulation with that taken from the \textsc{PLUTO} simulation. Both simulations exhibit exponential growth followed by non-linear saturation, together with the development of convective cells (not shown).
These results also contrast with the SB96 runs with ZEUS and thus demonstrate that the positive transport reported in previous sections
is not special to the Gaussian temperature profile.

Although for this particular combination of schemes the overall result
is the same, we noticed that different schemes resulted in qualitative
differences between the two codes. For example, when we use the HLLC
solver, piecewise parabolic reconstruction (PPM) and
Corner-Transport-Upwind (CTU) integration, \textsc{ATHENA} exhibits
delayed onset of instability, a more gradual drop in kinetic energy
density following non-linear saturation, and a slower drop in angular
momentum transport compared to \textsc{PLUTO}. When we use the
\text{Roe} solver, PPM reconstruction and CTU integration, the
Reynolds stress is highly oscillatory in time in both codes,
indicative, perhaps, of numerical instability. The different behavior
(in both codes) based on which combination of numerical schemes is
chosen is worrying, although we emphasize that the differences are  
probably accentuated by the transient nature of the evolution and
its sensitivity to the initial conditions, and the fact that the
linear phase is partially controlled by grid diffusion. The robustness 
of the results to changes is numerical scheme is investigated in more detail in the following section. 

\subsection{Sensitivity of sign of angular momentum transport to numerical scheme}
\label{AMtransportnumericaldiffusivity}

An important result of Sections
\ref{fiducialsimulations}-\ref{athenaplutocomparison} is that purely
hydrodynamic convection in $\textsc{PLUTO}$ and in \textsc{ATHENA}
resulted in $R_{xy} > 0$, i.e. outward angular momentum
transport. This is in disagreement 
 with the \textsc{ZEUS} results of SB96, who reported a Reynolds stress of $R_{xy} < 0$. 
Given that \textsc{ZEUS} is a non-conservative, finite-difference
code, our hypothesis for explaining the discrepancy is that
\textsc{ZEUS} run at comparatively low resolution (as in SB96) 
is sufficiently diffusive that it imposes an
artificial near-axisymmetry on the flow (which will send angular
momentum inward). In this section we report our attempts to assess 
the numerical diffusivity of various algorithms and their impact on convection.

First we reran the fiducial simulations described in Section
\ref{setupandinitializationoffiducialsimulations} but with different
combinations of numerical schemes. We found that the
sign of angular momentum transport is indeed sensitive to our choice
of scheme. This is best illustrated in Figure \ref{HLLHLLCfigure}: the
left-hand panel shows the time-evolution of the volume-averaged
$R_{xy}$ taken from a simulation run with
the HLLC Riemann solver, while the right-hand panel shows the same
quantity but taken from a simulation run with the (more diffusive)
 HLL solver. Both
simulations exhibit similar exponential growth in the vertical
kinetic energy during the linear phase of the instability, and the
velocity field shows the development of convective cells in both
cases, but the sign of the Reynolds stress is radically
different. The HLL run is also far more laminar and axisymmetric.
We next repeated the HLL runs with higher
   resolutions, up to $256^3$, but with no change in the sign of
   $R_{xy}$. It must be stressed that going to higher resolutions does
   not necessarily help in the
   problem of convection;
this is because the fastest growing linear modes are always near the
grid scale and hence it is impossible (in the linear phase at least)
to escape grid effects.

The results of different combinations of schemes is summarized in
Table \ref{tablenumericalschemes}. They indicate that the sign of $R_{xy}$
appears to be robust to changes in the interpolation and time-stepping
schemes, but sensitive to the Riemann solver. In particular, the less
diffusive Riemann solvers (Roe and HLLC) gave $R_{xy} > 0$, while the more diffusive Riemann solvers (HLL and a simple Lax-Friedrichs solver) gave $R_{xy} < 0$.
Altogether we explored twelve different configurations of Riemann
solver, interpolation scheme, and time-stepping method. These ranged
from the most accurate (least diffusive) set-up which was identical to
that employed in the simulations described in the previous section (a
Roe Riemann solver, third-order-in-space WENO interpolation and
third-order-in-time Runge-Kutta time-stepping), to the least accurate
(most diffusive) set-up (a simple Lax-Friedrich's Riemann solver,
second-order-in-space linear interpolation and second-order-in-time
Runge-Kutta time-stepping). 

Thus we have compelling evidence that, as suspected, angular momentum
transport due to convection can be sensitive to the diffusivity of the
underlying numerical scheme. It appears
that over-smoothing of the flow by diffusive Riemann solvers, such as
HLL, or by codes such as \textsc{ZEUS} that employ artificial
viscosity and finite-differencing of the pressure terms, impose
a spurious axisymmetry on the flow.  
This axisymmetry picks out the axisymmetric convective modes, which in turn transport angular momentum inwards.

\subsection{Viscous simulations}

\label{simswithexplicitdiffusioncoeffs}

Given the concerns raised in the last section regarding numerical schemes,
as well as the fact that the fastest growing inviscid modes are on the
shortest scales, we expand our study to include
explicit viscosity (and thermal diffusivity). A properly resolved
viscous simulation should exhibit none of the numerical problems encountered above, and the fastest
growing mode occurs on a well defined scale above the (resolved) viscous scale.
Our main aim in this section is to test
 whether the results of our fiducial simulations are solid: mainly,
 if angular momentum transport can be positive in the presence of
 viscosity.
 Additionally,
 the inclusion of explicit diffusion coefficients enables us 
 to investigate the Rayleigh number dependence of
 fully compressible hydrodynamic convection in the shearing box, and
 thus connect to previous work by Lesur and Ogilvie (2010).

We carry out a suite of simulations at a resolution mainly of $256^3$
 investigating the effects on hydrodynamic convection when the
Rayleigh number $\text{Ra}$ is increased, but the Prandtl number
$\text{Pr}$ is fixed at 2.5.\footnote{We have also repeated some of the simulations at a Prandtl number of unity. The differences with the $\text{Pr}=2.5$ simulations are nominal.} Thus viscosity and thermal diffusivity each have
the same magnitude in any given simulation, and we decrease both in
order to increase the Rayleigh number. 
The Rayleigh numbers of the simulations are
$\text{Ra} = 10^{5}, 10^{6}, 10^{7}, 10^{8}, 10^{9}$, and $10^{10}$.
(Note that the simulations undertaken at the two highest Ra are probably
underresolved,
as explored later.)
The Richardson number at
initialization is fixed by the initial vertical
profile,
which is described in
detail in Section \ref{setupandinitializationoffiducialsimulations}
(and shown in Figure \ref{Gaussianprofileverticalstructure1}). For the
profile parameters chosen, the Richardson
number at initialization is $\text{Ri} \sim 0.05$. Further details of the
simulations are given in  
Table \ref{simsexplicitdiffusioncoeffs} in Appendix \ref{tables}. 

\subsubsection{Rayleigh number dependence} 

As the Rayleigh number is increased from low to high values 
the system proceeds through the same sequence of states found
by Lesur and Ogilvie (2010).
We observe no instability for $\text{Ra} = 10^{5}$. At $\text{Ra}
= 10^{6}$ instability occurs but the flow appears relatively laminar
and axisymmetric; in particular the Reynolds stress is \emph{negative}
througout the linear and nonlinear phases.
We conclude that the critical Rayleigh number for the onset of convection
lies in the range
$10^{5} < \text{Ra}_\text{c} < 10^{6}$. At $\text{Ra}= 10^{7}$ the
instability is more vigorous and the flow field significantly more
chaotic and non-axisymmetric in the nonlinear phase, at which point the
Reynolds stress has become positive. We conclude that the critical
Ra at which the sign of $R_{xy}$ flips lies between $10^6$ and
$10^7$. At higher Ra the flow appears even more turbulent and
non-axisymmetric. It is a relief that in the nonlinear phase of the 
instability we find agreement with the inviscid simulations of
previous sections at sufficiently high Ra.

Our critical Rayleigh number for the onset of convection is considerably higher than
in \cite{lesur2010angular}, who report a value of $\text{Ra}_c =
6900$, but this is perhaps explained by their much larger
Richardson number
$=0.2$ which remains constant throughout their box. 
On the other hand, the critical Ra at which $R_{xy}$ changes sign
is much closer to ours, which suggests that the breakdown into
non-axisymmetry may be controlled by secondary shear
instabilities that are more sensitive to the viscosity than the
convective driving. 

\begin{figure}
\centering
\includegraphics[scale=0.11]{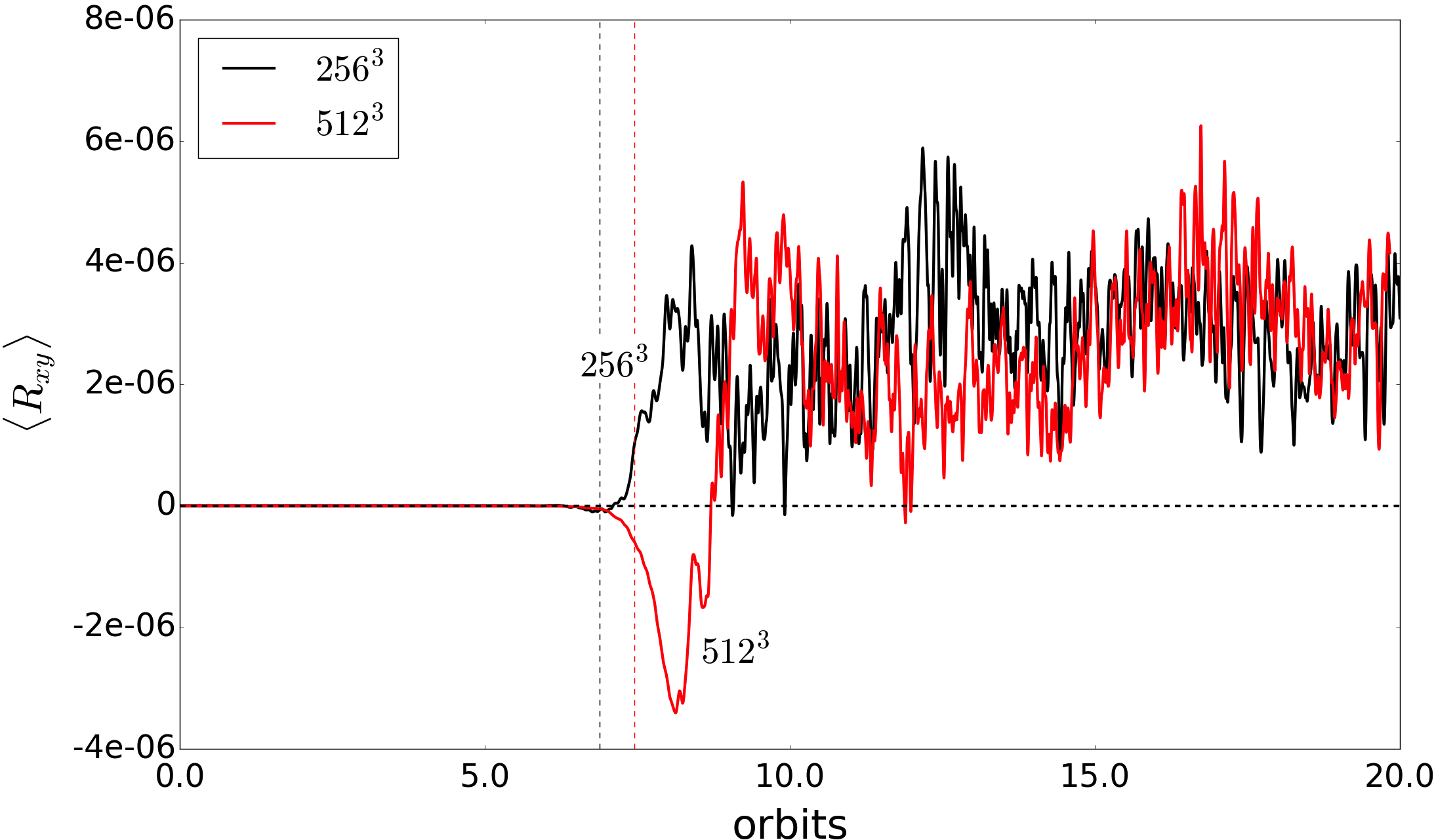}
\caption{Time-evolution of volume-averaged $xy$-component of Reynolds stress tensor from two simulations run at resolutions of $256^3$ (black curve) and $512^3$ (red) curve at a Rayleigh number of $\text{Ra} = 10^{9}$. The vertical dashed lines mark the end of the linear phase in the $256^3$ (black) and $512^3$ (red) simulations, respectively.}
\label{Rayleighnumbercomparison}
\end{figure}

While the transport of angular momentum is unambigously outward 
after nonlinear saturation when $\text{Ra} \geq 10^{7}$, it is almost
always inwards
during the linear phase, as is illustrated by the red curve in
Fig.~\ref{Rayleighnumbercomparison}.
Moreover, during this early stage, the
simulation is dominated by the axisymmetric modes of
Section 3. This disagrees with our inviscid fiducial simulations (see
discussion in Section 4.1.2), which are non-axisymmetric from the
outset. We confess that our viscous simulations are
more in tune with our physical intuition: (a) in the linear phase 
growing axisymmetric modes outcompete shearing waves, (b) at
sufficiently large amplitudes the modes are subject to secondary shear
instabilities in the $xz$-plane that buckle the rising and falling
plumes, (c) perhaps concurrently or on a short time later (and viscosity permitting),
secondary \emph{non-axisymmetric}
instabilities also attack the plumes because they exhibit
significant shear in the $xy$-plane as well, (d) at this point, the flow
degenerates into something more disordered, and importantly,
non-axisymmetric
and
the Reynolds stress flips sign. We conclude that 
viscosity preferentially damps shearing waves vis-a-vis axisymmetric
modes, or effectively kills off
the artificial aliasing of shearing waves in inviscid simulations (if
this is present).  

\subsubsection{Convergence with resolution}


A final issue is whether our viscous simulations are
adequately resolved. More specifically: above what critical Ra is a grid of $256^3$ points
inadequate? We conducted
simulations at Ra$=10^8$ with
$256^3$ and $512^3$ grid points and found generally good agreement
between the two. 
The linear growth 
rates were almost identical and the ultimate nonlinear state statistically similar.
The only noticeable difference was in the peak Reynolds stress, which was somewhat larger
in the higher resolution run. Overall, we conclude that $256^3$ grid points are
adequate to resolve a simulation with Ra$=10^8$.

Things start to deteriorate at a Rayleigh number of $\text{Ra} = 10^{9}$.
In Figure \ref{Rayleighnumbercomparison} we plot
the time-evolution of the $xy$-component of the volume-averaged
Reynolds stress for two simulations run at $256^3$ (black) and $512^3$ (red). 
The lower resolution run possesses no
extended period of negative $R_{xy}$, in contrast to the runs at
Ra$=10^8$. 
We speculate that 
at this resolution physical viscosity is subdominant to the grid and
non-axisymmetric disturbances are artificially enhanced,
probably via aliasing. At $512^3$, the stress is definitely negative
in the linear phase and there is a strong negative peak shortly
afterward. 
The physical viscosity is now permitted to work properly
and appears to prohibit artificial non-axisymmetric disturbances. As a consequence,
the
axisymmetric modes preserve
their
control of the simulation for significantly longer and the simulation is
in accord with those at lower Ra. More
reassuring
is the nonlinear phase a little later in which the two flows closely
resemble each other. We conclude that in the linear phase $256^3$ is
insufficient to describe a simulation of $Ra=10^9$, 
but in the later nonlinear phase it is probably adequate.

\begin{figure*}
\captionsetup[subfigure]{labelformat=empty}
\centering
\subfloat[]{\includegraphics[scale=0.21]{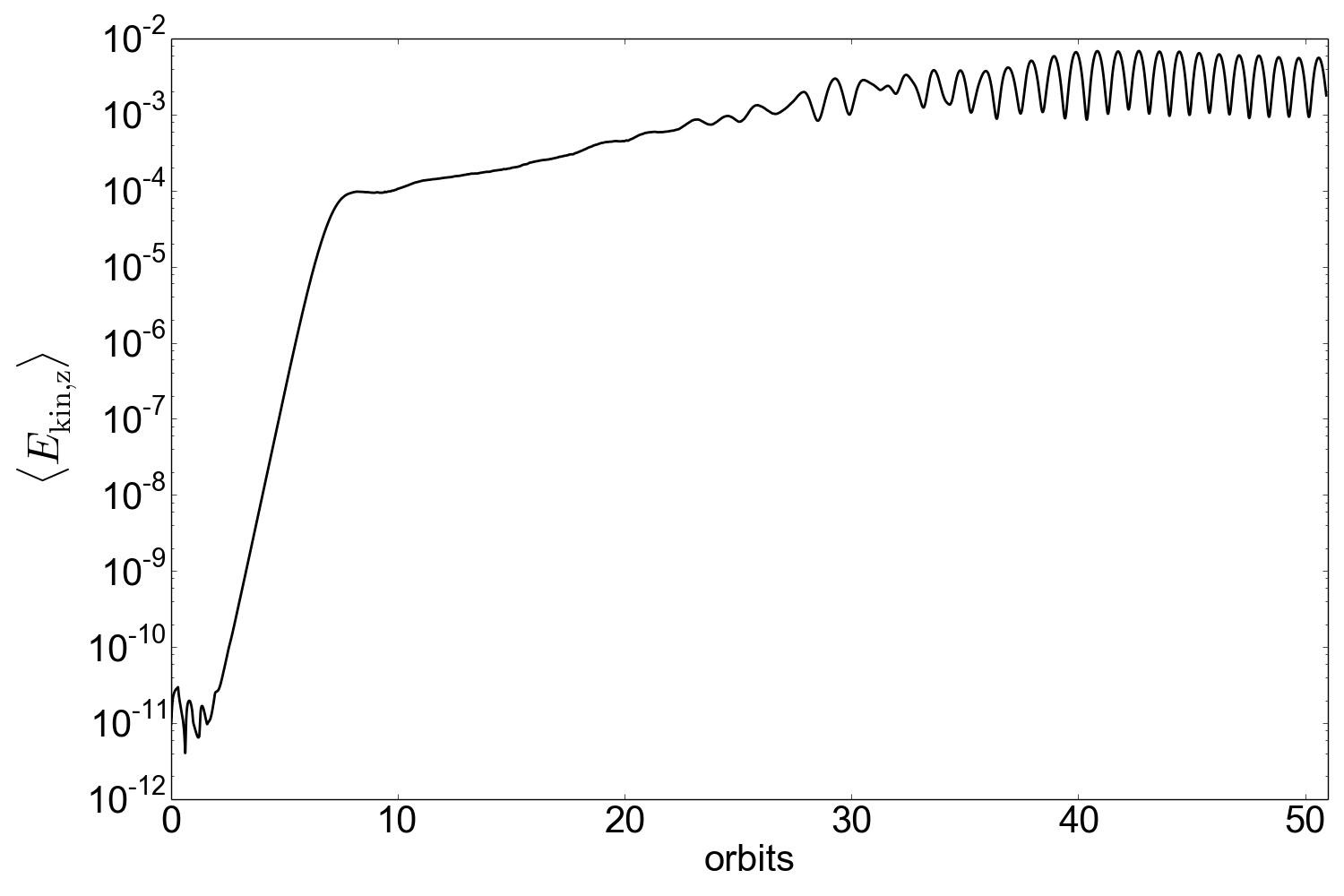}}\hspace{0.2em}
\subfloat[]{\includegraphics[scale=0.15]{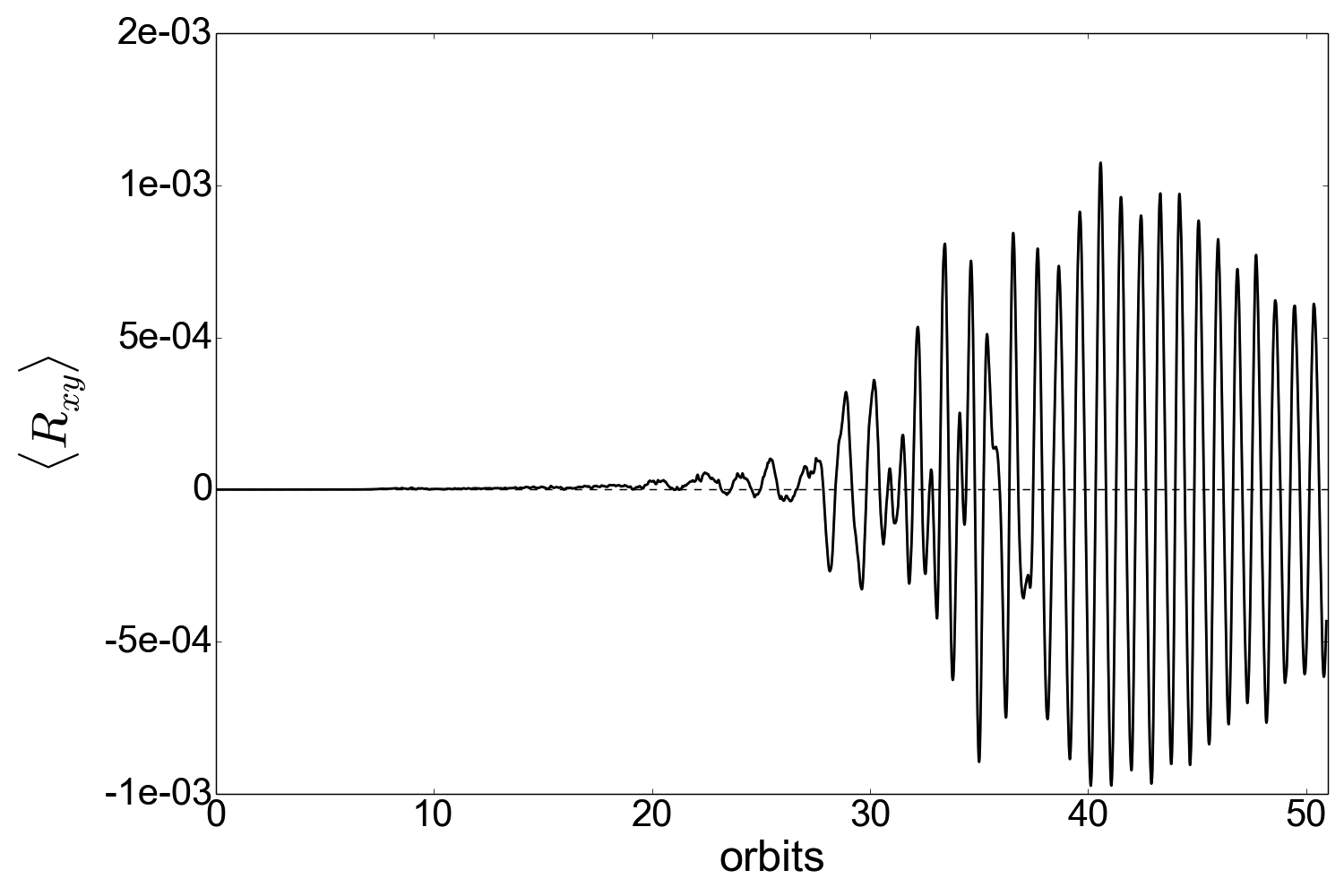}}\hspace{0.2em}
\caption{Left: semi-log plot of time-evolution of volume-averaged vertical kinetic energy density in a simulation with thermal relaxation. Right: time-evolution of volume-averaged $xy$-component of the Reynolds stress tensor. The thermal relaxation time was set to $\tau_\text{relax} = 4.2\, \Omega^{-1}$.}\label{thermalrelaxfig01}
\end{figure*}

\section{Structures in forced compressible convection}
\label{forcedsims}
In Section \ref{unforcedsims} we initialized our simulations with a
convectively unstable vertical profile, but otherwise did not
include any source of heating or cooling. Convection (and to a much
lesser extent, conduction) transferred heat and mass vertically so as
to zero
the buoyancy frequency and send the box into a
convectively stable equilibrium.  
As a result, we were only able to probe non-linear convection for a short period of time.
Now we aim to continually sustain convection in order to explore this phase in greater depth. In the absence
of self-sustaining convection, this means we have to maintain the
convectively unstable profile artificially. 
\subsection{Set-up}
\label{thermalrelaxation}

SB96 perpetuated convection by forcing the temperature at the mid-plane to
adjust to its value at initialization. This strategy, however, raises
problems in conservative codes, such as \textsc{PLUTO} and
\textsc{ATHENA}, because the energy injection at the midplane has no
way to leave the box, except through the distant vertical boundaries
(in \textsc{ZEUS} numerical losses on the grid supply a
turbulence-dependent `cooling', in addition to the cooling facilitated through the fixed-temperature boundaries).
In practice, we found that the disk heats up to an inordinate level and, more
importantly, it settles into a marginally unstable state,
rather than a driven convective state.

Rather than forcing the code in this way,
we mimic the effects of \emph{both} heating and cooling through thermal
relaxation. The idea is to add a source term to the energy equation
such that the vertical internal energy profile relaxes to its value
at equilibrium on a time-scale $\tau_\text{relax}$.  
Although this technique is artificial, it serves as a very basic tool
for approximating the effects of realistic heating and cooling, as
might be supplied by MRI turbulence and radiative transfer. The timescale
$\tau_\text{relax}$ then would be the characteristic time that the radiative
MRI 
system achieves thermodynamical quasi-equilibrium. In our simulations,
however, we take the relaxation timescale to be equal to the linear growth
time of convection. 

In addition, to mitigate the effects of
 mass outflows, we incorporate a mass source term to the
simulations. At the end of the
$n$th time-step we calculated the total vertical mass loss through the
upper and lower vertical boundaries. We then added this mass back into
the box in
the $(n+1)$th time-step with the same vertical profile used to
 initialize the density (c.f.\ Equation \eqref{gausstempdensity}).

We implement the thermal relaxation term by making slight
modifications 
to \textsc{PLUTO}'s built-in optically thin cooling module. 
The thermal energy equation is updated during a substep to 
take into account user-defined sources of heating and cooling. 
The resolution employed in the simulation described in this section is
$256\times256\times256$ in a box of size $6H\times6H\times4H$,
corresponding to about $43$ grid cells per scaleheight in the $x$- and
$y$-directions and $64$ grid cells per scaleheight in the vertical
direction. The numerical set-up, boundary conditions and initial
conditions are the same as those describe in Section
\ref{setupandinitializationoffiducialsimulations}. The thermal relaxation
time is taken to be equal to the inverse of the growth rate of the
convective instability which we measured to be $\sigma =
0.2404\,\Omega$. Thus the thermal relaxation time is
$\tau_\text{relax} = 4.2\, \Omega^{-1}$, i.e. the internal energy is
relaxed back to its equilibrium profile on about  0.7
orbits. We have also included  an explicit viscosity and thermal 
diffusivity of $\nu = \chi = 1.075\times10^{-5}$ corresponding to a Rayleigh number of $\text{Ra} = 10^{9}$ and a Prandtl number of $\text{Pr} = 1$.

Finally, we have repeated the simulations in a cube of resolution
$64^{3}$ and size $4H\times4H\times4H$, as well as at a resolution of
$128\times128\times64$ in a box of size $6H\times6H\times4H$, and also
with periodic boundary conditions in  
the vertical direction, and observed similar results. We have also run
a simulation with the HLL solver to partially explore any code
dependence.

\begin{figure}
\centering
\subfloat[]{\includegraphics[scale=0.15]{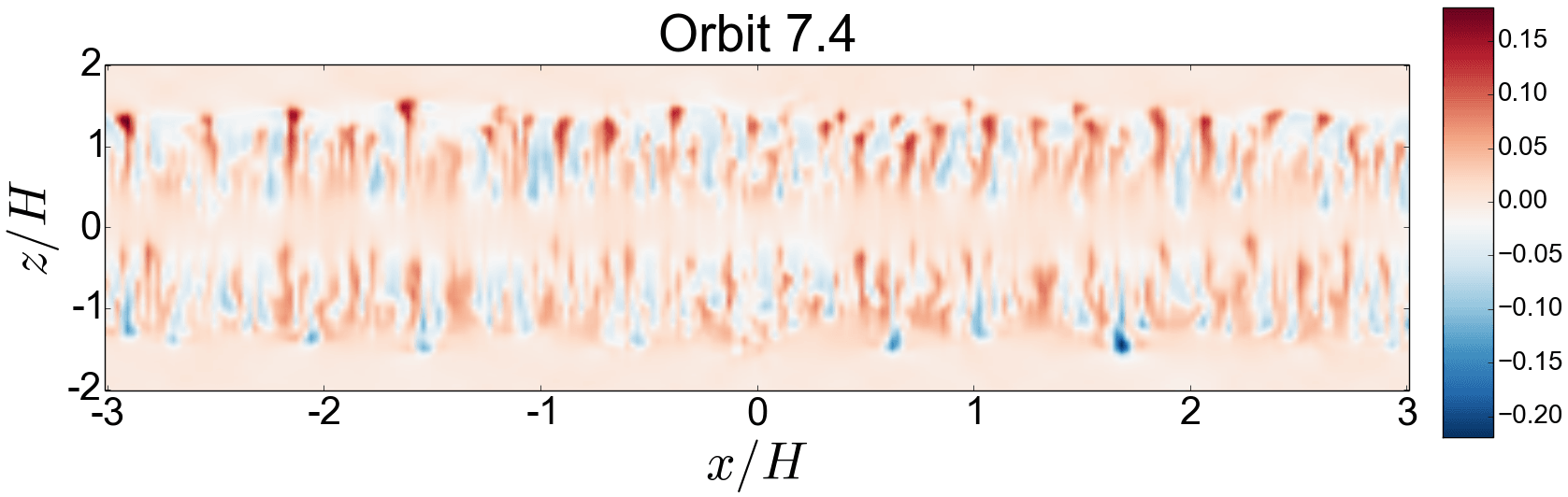}}\vspace{0.04em}
\subfloat[]{\includegraphics[scale=0.15]{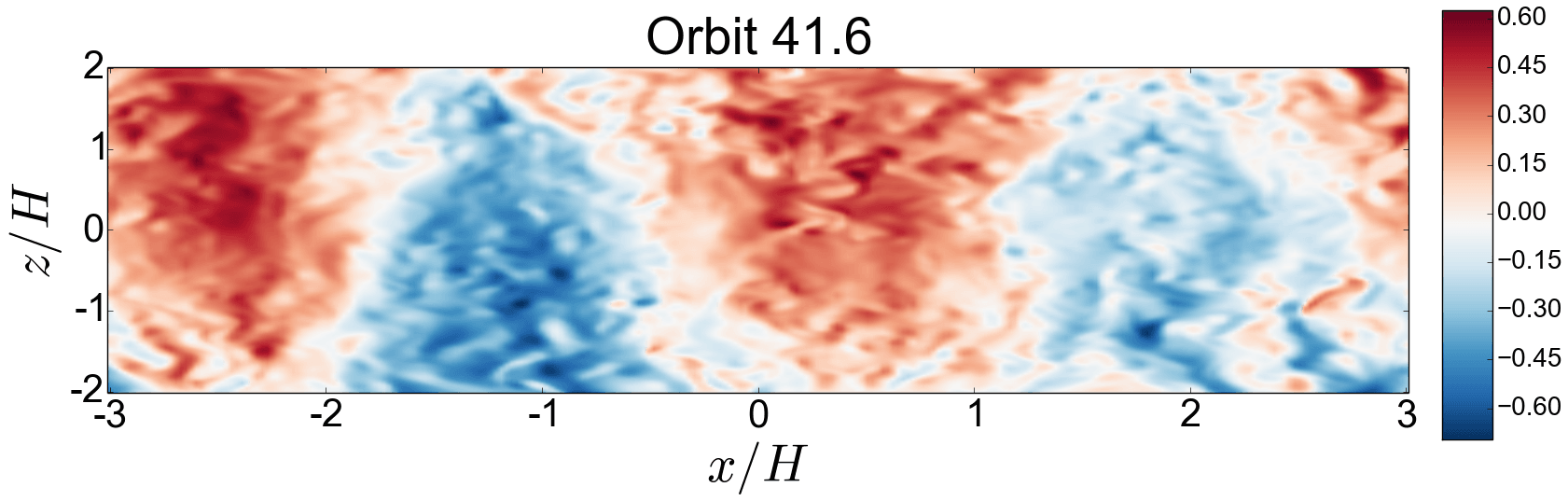}}\vspace{0.04em}
\qquad
\subfloat[]{\includegraphics[scale=0.15]{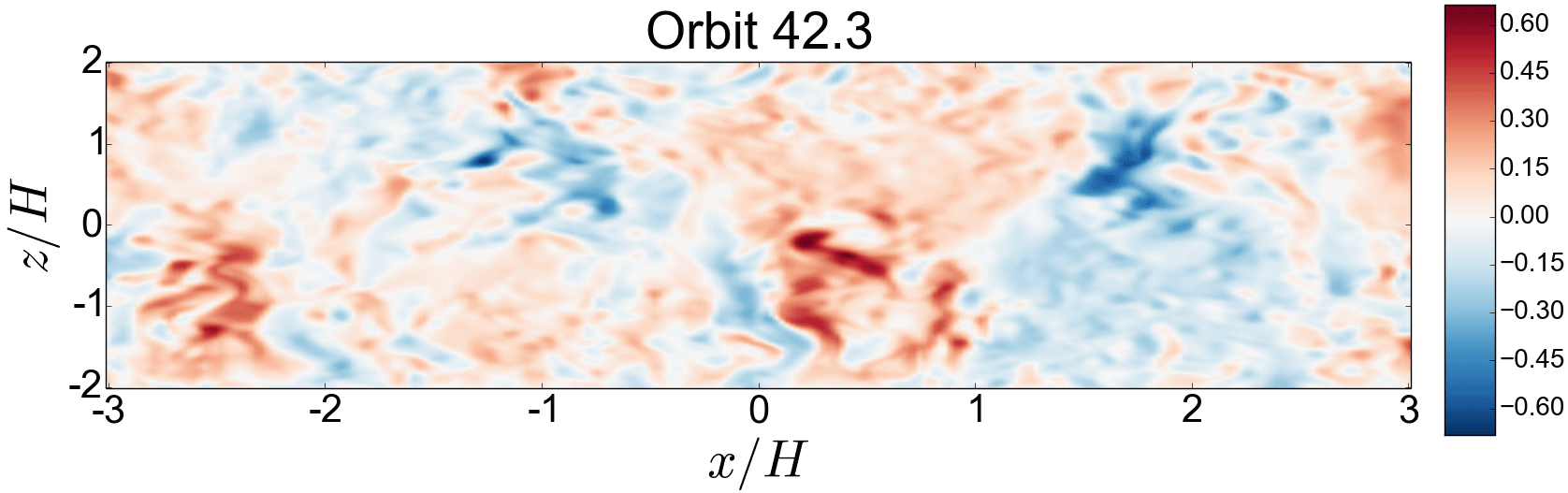}}\vspace{0.04em}
\subfloat[]{\includegraphics[scale=0.15]{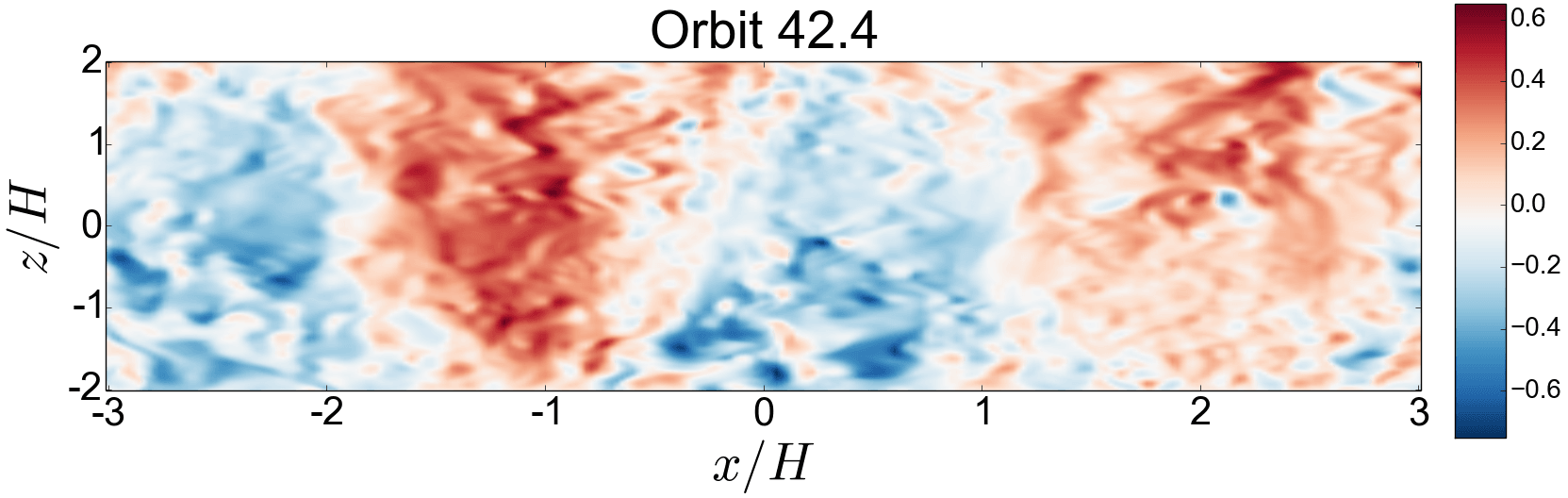}}\vspace{0.04em}
\qquad
\subfloat[]{\includegraphics[scale=0.15]{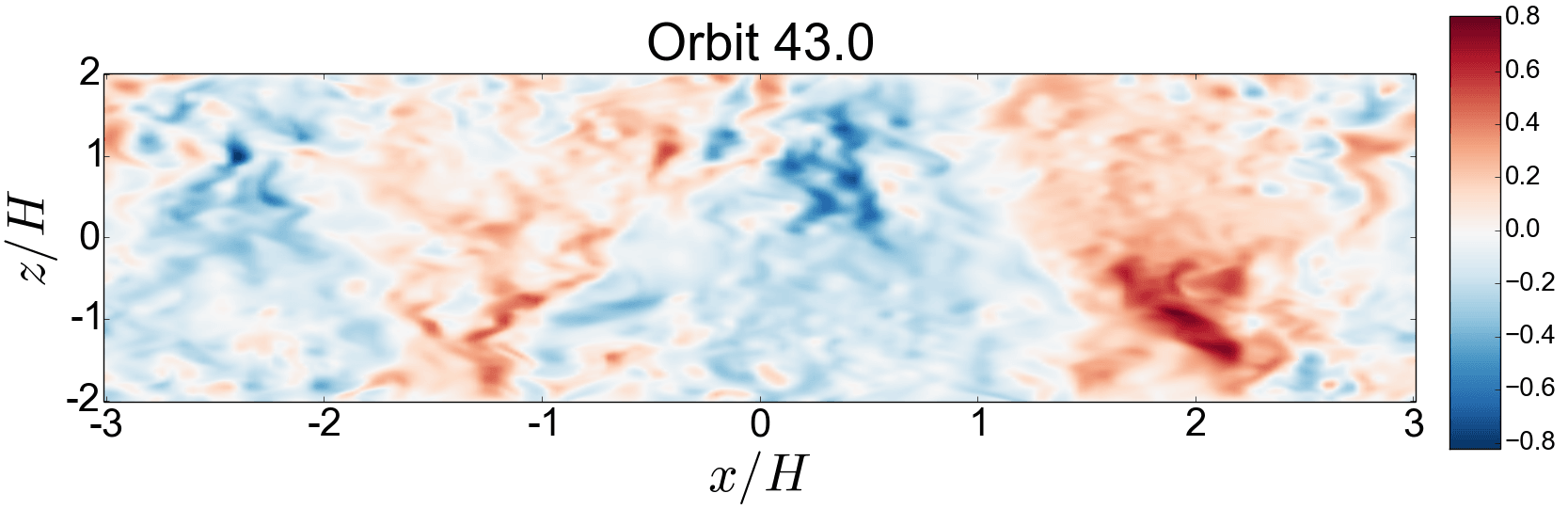}}\vspace{0.04em}
\subfloat[]{\includegraphics[scale=0.15]{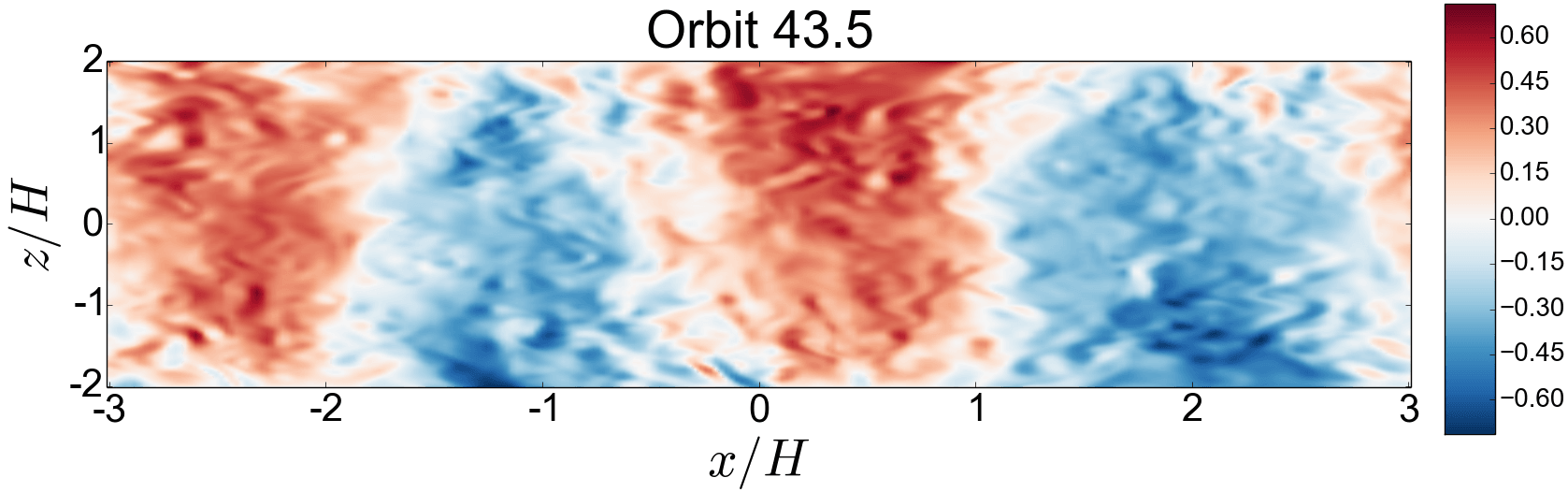}}\vspace{0.04em}
\caption{Snapshots at different times of the $z$-component of the velocity in the $xz$-plane taken from a simulation in which convection was sustained using thermal relaxation. Narrow convective cells shown in (a) just after non-linear saturation merge to form large scale structures shown in (b), which are destroyed (c) and recreated (d) in a cyclical manner with the opposite rotation.}
    \label{CyclicalConvectionFig}
\end{figure}

\subsection{Large-scale oscillatory cells}
\label{LSCcells}

In the left-hand panel of Figure \ref{thermalrelaxfig01} we show the
time-evolution of the volume-averaged kinetic energy density
associated with the vertical velocity. As in Section
\ref{unforcedsims},
exponential growth in the linear phase is followed by non-linear
saturation. The forced simulations, however, do not subsequently
decay. Instead the vertical kinetic energy
 increases at a slower rate until about orbit 20, at which point oscillations in the
kinetic energy density begin to develop.  The cycles
increase in frequency and amplitude until about orbit 41 at which
point the system settles into a quasi-equilibrium, with
the volume-averaged vertical kinetic energy density oscillating
about $\langle E_\text{kin,z} \rangle  \sim 2\times10^{-3}$ and
the period remaining
steady at $\Delta t = 0.88$ orbits.

\begin{figure*}
\centering
\subfloat[]{\includegraphics[scale=0.13]{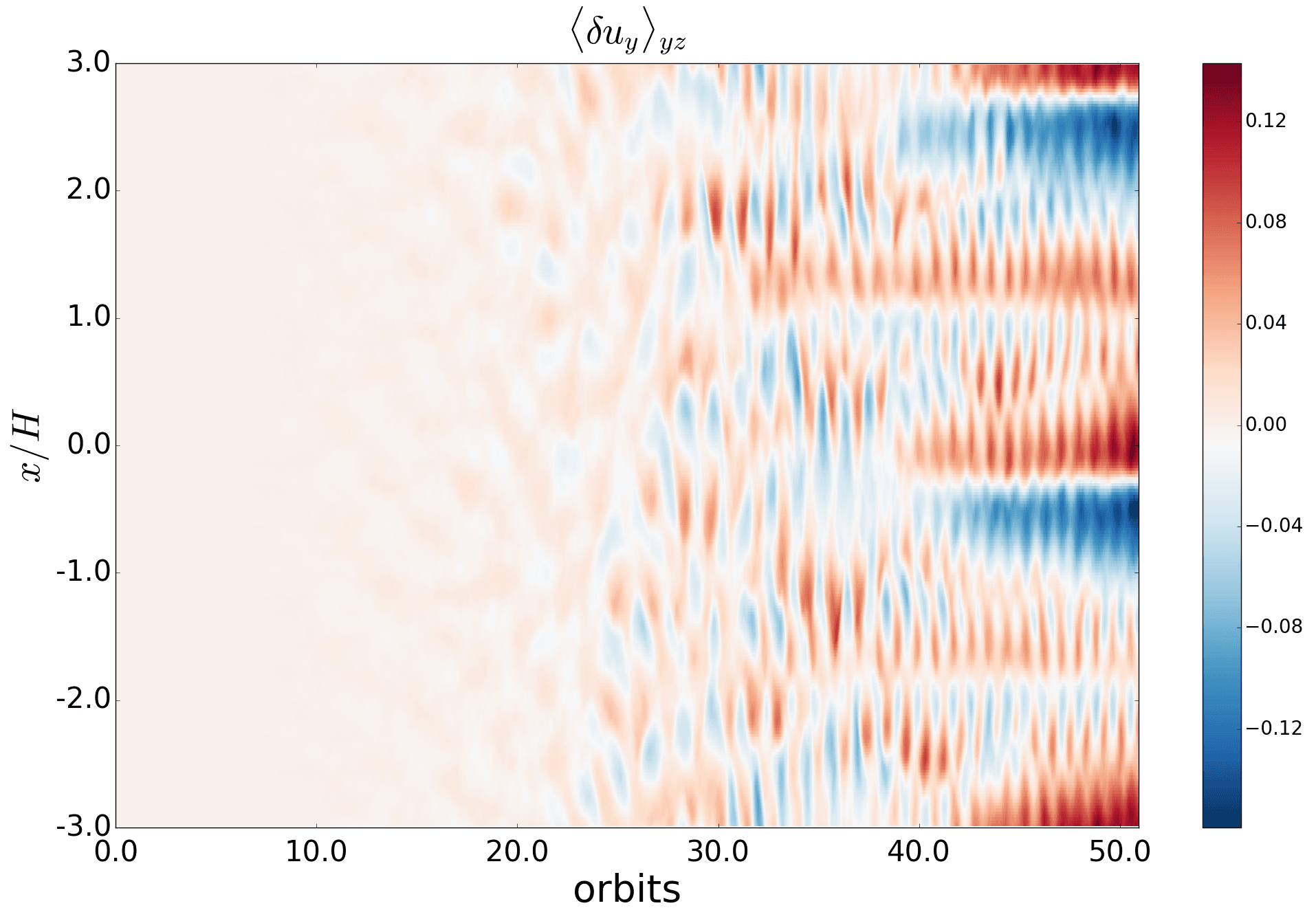}}\hspace{0.04em}
\subfloat[]{\includegraphics[scale=0.13]{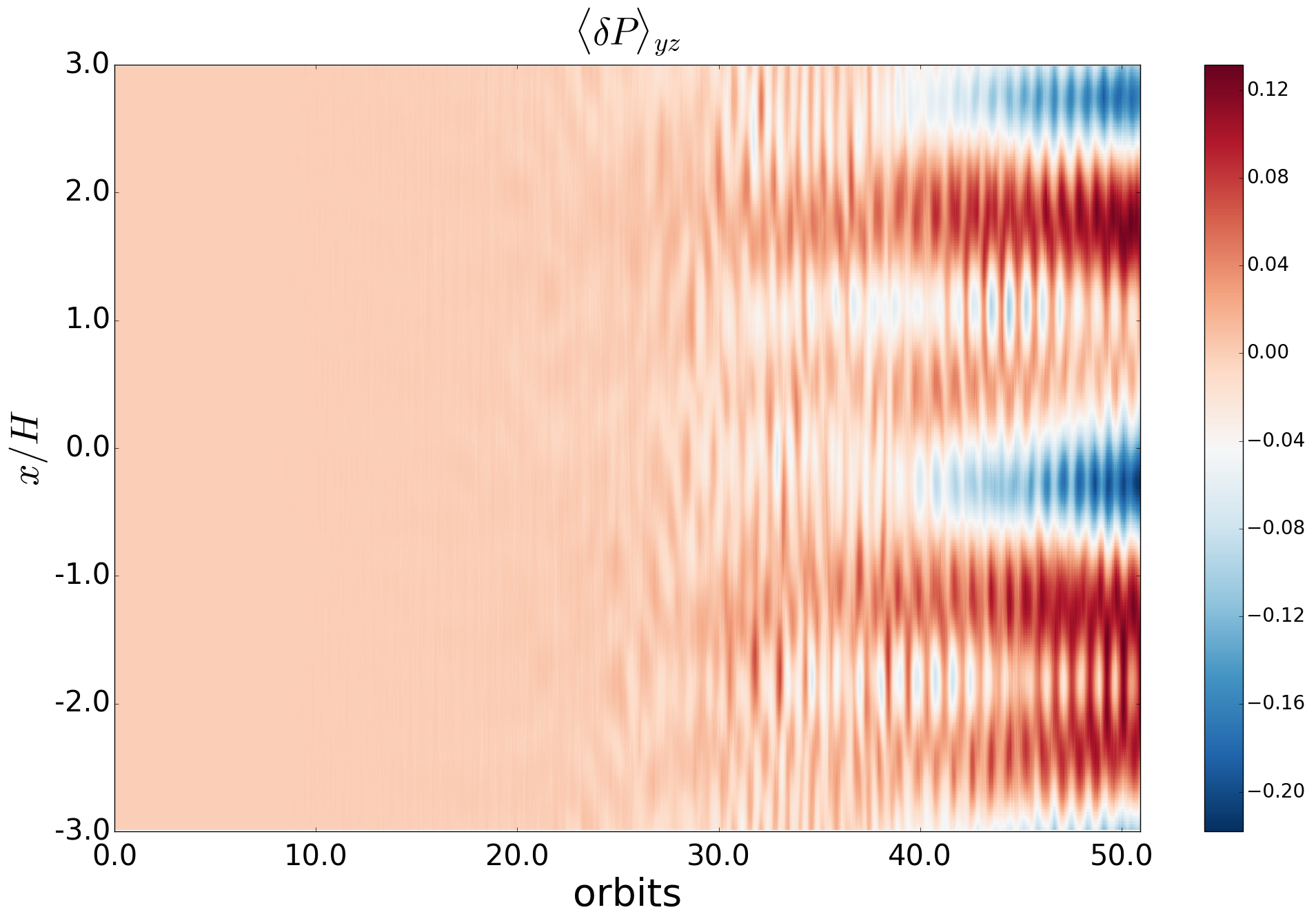}}\vspace{0.04em}
\caption{Space-time diagrams of $yz$-averaged $y$-component of the perturbed velocity $\delta u_y$ (left) and of the pressure perturbation $\delta P$ (right). (The perturbed pressure is defined in the first paragraph of Section \ref{vorticesandzonalflows}).}
	\label{zonalflowsfigure1}
\end{figure*}

Associated with the oscillations are large fluctuations in the
$xy$-component of the Reynolds stress tensor (shown in the right-hand
panel of Figure \ref{thermalrelaxfig01}). Instantaneous fluctuations
in $\langle R_{xy} \rangle$ and in $\langle \alpha \rangle$ may be
either positive or negative, but the time-averaged values over this
cyclical phase (orbit 20 to the end of the simulation) are $R_{xy}
\sim +9.9 \times 10^{-6}$ and $\alpha \sim +3.9 \times 10^{-5}$,
respectively. Furthermore, comparing the oscillations in the vertical
kinetic energy density to the fluctuations in $\langle R_{xy} \rangle$
and also to the volume-averaged gas pressure $\langle P \rangle$ (not shown
here) it is apparent that \textit{peaks} in the kinetic energy density
are correlated with both $\langle R_{xy} \rangle < 0$ and troughs 
 in $\langle P \rangle$, while \emph{troughs} in the kinetic energy density 
are correlated with $\langle R_{xy} \rangle > 0$ and peaks in $\langle P \rangle$.

\begin{figure*}
\centering
\subfloat[]{\includegraphics[scale=0.13]{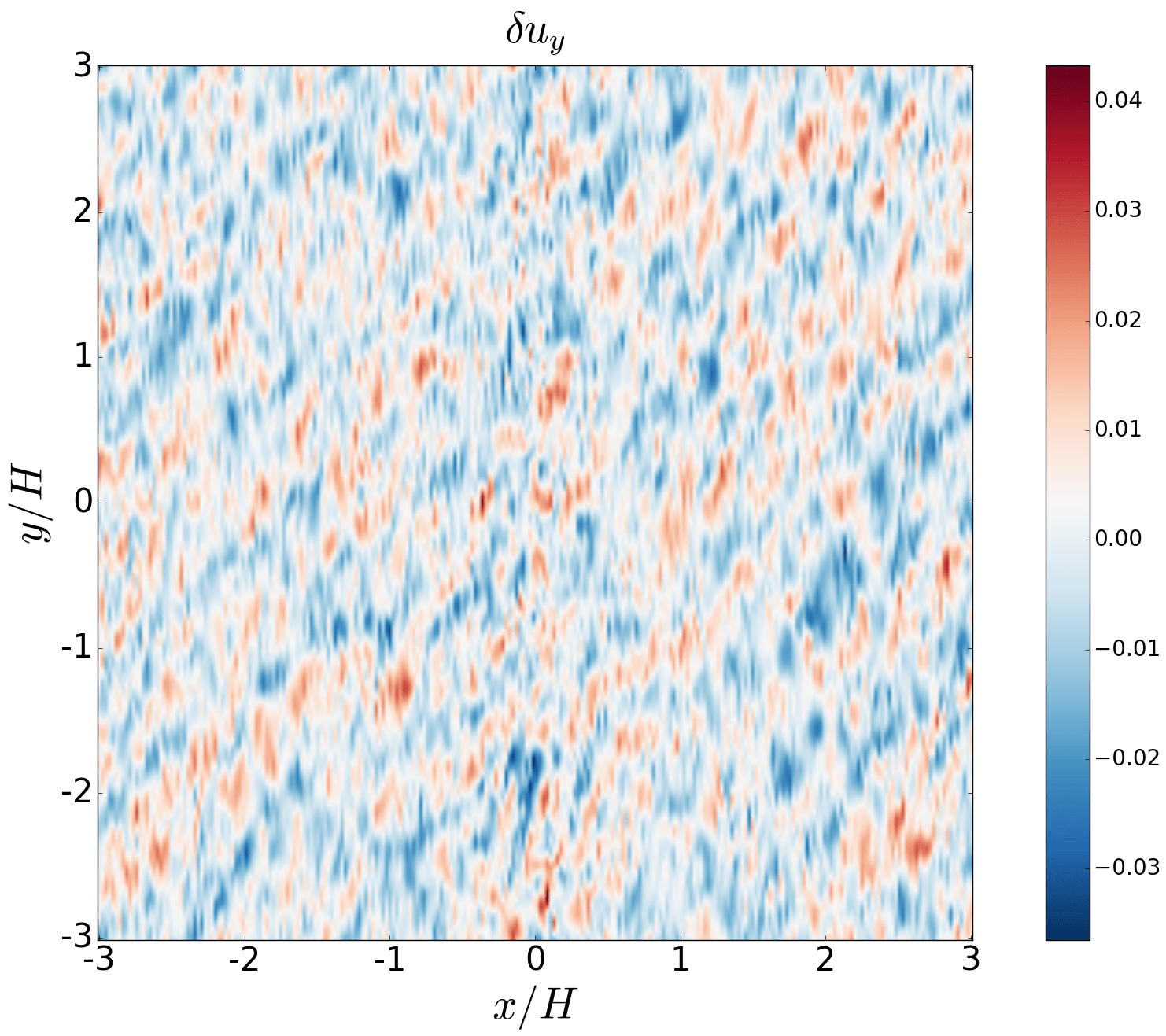}}\hspace{0.04em}
\subfloat[]{\includegraphics[scale=0.13]{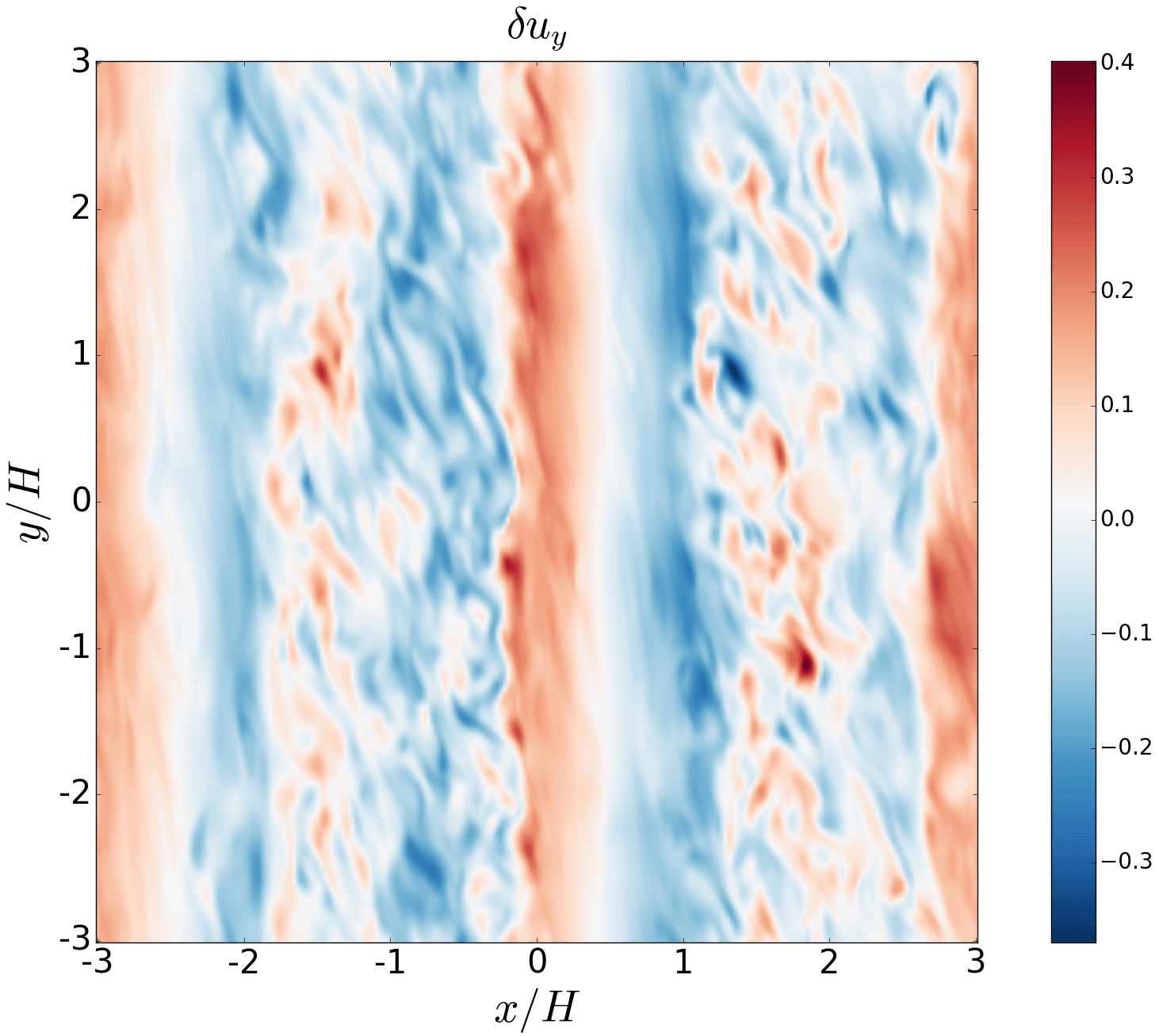}}\vspace{0.04em}
\qquad
\subfloat[]{\includegraphics[scale=0.13]{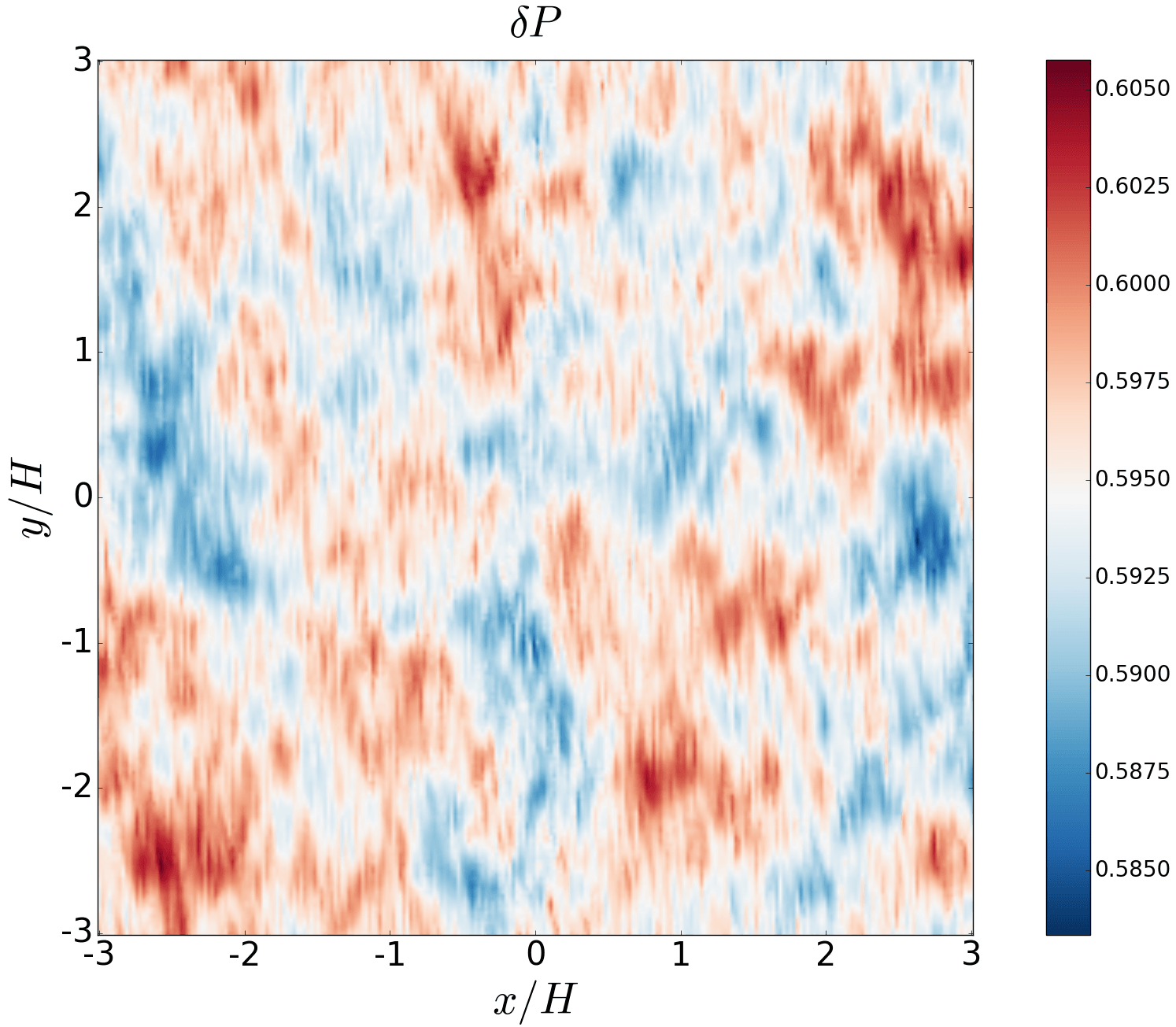}}\hspace{0.04em}
\subfloat[]{\includegraphics[scale=0.13]{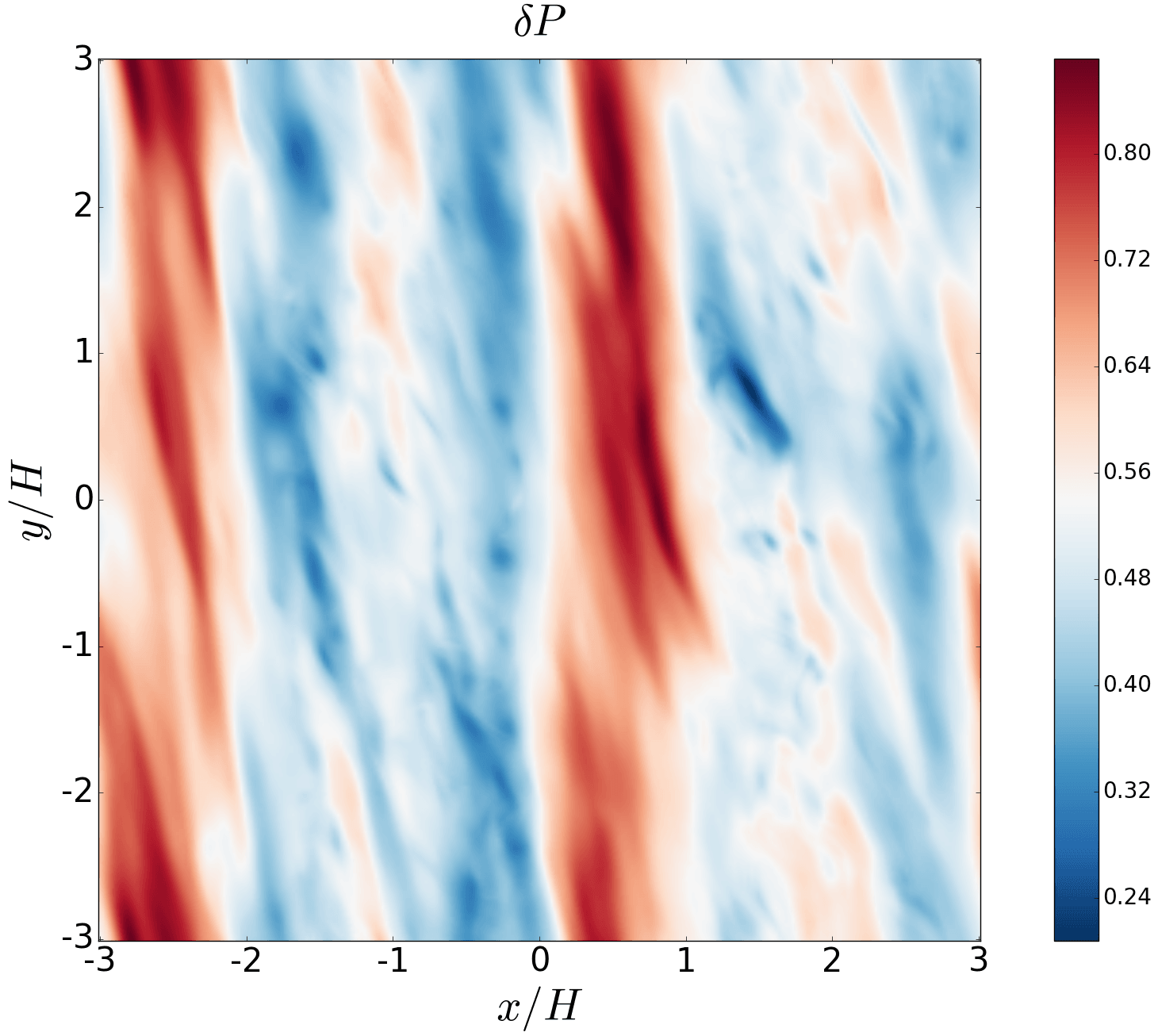}}\vspace{0.04em}
\caption{Two-dimensional slices in the $xy$-plane of the perturbation
  to the $y$-component of the velocity $\delta u_y$, and of the
  fractional pressure perturbation $\delta P$ from a simulation of
  forced compressible convection at $z\approx 0.5H$. 
 Left column: slices taken from a snapshot just after non-linear saturation (orbit 7.4). Right column: slices taken from snapshot generated during the cyclical phase (orbit 41.6).}
	\label{zonalflowsfigure}
\end{figure*}

A clearer picture of the behaviour of the system emerges when we study
the structure of the flow during the cyclical phase. Figure
\ref{CyclicalConvectionFig} shows snapshots of the $z$-component of
the velocity in the $xz$-plane. The first panel is taken just after
the end of the linear phase (orbit 7.4); the subsequent panels are
 taken at five successive peaks and troughs in the kinetic energy. As the
kinetic energy rises in the non-linear phase, the thin convective
cells with radial wavelengths $\lambda_x \sim 5\, \Delta x\sim 0.177H$, where
$\Delta x$ is the size of grid a cell in the $x$-direction, slowly
begin to merge into larger coherent structures which couple the two
halves of the disc together. By orbit 20 (start of the cyclical
phase), the radial wavelength of the convective cells has increased to
$\lambda_x \sim H$, and by the time the quasi-steady
equilibrium  state sets in (at around orbit 40) the radial wavelength 
of the convective cells is $\lambda_x \sim 3.4H$.

Comparing the snapshots of $u_z$ to the peaks and troughs in the kinetic energy, it becomes apparent that peaks in the convective energy are associated with the large-scale axisymmetric convective cells, hence $\alpha < 0$. Troughs in the kinetic energy are associated with destruction of those convective cells and with positive stress (outward angular momentum transport). Thus we are observing large-scale convective eddies that appear to be created and destroyed cyclically. Furthermore, it is evident from Figure \ref{CyclicalConvectionFig} that after the eddies are destroyed, they are re-formed with the opposite rotation.

The reader may be alarmed that the large eddies extend all the way
to the vertical boundaries of the domain, and indeed there is an uncomfortable level
of mass loss during this phase. To check that the flows are not an artefact of our box size, we ran additional simulations with a vertical domain of $\pm 3H$. A density floor 
had to be activated in such runs, which unfortunately compromised our thermal equilibrium
and complicated the onset of convection early in the simulation. However,
they ultimately 
settled onto the cyclical state described above, but now the large-scale
eddies peter out before reaching the vertical boundaries. As a result, the mass loss 
drops to negligible amounts. This confirms that these flows are
physical and robust, though only
marginally contained within our smaller boxes.

\begin{figure}
\includegraphics[scale=0.15]{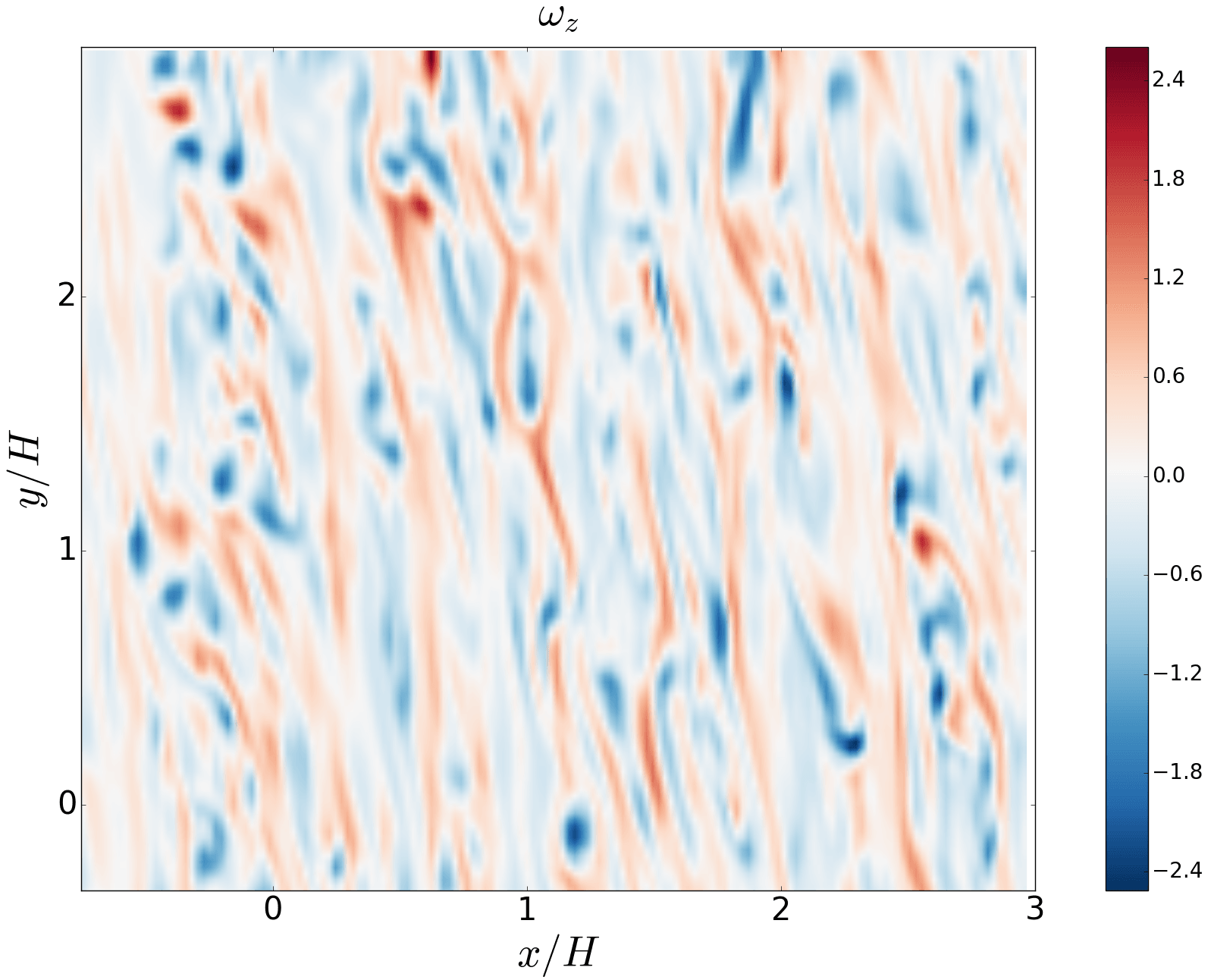}
\caption{Two-dimensional slice in the $xy$-plane of the $z$-component
  of the vorticity $\omega_z$ taken from snapshot generated at the
  start of the cyclical phase (orbit 20) and at $z\approx 0.5H$. 
For clarity we have zoomed in on the upper-right quadrant of the $xy$-plane.}
\label{vorticesfigure}
\end{figure}

Next, to explore any code dependence of this final
  outcome, we have rerun the simulation with the more
  diffusive HLL solver. We observe a
  negative Reynolds stress during the linear phase and well into the
  non-linear phase. Associated with this is a remarkably axisymmetric
  flow field, as confirmed by viewing slices of the pressure
  perturbation $\delta P$ in the $xy$-plane at different times. The
  simulation, nonetheless, enters the cyclical phase during which this axisymmetry is broken.
  As expected there is a flip in the sign of the Reynolds stress
  from negative to positive. The behavior thereafter mirrors that of
  the simulation of forced 
compressible convection run with the Roe solver: 
 the cyclical creation and destruction of large scale convective cells
  and an oscillatory Reynolds stress.
  We conclude that for forced runs, the ultimate quasi-steady state
  depends negligibly on the algorithm.

\subsection{Zonal Flows}
\label{vorticesandzonalflows}
In Figure \ref{zonalflowsfigure1} we plot, side-by-side, space-time
diagrams in the $xt$-plane of the $yz$-averaged pressure perturbation
$\delta P(x, t) \equiv (\langle P \rangle_{yz} - \langle P \rangle) /
\langle P \rangle$ and of the $yz$-averaged perturbation to the
$y$-component of the velocity $\langle \delta u_y \rangle_{yz} (x,t)
\equiv \langle u_y - 1.5\,\Omega\,x \rangle_{yz}$. It is immediately
evident from Figure \ref{zonalflowsfigure1} that around the onset of
the cyclical phase (orbit 20) alternating streaks in $\delta P$ and
$\delta u_y$ begin to develop. Respectively, these mark alternating
bands (in $x$) of high ($\delta P > 0$)  
and low ($\delta P < 0$) pressure, and of super-Keplerian
($\delta u_y > 0$) and sub-Keplerian ($\delta u_y < 0$) flow. 

From Figure \ref{zonalflowsfigure1} it is clear that the pressure and
velocity perturbations are 90 degrees out of phase, which is
characteristic of \textit{zonal flows}. We hesitate, however, to claim
that these structures are in geostrophic balance (i.e. that
$\partial_x P' \sim\, \rho_0 \,\Omega\, \delta u_y$) because, as is
evident from the space-time diagrams, the flows are not stationary in
time but appear to fluctuate over about one orbit -- the same
timescale over which the large convective cells are created and
destroyed. Indeed the creation and destruction of the zonal flows
tracks precisely that of the large-scale convective cells, showing
clearly
that the two phenomena are connected.
The axisymmetry observed during the formation of large-scale
convective cells and zonal flows is consistent with the inward
transport of angular momentum observed while these structures remain
coherent:  axisymmetric convective modes dominate during the
 lifetime of these structures and transport angular momentum inwards.

In Figure \ref{zonalflowsfigure} (right-hand panels) we plot snapshots in the $xy$-plane taken during the cyclical phase of the perturbation of the $y$-component of the velocity $\delta u_y$ and of the perturbed pressure $\delta P$. The axisymmetric structure of the zonal flows is clearly visible in panels (b) and (d), as is the $\pi/2$ phase difference between the pressure perturbation $\delta P$ and the perturbed $y$-component of the velocity $\delta u_y$.

\subsection{Vortices}

Because the zonal flows consist of alternating bands of sub- and
super-Keplerian motion -- with strong shear between the bands -- we
expect that they could give rise to vortices via the Kelvin-Helmholtz
instability (modified by rotation and stratification). In Figure
\ref{zonalflowsfigure} we plot a snapshot in the $xy$-plane taken
during the cyclical phase of the the $z$-component of the vertical
component of vorticity $\omega_z$. Small but coherent anti-cyclonic blobs of vorticity are observed during the cyclical phase, and these occur precisely where the flow transitions from sub-Keplerian to super-Keplerian (i.e. at the edges of the zonal flows). Typically only anti-cyclonic vortices are observed, which is consistent with the fact that cyclonic vortices tend to be more unstable in Keplerian shear flows.

The vortices tend to be elongated at the start of the cyclical phase (around orbit 20), with an aspect ratio of about 0.4, and grow increasingly circular with time; in fact, once quasi-steady equilibrium has been reached at around orbit 41, many of the vortices have aspect ratios approaching unity. They appear to have limited three-dimensional extent, and we did not observe any vortices that remained coherent for depths exceeding half a scale-height.

\subsection{Discussion}
\label{Discussion}

It is tempting to link our results to the large-scale emergent structures in recent simulations of rotating hydrodynamic convection with uniform (rather than Keplerian) rotation and in the Boussinesq (rather than compressible) regime \citep{favierproctor, guervilly2014large}. Both \cite{favierproctor} and \cite{guervilly2014large} observe growth in the vertical component of the kinetic energy at high Rayleigh numbers ($\text{Ra} \sim 10^{7}-10^{9}$), as we do. In addition,  \cite{favierproctor} and \cite{guervilly2014large} notice the formation of depth-invariant large-scale vortices in the $xy$-plane of their simulations (i.e. in the plane perpendicular to the rotation axis).

There are both qualitative and quantitative differences between our
results and those of \cite{favierproctor} and
\cite{guervilly2014large}. Although the Rayleigh number of
our simulation with thermal relaxation was $\text{Ra} = 10^{9}$, which
is consistent with the Rayleigh numbers at which large-scale vortices
were observed in the simulations of  \cite{favierproctor} and
\cite{guervilly2014large}, we observe large-scale \textit{convective
  cells} (in the $xz$-plane) rather than large-scale vortices (in the
$xy$-plane). We, too, observe anti-cyclonic vortices in the
$xy$-plane, but these are small and do not appear to have any three-dimensional extent
compared to the depth invariant vortices observed in the Boussinesq
simulations with uniform rotation. It is possible that they 
are prevented from merging and thus growing in size
because of the turbulence associated with repeated destruction of the
large-scale convective cells. The strong shear might also inhibit their
growth.

Due to the cyclical nature of the large-scale convective cells, 
it is tempting
to link our results to the intermittent convection reported in
the MRI shearing box simulations of \cite{hirose2014convection} and \cite{coleman2018convection}). 
In our runs
the forcing is due to explicit thermal relaxation, whereas in the
runs of \cite{hirose2014convection} the forcing is
self-consistently provided by MRI heating and radiative cooling. Thus the
forcing, and in particular the time-scales associated with the
forcing are rather different. However, both
thermal relaxation and the MRI limit cycles are similar in that they
lead to a cyclical build up of heat and its subsequent purge through vertical
convective transport, with the role of opacity in the simulations of
\cite{hirose2014convection} being simply to modulate this
cycle. Thus our results demonstrate that strongly non-linear 
 hydrodynamic turbulent convection has a cyclical nature that might 
be generic, and that might therefore be robust to the inclusion of more realistic thermodynamics.

\section{Conclusions}
\label{conclusions}

Motivated by recent radiation magnetohydrodynamic shearing box simulations that indicate that an interaction between convection and the magnetorotational instability in dwarf novae can enhance angular momentum transport, we have studied the simpler case of purely hydrodynamic convection, both analytically and through three-dimensional, fully-compressible simulations in \textsc{PLUTO}.

For the linear phase of the instability, we find agreement between the
growth rates of axisymmetric modes calculated theoretically and those
measured in the simulations to within a percentage error of $<1\%$,
thus providing a useful
check on our \textsc{PLUTO} code. The linear eigenmodes are worth
examining not only to help understand the physical nature of
convection in disks, but also because
they may appear in some form during the nonlinear phase of the
evolution, especially on large-scales, and during intermittent or
cyclical convection. 

We then explored the nonlinear saturation of the instability, both
when convection is continually forced and when it is allowed to reshape the
background gradients so that it ultimately dies out. We focussed especially
on the old problem of whether
hydrodynamic convection in a disc leads to inward ($\alpha < 0$) or
outward ($\alpha > 0$) angular momentum transport. In both forced and
unforced convection we found $\alpha>0$ in general in the
nonlinear phase. These results were confirmed by a separate run using
the code \textsc{ATHENA}, but contradict the classical simulations of
SB96 who
reported inward transport in both cases using the code \textsc{ZEUS}.

This discrepancy reveals a set of unfortunate numerical difficulties that complicate
the simulation of convection in disks. These, in large
part, issue from the fact that the inviscid linear modes of convection
grow fastest on the shortest possible scales. Thus, no
matter the resolution, the nature of the code's grid dissipation
will always impact on the system's evolution, certainly in the linear phase
and possibly afterwards. We argue that a more diffusive 
numerical set-up, such as supplied by \textsc{ZEUS} at low resolution
or Riemann solvers such as HLL, impose an axisymmetry
on the flow which leads to generally inward transport of angular
momentum.
But, on the other hand, we suspect that less diffusive
solvers such as Roe and HLLC artificially alias shearing waves in the
linear phase of the evolution, leading to
spurious non-axisymmetric flow early in a simulation. Though
concerning, we believe this is only a problem in the low amplitude
linear phase because physical mode-mode interactions will dominate once the
perturbations achieve larger amplitudes. Nonetheless, 
shearing wave aliasing certainly deserves a separate study.

To properly dispense with these numerical issues one must add explicit viscosisty (and
thermal diffusivity), as this regularises the linear problem.   
 We find that at a Richardson number of $\text{Ri} \sim 0.05$, onset
 of convection is observed for a 
critical Rayleigh number $10^{5} < \text{Ra}_c \leq 10^{6}$. Just above
this value convection is largely axisymmetric and $\alpha<0$.
At a larger second critical Ra between $10^6$ and $10^7$, the sign of
$\alpha$ switches and the flow becomes more turbulent and
nonaxisymmetric. This sequence of states mirrors that simulated by
Lesur and Ogilvie (2010). At large (resolved) Rayleigh numbers, viscous
simulations are initially controlled by the axisymmetric modes; these
are then attacked by secondary shear instabilities in both the $xz$
and $xy$-planes, which break the axisymmetry and order of these
structures, leading to a more chaotic state. At lower Ra, viscosity
suppresses the non-axisymmetric shear instabilities and axisymmetry
is never broken. (At even lower Ra, convection never begins, of course.)

In forced convective runs,  rather than maintaining the convection by
a fixed heating source at the mid-plane, we instead
allowed the vertical equilibrium to relax to its initial, convectively
unstable, state.  Our thermal
relaxation is artificially imposed, but its overall effect is to mimic
the heating of the mid-plane and cooling of the corona due to physical
mechanisms that maintain the
convectively unstable entropy profile, such as the MRI and radiative losses present in the
simulations of \cite{hirose2014convection}.
We observed in the non-linear stage
now the formation of large-scale convective cells (similar in some
respects to elevator flow) that emerge and break down cyclically,
in addition to zonal flows and vortices. Although further checks are
required, it is tempting to link this cyclical convection to the
convective limit cycle observed in the radiative magnetohydrodynamic 
 simulations of \cite{hirose2014convection}, since both processes 
rely on the build-up and rapid evacuation of heat.

Despite our
demonstration that hydrodynamic convection can lead to positive stress
and outward transport of angular momentum, the fact remains that the
stresses are small (typically we
measured $\alpha \sim 10^{-6}-10^{-5}$). Having
said that, the
magnitude of $\alpha$ is sensitive to the depth of the buoyancy
frequency profile, and a deeper profile could increase $\alpha$ by an order
of magnitude or more.

Finally we have not observed self-sustaining hydrodynamic convection
in any of our simulations. By self-sustaining convection we mean
that (when $\alpha > 0$) energy extracted from the shear by convection
might itself cause convective
motions, which in turn extract energy energy from the shear, closing
the loop. It is more likely that if convection is to occur in disks
it will be as a 
byproduct of other processes, such as heating by density waves,
emitted in the presence of a planet, or by dissipation of
magnetorotational turbulence. We intend to investigate both mechanisms
and their instigation of convection in future work.

\section*{Acknowledgements}
This work was funded by a Science and Technologies Facilities Council
(STFC) studentship. The authors acknowledge useful input
from John Papaloizou, Doug Lin, Jim Stone, and Steve Balbus. 
LEH  would like to thank Antoine Riols and William Bethune for stimulating conversations and advice on using the \textsc{PLUTO} code.



\bibliography{LEH_HydroConvectionPaper_FINALJULY302018} 

\begin{thebibliography}{}
\makeatletter
\relax
\def\mn@urlcharsother{\let\do\@makeother \do\$\do\&\do\#\do\^\do\_\do\%\do\~}
\def\mn@doi{\begingroup\mn@urlcharsother \@ifnextchar [ {\mn@doi@}
  {\mn@doi@[]}}
\def\mn@doi@[#1]#2{\def\@tempa{#1}\ifx\@tempa\@empty \href
  {http://dx.doi.org/#2} {doi:#2}\else \href {http://dx.doi.org/#2} {#1}\fi
  \endgroup}
\def\mn@eprint#1#2{\mn@eprint@#1:#2::\@nil}
\def\mn@eprint@arXiv#1{\href {http://arxiv.org/abs/#1} {{\tt arXiv:#1}}}
\def\mn@eprint@dblp#1{\href {http://dblp.uni-trier.de/rec/bibtex/#1.xml}
  {dblp:#1}}
\def\mn@eprint@#1:#2:#3:#4\@nil{\def\@tempa {#1}\def\@tempb {#2}\def\@tempc
  {#3}\ifx \@tempc \@empty \let \@tempc \@tempb \let \@tempb \@tempa \fi \ifx
  \@tempb \@empty \def\@tempb {arXiv}\fi \@ifundefined
  {mn@eprint@\@tempb}{\@tempb:\@tempc}{\expandafter \expandafter \csname
  mn@eprint@\@tempb\endcsname \expandafter{\@tempc}}}

\bibitem[\protect\citeauthoryear{Alexiades, Amiez  \& Gremaud}{Alexiades
  et~al.}{1996}]{alexiades1996super}
Alexiades V.,  Amiez G.,   Gremaud P.-A.,  1996, Communications in numerical
  methods in engineering, 12, 31

\bibitem[\protect\citeauthoryear{Armitage}{Armitage}{2011}]{armitage2011dynamics}
Armitage P.~J.,  2011, Annual Review of Astronomy and Astrophysics, 49, 195

\bibitem[\protect\citeauthoryear{Aurnou, Calkins, Cheng, Julien, King, Nieves,
  Soderlund  \& Stellmach}{Aurnou et~al.}{2015}]{aurnou2015rotating}
Aurnou J.,  Calkins M.,  Cheng J.,  Julien K.,  King E.,  Nieves D.,  Soderlund
  K.,   Stellmach S.,  2015, Physics of the Earth and Planetary Interiors, 246,
  52

\bibitem[\protect\citeauthoryear{Bell \& Lin}{Bell \& Lin}{1994}]{bell427jun}
Bell K.,  Lin D.,  1994, ApJ, 427, 987

\bibitem[\protect\citeauthoryear{Bodo, Cattaneo, Mignone  \& Rossi}{Bodo
  et~al.}{2013}]{bodo2013fully}
Bodo G.,  Cattaneo F.,  Mignone A.,   Rossi P.,  2013, The Astrophysical
  Journal Letters, 771, L23

\bibitem[\protect\citeauthoryear{Boley \& Durisen}{Boley \&
  Durisen}{2006}]{boley2006hydraulic}
Boley A.~C.,  Durisen R.,  2006, The Astrophysical Journal, 641, 534

\bibitem[\protect\citeauthoryear{Boyd}{Boyd}{2001}]{boyd2001chebyshev}
Boyd J.~P.,  2001, Chebyshev and Fourier spectral methods.
Courier Corporation

\bibitem[\protect\citeauthoryear{Cabot}{Cabot}{1996}]{cabot1996numerical}
Cabot W.,  1996, The Astrophysical Journal, 465, 874

\bibitem[\protect\citeauthoryear{Cannizzo}{Cannizzo}{2011}]{2011cannizzo}
Cannizzo J.~K.,  2011, The Limit Cycle Instability In Dwarf Nova Accretion
  Disks.
World Scientific, pp 6--40, \mn@doi{10.1142/9789814350976_0002}, \url
  {http://www.worldscientific.com/doi/abs/10.1142/9789814350976_0002}

\bibitem[\protect\citeauthoryear{Chiang \& Goldreich}{Chiang \&
  Goldreich}{1997}]{chiang1997spectral}
Chiang E.,  Goldreich P.,  1997, The Astrophysical Journal, 490, 368

\bibitem[\protect\citeauthoryear{Coleman, Blaes, Hirose  \& Hauschildt}{Coleman
  et~al.}{2018}]{coleman2018convection}
Coleman M.~S.,  Blaes O.,  Hirose S.,   Hauschildt P.~H.,  2018, The
  Astrophysical Journal, 857, 52

\bibitem[\protect\citeauthoryear{D'Alessio, Cant{\"o}, Calvet  \&
  Lizano}{D'Alessio et~al.}{1998}]{d1998accretion}
D'Alessio P.,  Cant{\"o} J.,  Calvet N.,   Lizano S.,  1998, The Astrophysical
  Journal, 500, 411

\bibitem[\protect\citeauthoryear{Favier, Silvers  \& Proctor}{Favier
  et~al.}{2014}]{favierproctor}
Favier B.,  Silvers L.~J.,   Proctor M. R.~E.,  2014, \mn@doi [Physics of
  Fluids] {10.1063/1.4895131}, 26, 096605

\bibitem[\protect\citeauthoryear{Goldreich \& Lynden-Bell}{Goldreich \&
  Lynden-Bell}{1965}]{goldreichlyndenbell1965}
Goldreich P.,  Lynden-Bell D.,  1965, \mn@doi [Monthly Notices of the Royal
  Astronomical Society] {10.1093/mnras/130.2.125}, 130, 125

\bibitem[\protect\citeauthoryear{Guervilly, Hughes  \& Jones}{Guervilly
  et~al.}{2014}]{guervilly2014large}
Guervilly C.,  Hughes D.~W.,   Jones C.~A.,  2014, Journal of Fluid Mechanics,
  758, 407

\bibitem[\protect\citeauthoryear{Hawley, Gammie  \& Balbus}{Hawley
  et~al.}{1995}]{hawley1995local}
Hawley J.~F.,  Gammie C.~F.,   Balbus S.~A.,  1995, The Astrophysical Journal,
  440, 742

\bibitem[\protect\citeauthoryear{Hirose}{Hirose}{2015}]{hirose2015magnetic}
Hirose S.,  2015, Monthly Notices of the Royal Astronomical Society, 448, 3105

\bibitem[\protect\citeauthoryear{Hirose, Blaes, Krolik, Coleman  \&
  Sano}{Hirose et~al.}{2014}]{hirose2014convection}
Hirose S.,  Blaes O.,  Krolik J.~H.,  Coleman M.~S.,   Sano T.,  2014, The
  Astrophysical Journal, 787, 1

\bibitem[\protect\citeauthoryear{Klahr, Henning  \& Kley}{Klahr
  et~al.}{1999}]{klahr1999azimuthal}
Klahr H.,  Henning T.,   Kley W.,  1999, The Astrophysical Journal, 514, 325

\bibitem[\protect\citeauthoryear{Kley, Papaloizou  \& Lin}{Kley
  et~al.}{1993}]{kley1993angular}
Kley W.,  Papaloizou J.,   Lin D.,  1993, The Astrophysical Journal, 416, 679

\bibitem[\protect\citeauthoryear{Latter \& Papaloizou}{Latter \&
  Papaloizou}{2017}]{latter2017local}
Latter H.~N.,  Papaloizou J.,  2017, Monthly Notices of the Royal Astronomical
  Society

\bibitem[\protect\citeauthoryear{Lesur \& Ogilvie}{Lesur \&
  Ogilvie}{2010}]{lesur2010angular}
Lesur G.,  Ogilvie G.~I.,  2010, Monthly Notices of the Royal Astronomical
  Society: Letters, 404, L64

\bibitem[\protect\citeauthoryear{Lin \& Papaloizou}{Lin \&
  Papaloizou}{1980}]{lin1980structure}
Lin D.,  Papaloizou J.,  1980, Monthly Notices of the Royal Astronomical
  Society, 191, 37

\bibitem[\protect\citeauthoryear{Lin, Papaloizou  \& Kley}{Lin
  et~al.}{1993}]{lin1993nonaxisymmetric}
Lin D.,  Papaloizou J.,   Kley W.,  1993, The Astrophysical Journal, 416, 689

\bibitem[\protect\citeauthoryear{Lyra, Richert, Boley, Turner, Mac~Low, Okuzumi
   \& Flock}{Lyra et~al.}{2016}]{lyra2016shocks}
Lyra W.,  Richert A.~J.,  Boley A.,  Turner N.,  Mac~Low M.-M.,  Okuzumi S.,
  Flock M.,  2016, The Astrophysical Journal, 817, 102

\bibitem[\protect\citeauthoryear{Mignone, Bodo, Massaglia, Matsakos, Tesileanu,
  Zanni  \& Ferrari}{Mignone et~al.}{2007}]{mignone2007pluto}
Mignone A.,  Bodo G.,  Massaglia S.,  Matsakos T.,  Tesileanu O.,  Zanni C.,
  Ferrari A.,  2007, The Astrophysical Journal Supplement Series, 170, 228

\bibitem[\protect\citeauthoryear{Mignone, Flock, Stute, Kolb  \&
  Muscianisi}{Mignone et~al.}{2012}]{mignone2012conservative}
Mignone A.,  Flock M.,  Stute M.,  Kolb S.,   Muscianisi G.,  2012, Astronomy
  \& Astrophysics, 545, A152

\bibitem[\protect\citeauthoryear{Ruden, Papaloizou  \& Lin}{Ruden
  et~al.}{1988}]{ruden1988axisymmetric}
Ruden S.~P.,  Papaloizou J.,   Lin D.,  1988, The Astrophysical Journal, 329,
  739

\bibitem[\protect\citeauthoryear{Ryu \& Goodman}{Ryu \&
  Goodman}{1992}]{ryu1992convective}
Ryu D.,  Goodman J.,  1992, The Astrophysical Journal, 388, 438

\bibitem[\protect\citeauthoryear{Spruit, Nordlund  \& Title}{Spruit
  et~al.}{1990}]{spruit1990solar}
Spruit H.~C.,  Nordlund A.,   Title A.,  1990, Annual review of astronomy and
  astrophysics, 28, 263

\bibitem[\protect\citeauthoryear{Stone \& Balbus}{Stone \&
  Balbus}{1996}]{stone1996angular}
Stone J.~M.,  Balbus S.~A.,  1996, The Astrophysical Journal, 464, 364

\bibitem[\protect\citeauthoryear{Stone \& Gardiner}{Stone \&
  Gardiner}{2010}]{stone2010implementation}
Stone J.~M.,  Gardiner T.~A.,  2010, The Astrophysical Journal Supplement
  Series, 189, 142

\bibitem[\protect\citeauthoryear{Stone, Gardiner, Teuben, Hawley  \&
  Simon}{Stone et~al.}{2008}]{stone2008athena}
Stone J.~M.,  Gardiner T.~A.,  Teuben P.,  Hawley J.~F.,   Simon J.~B.,  2008,
  The Astrophysical Journal Supplement Series, 178, 137

\bibitem[\protect\citeauthoryear{Zingale et~al.,}{Zingale
  et~al.}{2002}]{zingale2002mapping}
Zingale M.,  et~al., 2002, The Astrophysical Journal Supplement Series, 143,
  539

\makeatother
\end{thebibliography}
\bibliographystyle{mnras}


\appendix

\section{Convectively unstable vertical disc profiles}
\label{verticaldiskprofiles}
We describe the two convectively unstable vertical
profiles that are used to initialize our simulations. The reader should
note that these profiles may or may not correspond to those in \textit{real}
astrophysical discs, which will be determined by several sources of
heating and cooling, none considered in this paper. Here we simply present
convectively unstable disc profiles that satisfy hydrostatic
equilibrium, are convectively unstable, and can conveniently initialize simulations.

\subsection{Stone and Balbus (1996) profile}
\label{SB96profile}
The SB96 profile employs a power law profile in
temperature, see \ Eq.~\eqref{Tpower}. For $p = 3/2$, the density can be obtained analytically
$$\rho= \rho_0 (1 - s^3)^{-(1+g/3)}(1-s)^g\exp{\left\{2g\left[s - \frac{1}{\sqrt{3}}\text{tan}^{-1}\left(\frac{\sqrt{3} s}{s+2)}\right)\right]\right\}}
$$
where $\rho_0$ is the midplane density,  $s = (z/H)^{1/2} f^{1/3}$,
and $g = f^{-4/3}$, in which $f=H^{3/2}A/T_0$, and $H=c_s/\Omega$ is
the midplane scaleheight. The pressure is
obtained from the ideal gas equation of state.
Note that for $|z|> (T_0/A)^{1/p}$ we must have vacuum, which ties the
maximum numerical domain to the ratio $T_0/A$.

The buoyancy frequency for $p=3/2$ is given by
\begin{equation}
\frac{N_{B}^2}{\Omega^2} = \frac{1}{1-\left(\frac{z}{H}\right)^{3/2}f} \left[\left(1-\frac{1}{\gamma}\right)\left(\frac{z}{H}\right)^2 - \frac{3}{2}\left(\frac{z}{H}\right)^{3/2}f\right]
\label{SB96profile5}
\end{equation}
which corresponds to a profile that is negative (convectively unstable) within some region $|L_c| < 0$ about the mid plane and positive (convectively unstable) outside of this region. The width of the convectively unstable region is given by
\begin{equation}
\lvert L_c \rvert= \left[\frac{3}{2}\left(1-\frac{1}{\gamma}\right)^{-1} H^2 \frac{A}{T_0}\right]^2.
\label{SB96profile6}
\end{equation}

 Although convenient within a limited choice of parameters, the SB96
 profile suffers from the drawback that  the width of the convectively
 unstable region is sensitive to the size of the box through the ratio
 $T_0 / A$. Increasing the vertical box size necessarily decreases the size of the convectively unstable region.

\subsection{Gaussian temperature profile}
\label{gaussiantempprofile}
The drawbacks of the SB96 profile motivated us to search for a more
convenient unstable profile,
one that would leave the size and depth of the convectively unstable
region independent of the vertical extent of the box. Setting
the temperature to a Gaussian, cf.\ Eq.~\eqref{Tgauss} provided such a
profile.

 The associated density is
\begin{equation}
\rho = \rho_0 e^{z^2/\beta H^2} \text{exp}\left(-\frac{\beta H^2\Omega^2}{2T_0} e^{z^2/\beta H^2}\right),
\label{gausstempdensity}
\end{equation}
where $\rho_0$ is midplane density and $H$ the midplane scale height. 
Pressure is obtained from the
ideal gas equation of state, as above. 

The buoyancy frequency is given by 
\begin{equation}
\frac{N_B^2}{\Omega^2} = \frac{2}{\beta H^2} z^2 \left[\left((1-\frac{1}{\gamma}\right) \frac{\beta H^2 \Omega^2\mu}{2 T_0\mathcal{R}} e^{z^2/\beta H^2} - 1\right].
\label{gausstempbuoyancy}
\end{equation}
The boundary of the convectively unstable region is hence given by
\begin{equation}
\lvert L_c \rvert \equiv \sqrt{\beta H^2 \ln{\left[\left(1-\frac{1}{\gamma}\right)^{-1} \frac{2 T_0\mathcal{R}}{\beta\mu H^2 \Omega^2}\right]}}.
\label{LcGaussianTemp}
\end{equation}


\section{Tables of simulations}
\label{tables}
\begin{table*}
	\centering
	\caption{Simulations of unforced compressible hydrodynamic convection in the shearing box carried out in \textsc{PLUTO}. No explicit viscosity or heating/cooling/thermal relaxation were included. The box size in all simulations is $4H\times4H\times4H$. The simulations were initialized either with a Gaussian temperature convectively unstable vertical disc profile (see Appendix \protect\ref{gaussiantempprofile}), or with the convectively unstable vertical disc profile used in \protect\cite{stone1996angular} (SB96; see Appendix \protect\ref{SB96profile}). $\langle \langle . \rangle \rangle_t$ denotes a time- and volume-averaged quantity. $E_\text{kin, $z$}$ is the vertical kinetic energy density, which we have measured both at non-linear saturation (where it is maximum) and at the end of the simulation. $\Delta \rho$ is the percentage change in density over the course of the simulation, $R_{xy}$ is the Reynolds stress, $\alpha$ is the alpha viscosity parameter, and $\sigma$ is the linear phase growth rate given in units of $\Omega$. All simulations were run for 38 orbits.}
		\label{fiducialKERxy}
			\begin{tabular}{lcccccccr}
				\hline
				Run	&	Resolution	&	Profile	& max$\left(\langle E_\text{kin, $z$}\rangle\right)$ & $\langle E_\text{kin, $z$}\rangle |_{\text{end}}$  &  $\Delta\left(\langle \rho \rangle\right)\,(\%)$	&$\langle\langle R_{xy}\rangle\rangle_t$	&	$\langle\langle\alpha\rangle\rangle_t$	 & $\sigma\,(\Omega)$\\
				\hline
				NSTR22c01	&	$64^{3}$	&	Gaussian Temp.	& $1.7\times10^{-4}$	&	$2.0\times10^{-6}$	& $-1.4$ & $+4.8\times10^{-6}$ & $+2.3\times10^{-5}$& 0.2404\\
				NSTR22c02	&	$128^{3}$	&	Gaussian Temp.	& $1.1\times10^{-4}$	&	$3.7\times10^{-6}$    & $-3.4$& $+2.7\times10^{-6}$ & $+1.5\times10^{-5}$& 0.2707\\
				NSTR22c03	&	$256^{3}$	&	Gaussian Temp.	& $7.3\times10^{-5}$	&	$9.8\times10^{-6}$	& $-8.8$& $+1.7\times10^{-6}$ & $+9.6\times10^{-6}$& 0.2853\\
				NSTR22c04	&	$512^3$   &	Gaussian Temp.	& $6.4\times10^{-5}$&	$1.9\times10^{-5}$	& $-12.7$& $+1.3\times10^{-6}$ & $+7.1\times10^{-6}$& 0.2811\\
				\hline
				NSTR21c01	&	$64^3$	&	SB96			& $9.4\times10^{-4}$&	$9.0\times10^{-6}$	& $-2.1$ & $+1.9\times10^{-5}$ & $+2.4\times10^{-5}$ & 0.2059\\
				NSTR21c03	&	$256^3$	&	SB96			& $3.6\times10^{-6}$&	$3.9\times10^{-5}$	& $-10.3$ & $+6.0\times10^{-6}$ & $+7.7\times10^{-6}$ & 0.2424\\
				\hline
			\end{tabular}
\end{table*}

\begin{table*}
	\centering
	\caption{Simulations of unforced compressible hydrodynamic convection in the shearing box with different numerical schemes in \text{PLUTO}. No explicit viscosity or heating/cooling/thermal relaxation were included. All simulations were run in a box of size $4H\times4H\times4H$, and initialized with the Gaussian temperature profile discussed in Appendix \protect\ref{gaussiantempprofile}. The time-averages of the volume-averaged Reynolds stress $\langle R_{xy} \rangle$ and of the volume-averaged angular momentum transport parameter $\langle \alpha \rangle$ have been taken over an interval spanning non-linear saturation. All simulations were run for 38 orbits.}
	\label{tablenumericalschemes}
	\begin{tabular}{lcccccccr} 
		\hline
		Run & Resolution &  Solver & Interpolation & Timestepping & $\langle\langle R_{xy}\rangle\rangle_t$& $\langle\langle\alpha\rangle\rangle_t$& Comments &\\
		\hline
		NSTR22e01a & $64^{3}$ &  HLLC & WENO3 & RK3 & $+4.5\times10^{-6}$ & $+2.5\times10^{-5}$\\
		NSTR22e01c & $256^{3}$ &  HLLC & WENO3 & RK3 & $+1.8\times10^{-6}$ & $+1.0\times10^{-5}$\\
		NSTR22e02a & $64^{3}$ &  HLL & WENO3 & RK3 &  $-1.3\times10^{-7}$ & $-7.4\times10^{-7}$& $|\delta \mathbf{u}| = \mathcal{O}(10^{-3})$\\
		NSTR22e02c & $256^{3}$ &  HLL & WENO3 & RK3 &  $-1.1\times10^{-5}$ & $-6.1\times10^{-5}$& \\
		NSTR22e03a & $64^{3}$ & TVDLF & WENO3 & RK3 & - & -& No instability.\\
		\hline
		NSTR22e11a & $64^{3}$ & HLLC & Linear TVD & Hancock & $+3.4\times10^{-6}$ & $+2.2\times10^{-5}$\\
		NSTR22e11i & $256^{3}$ & HLLC & Linear TVD & Hancock & $+1.5\times10^{-6}$ & $+9.2\times10^{-6}$& UMIST limiter\\
		NSTR22e12a &$64^{3}$ & HLL & Linear TVD & Hancock & $-2.2\times10^{-5}$   &  $-1.4\times10^{-4}$\\
		NSTR22e12b & $128^{3}$ & HLL & Linear TVD & Hancock &  $-1.6\times10^{-5}$ & $-1.0\times10^{-4}$\\
		NSTR22e12c & $256^{3}$ & HLL & Linear TVD & Hancock &  $-9.5\times10^{-6}$ &$-6.2\times10^{-5}$\\
		NSTR22e12i & $256^{3}$ & HLL & Linear TVD & Hancock &  $-1.2\times10^{-5}$ &$-8.2\times10^{-5}$& UMIST limiter\\
		NSTR22e13a & $64^{3}$ & TVDLF & Linear TVD & Hancock & $-3.5\times10^{-5}$ & $-2.3\times10^{-4}$\\
		NSTR22e13b & $128^{3}$ & TVDLF & Linear TVD & Hancock & $-1.4\times10^{-5}$  & $-9.4\times10^{-5}$\\
		\hline
		NSTR22e21a & $64^{3}$ & HLLC & Linear TVD & RK2 & $+4.0\times10^{-6}$ & $+2.6\times10^{-5}$\\
		NSTR22e22a & $64^{3}$& HLL & Linear TVD & RK2 & $-2.2\times10^{-5}$ & $-1.3\times10^{-4}$\\
		NSTR22e22b & $128^{3}$ & HLL & Linear TVD & RK2 & $-2.0\times10^{-5}$  & $-1.2\times10^{-4}$\\
		NSTR22e23a & $64^{3}$ & TVDLF & Linear TVD & RK2 & $-1.9\times10^{-5}$ & $-1.2\times10^{-4}$\\
		NSTR22e23b & $128^{3}$ & TVDLF & Linear TVD & RK2 & $-1.3\times10^{-5}$ & $-7.6\times10^{-4}$\\

		\hline
	\end{tabular}
\end{table*}

\begin{table*}
	\centering
	\caption{Simulations of unforced compressible hydrodynamic convection in the shearing box with explicit kinematic viscosity $\nu$ and thermal diffusivity $\chi$ included. No thermal relaxation was included. All simulations were run in a box of size $4H\times4H\times4H$. $\langle\langle\alpha\rangle\rangle_t|_{\text{linear}}$ is the volume-averaged alpha viscosity parameter time-averaged over the linear phase, $\langle\langle\alpha\rangle\rangle_t|_{\text{NL}}$ is the same quantity time-averaged over the non-linear phase, and $\text{min}\left(\langle R_{xy} \rangle \right)$ is the minimum value of the $xy$-component of the volume-averaged Reynolds stress.}
		\label{simsexplicitdiffusioncoeffs}
			\begin{tabular}{lcccccccccr}
				\hline
				Run	& Resolution & Instability? & $\nu$ & $\chi$ & Ra & Ri & Pr & $\langle\langle\alpha\rangle\rangle_t|_{\text{linear}}$ & $\langle\langle\alpha\rangle\rangle_t|_{\text{NL}}$ & $\text{min}\left(\langle R_{xy} \rangle \right)$\\
				\hline
				NSTR22Ra1     & $256^3$  & N & $1.075\times10^{-3}$ &  $4.300\times10^{-4}$  & $10^{5}$   & 0.05 & 2.5 & - & - & -\\
				NSTR22Ra2     & $256^3$  & Y & $3.400\times10^{-4}$ & $1.360\times10^{-4}$ & $10^{6}$   & 0.05 & 2.5 & $-5.1\times10^{-6}$  &$-3.9\times10^{-5}$ & $-2.1\times10^{-5}$ \\
				NSTR22Ra3     & $256^3$  & Y & $1.075\times10^{-4}$ &  $4.300\times10^{-5}$ & $10^{7}$   & 0.05 & 2.5 & $-3.5\times10^{-6}$  &$+3.9\times10^{-6}$ & $-2.7\times10^{-5}$ \\
				NSTR22Ra4a     & $256^3$  & Y & $3.400\times10^{-5}$ & $1.360\times10^{-5}$ & $10^{8}$   & 0.05 & 2.5 & $-1.0\times10^{-6}$  &$+1.4\times10^{-5}$ & $-1.1\times10^{-5}$ \\
				NSTR22Ra4b     & $512^3$  & Y & $3.400\times10^{-5}$ & $1.360\times10^{-5}$ & $10^{8}$   & 0.05 & 2.5 & $-2.8\times10^{-6}$  &$+7.1\times10^{-6}$ & $-1.4\times10^{-5}$ \\
				NSTR22Ra5a     & $256^3$  & Y & $1.075\times10^{-5}$ &  $4.300\times10^{-6}$ & $10^{9}$   & 0.05 & 2.5 & $-3.5\times10^{-8}$  &$+1.5\times10^{-5}$ & $-9.9\times10^{-8}$ \\
				NSTR22Ra5b & $512^3$  & Y & $1.075\times10^{-5}$ &  $4.300\times10^{-6}$ & $10^{9}$   & 0.05 & 2.5 & $-1.6\times10^{-7}$  &$+1.3\times10^{-5}$ & $-3.4\times10^{-6}$ \\
				NSTR22Ra6a    & $256^3$  & Y & $3.400\times10^{-6}$ & $1.360\times10^{-6}$ & $10^{10}$ & 0.05 & 2.5 & $+7.6\times10^{-8}$  &$+1.5\times10^{-5}$ & $+2.2\times10^{-13}$ \\
				NSTR22Ra6b & $512^3$  & Y & $3.400\times10^{-6}$  & $1.360\times10^{-6}$  & $10^{10}$ & 0.05 & 2.5 & $-1.1\times10^{-8}$  & $+1.6\times10^{-5}$ & $-4.9\times10^{-8}$ \\
				\hline
				\hline
			\end{tabular}
\end{table*}


\bsp	
\label{lastpage}
\end{document}